\documentclass[twocolumn, times,tighten]{aastex62} 

\def\lsol       {\ensuremath{\mathrm{L}_{\odot}}}
\def\msol       {\ensuremath{\mathrm{M}_{\odot}}}
\def\msolyr     {\ensuremath{\mathrm{M}_{\odot}\,\mathrm{yr}^{-1}}}

\def\mujybeam {\ensuremath{\mu\mathrm{Jy}\,\mathrm{beam}^{-1}}}

\def\jykms {\ensuremath{\text{Jy\,km\,s}^{-1}}}
\def\kms {\ensuremath{\text{km\,s}^{-1}}}
\def\jybeam {\ensuremath{\text{Jy\,beam}^{-1}}}

\usepackage[caption=false]{subfig}
\usepackage{enumitem}

\shortauthors{Novak et al.}

\graphicspath{{./}{figures/}}

\def\cii     {\ensuremath{\text{\textsc{[C ii]}}}}

\begin{document}

\title{\Large \bf No evidence for \cii\ halos or high-velocity outflows in {\boldmath$z \gtrsim 6$} quasar host galaxies}

\correspondingauthor{Mladen Novak}
\email{novak@mpia.de}

\author[0000-0001-8695-825X]{Mladen Novak}
\affiliation{Max-Planck-Institut f\"{u}r Astronomie, K\"{o}nigstuhl 17, D-69117 Heidelberg, Germany}

\author[0000-0001-9024-8322]{Bram P. Venemans}
\affiliation{Max-Planck-Institut f\"{u}r Astronomie, K\"{o}nigstuhl 17, D-69117 Heidelberg, Germany}

\author[0000-0003-4793-7880]{Fabian Walter}
\affiliation{Max-Planck-Institut f\"{u}r Astronomie, K\"{o}nigstuhl 17, D-69117 Heidelberg, Germany}
\affiliation{National Radio Astronomy Observatory, Pete V. Domenici Array Science Center, P.O. Box O, Socorro, NM 87801, USA}

\author[0000-0002-9838-8191]{Marcel Neeleman}
\affiliation{Max-Planck-Institut f\"{u}r Astronomie, K\"{o}nigstuhl 17, D-69117 Heidelberg, Germany}

\author[0000-0002-1173-2579]{Melanie Kaasinen}
\affiliation{Max-Planck-Institut f\"{u}r Astronomie, K\"{o}nigstuhl 17, D-69117 Heidelberg, Germany}
\affiliation{Universit\"{a}t Heidelberg, Zentrum f\"{u}r Astronomie, Institut f\"{u}r Theoretische Astrophysik, Albert-Ueberle-Stra\ss{}e 2, D-69120 Heidelberg, Germany}

\author[0000-0001-9422-0095]{Lichen Liang}
\affiliation{Institute for Computational Science, University of Zurich, Zurich CH-8057, Switzerland}

\author[0000-0002-1109-1919]{Robert Feldmann}
\affiliation{Institute for Computational Science, University of Zurich, Zurich CH-8057, Switzerland}

\author[0000-0002-2931-7824]{Eduardo Ba\~nados}
\affiliation{Max-Planck-Institut f\"{u}r Astronomie, K\"{o}nigstuhl 17, D-69117 Heidelberg, Germany}

\author[0000-0001-6647-3861]{Chris Carilli}
\affil{National Radio Astronomy Observatory, Pete V. Domenici Array Science Center, P.O. Box O, Socorro, NM 87801, USA}

\author[0000-0002-2662-8803]{Roberto Decarli}
\affiliation{INAF - Osservatorio di Astrofisica e Scienza dello Spazio, via Gobetti 93/3, I-40129, Bologna, Italy}

\author[0000-0002-0174-3362]{Alyssa B. Drake}
\affiliation{Max-Planck-Institut f\"{u}r Astronomie, K\"{o}nigstuhl 17, D-69117 Heidelberg, Germany}

\author[0000-0003-3310-0131]{Xiaohui Fan}
\affiliation{Steward Observatory, The University of Arizona, 933 N.\ Cherry Ave., Tucson, AZ 85721, USA}

\author[0000-0002-6822-2254]{Emanuele P. Farina}
\affiliation{Max Planck Institut f\"ur Astrophysik, Karl--Schwarzschild--Stra{\ss}e 1, D-85748, Garching bei M\"unchen, Germany}

\author[0000-0002-5941-5214]{Chiara Mazzucchelli}
\affiliation{European Southern Observatory, Alonso de C\'ordova 3107, Vitacura, Regi\'on Metropolitana, Chile}

\author[0000-0003-4996-9069]{Hans--Walter Rix}
\affiliation{Max-Planck-Institut f\"{u}r Astronomie, K\"{o}nigstuhl 17, D-69117 Heidelberg, Germany}

\author[0000-0003-4956-5742]{Ran Wang}
\affiliation{Kavli Institute for Astronomy and Astrophysics, Peking University, Beijing 100871, People’s Republic of China}

\begin{abstract}
We study the interstellar medium in a sample of 27 high-redshift quasar host galaxies at $z\gtrsim6$, using the \cii\ 158\,$\mu$m emission line and the underlying dust continuum observed at $\sim1$\,kpc resolution with ALMA. By performing $uv$-plane spectral stacking of both the high  and  low spatial resolution data, we investigate the spatial and velocity extent of gas, and the size of the dust-emitting regions.
We find that the average surface brightness profile of both the \cii\ and the dust continuum emission can be described by a steep component within  a radius of 2\,kpc, and a shallower component with a scale length of 2\,kpc, detected up to $\sim$10\,kpc. 
The surface brightness of the extended emission drops below $\sim$1\% of the peak at radius of $\sim5$\,kpc, beyond which it constitutes 10 -- 20\% of the total measured flux density.
Although the central component of the dust continuum emission is more compact than that of the \cii\ emission, the extended components have equivalent profiles. The observed extended components are consistent with those predicted by hydrodynamical simulations of galaxies with similar infrared luminosities, where the dust emission is powered by star formation.
The \cii\ spectrum measured in the  mean $uv$-plane stacked data can be described by a single Gaussian, with no observable \cii\ broad-line emission (velocities in excess of $\gtrsim 500$\,\kms), that would be indicative of outflows.
Our findings suggest that we are probing the interstellar medium and associated star formation in the quasar host galaxies up to radii of 10\,kpc,
whereas we find no evidence for halos or outflows.
\end{abstract}

\keywords{
	AGN host galaxies 
	--- High-redshift galaxies 
	--- Dust continuum emission 
	--- Interstellar line emission
}

\section{Introduction} \label{sec:intro}

Quasars and their hosts are ideal targets to probe the properties of massive high-redshift galaxies.
Powered by the rapid accretion of material, near the Eddington limit, onto a supermassive black hole \citep[SMBH, see e.g.,][]{derosa11, derosa14, willott10a}, 
quasars within the first Gyr of the Universe ($z>6$)  are easily detected by current facilities.
Several hundred quasars have now been identified at $z>5.5$, owing to
both large surveys \citep[see][]{york00, arnaboldi07,lawrence07,chambers16}, and improved selection methods with  follow-up observations \citep[see e.g.,][]{fan06,venemans07,willott10a, mortlock11, morganson12, banados16, jiang16, mazzucchelli17, matsuoka18, banados18,wangfeige18,yang20}.

At $z\sim6$, the singly ionized carbon emission line at 158\,$\mu$m, arising from the ${}^2P_{3/2} \rightarrow {}^2P_{1/2}$  transition (hereafter referred to as the \cii\ line) falls conveniently within the atmospheric transition window of 
ground based interferometers such as the Atacama Large Millimeter Array (ALMA) and the NOrthern Extended Millimeter Array (NOEMA). This emission line is one of the brightest far-infrared (FIR) emission lines  and can be used to trace the cold molecular gas of the interstellar medium (ISM, see \citealt{carilliwalter13} for a review of high-redshift galaxies). The line  predominantly arises from within the photodissociation regions (PDRs) found around newly formed stars, although it can also stem from ionized regions \cite[e.g.,][]{hollenbach99, vallini17, ferrara19}. 
The \cii\ line has been used to probe the gas distribution and kinematics on kpc scales for dozens of quasar host galaxies, simultaneously providing precise redshift measurements \cite[e.g.,][]{walter09, maiolino12, wang13, willott15, venemans17b, venemans17c, decarli18, neeleman19}.

Detections of large-scale \cii\ emission and outflows in a high-redshift quasar host galaxy were first reported for the $z=6.4$ system SDSS J1148+5251 \citep{maiolino12, cicone15}. However, further statistical studies of \cii\ outflows in various, partly overlapping, samples of $z>4.5$ quasar host galaxies remain without consensus.
Some studies report tentative or strong evidence for outflows \citep[e.g.,][]{bischetti19,stanley19}, whereas other studies reported no outflows \citep[e.g.,][]{decarli18}. However, different image plane stacking techniques were used among the different studies.

We  revisit this topic by using the deepest available ALMA data and novel analysis techniques.
In this study, we conduct a multi-resolution analysis of 27 quasar host galaxies at $z \gtrsim 6$ observed with ALMA in order to search for signatures of extended \cii\ emission and outflows.
We require $\sim1\,$kpc imaging capability in order to accurately derive surface brightness profiles, but complement our data with lower spatial resolution observations of the same sources to minimize potential issues of missing flux and outresolving sources. For the spectral analysis, we use a novel technique of spectral $uv$-plane stacking.
By directly averaging observed visibilities of the sample of galaxies, in velocity bins of interest, we circumvent various problems present in interferometric image-based stacking (e.g.\ dirty beam residuals and beam matching), simultaneously ensuring that emission on multiple spatial scales can be recovered in the imaging step.

This is the third paper in a series of studies, in which different aspects of the quasar host galaxies observed in \cii\ at $\sim1$\,kpc resolution are discussed. The first paper, \cite{venemans20}, describes the sample in detail and provides the analysis of both the \cii\ and the dust continuum emission of individual galaxies. The second paper, \cite{neeleman20b}, capitalizes on the high resolution of these observations by modeling the gas kinematics of the host galaxies, yielding rotation and/or dispersion velocities, and estimates of the dynamical masses of the hosts.
These two studies investigate properties of individual galaxies, making use of the brightest emitting regions, where there is sufficient signal. In this paper, we perform stacking and $uv$-plane analysis to constrain the amount of faint, but extended (both spatially and spectrally), emission, that is below the detection threshold of individual objects.

The paper is structured as follows. In Section~\ref{sec:data}, we present the sample of $z\gtrsim6$ quasar host galaxies along with the new and archival ALMA data. We also describe the data reduction steps, along with the details of the $uv$-stacking procedure.
In Section~\ref{sec:methods}, we elaborate on the methods for accurately measuring fluxes, both from the image, and the $uv$-plane.
In Section~\ref{sec:res}, we present the results in three subsections, focusing on the
1) spatial extent of the \cii\ emission, 2) spatial extent of the dust continuum, and 3) spectral analysis of the \cii\ line.
We interpret and discuss our findings in the context of published studies in Section~\ref{sec:disc}, and summarize our main conclusions in Section~\ref{sec:conc}.

Throughout the paper we assume the concordance Lambda cold dark matter ($\Lambda$CDM) cosmology, defined by a Hubble constant  $H_0=70$\,km\,s$^{-1}$\,Mpc$^{-1}$, dark energy density $\Omega_\Lambda=0.7$, and matter density $\Omega_{\mathrm{m}}=0.3$. 
The mean redshift of our sample is $z=6.4$, at which an angular size of 1$\arcsec$ corresponds to a projected physical distance of 5.5\,kpc.

\section{Data and samples}
\label{sec:data}

The main sample for our study consists of 27 quasar host galaxies\footnote{Although
we happen to have the same number of galaxies, our sample is not the same as the one described in \cite{decarli18}, as only 15 objects are shared between the two samples. For our study, we require  data obtained at high resolution ($\sim0\farcs25$), in contrast to the $\sim1\arcsec$ used in  \cite{decarli18}. } at redshifts $z \gtrsim 6$ observed with ALMA  in the \cii\ emission at  $\sim1$\,kpc resolution ($\sim0\farcs25$). With the high-resolution criteria satisfied, we also supplemented the data with low-resolution archival observations of these sources, where available.
	Our sample spans the redshift range of $z=5.8 - 7.5$, and contains quasar host galaxies with \cii\ luminosities in the range of $(0.8 - 9)\times10^{9}\,\lsol$, and FIR luminosities in the range of $(0.5 - 12)\times10^{12}\,\lsol$ \citep[see][for further details on the sample selection]{venemans20}. 
	
\subsection{New and archival ALMA data}
 \label{sec:datalist}

For our analysis, we consider all available 12\,m and 7\,m ALMA observations performed in cycles 1 through 6, which were obtained at a resolution of $0\farcs1$ or coarser. The bulk of the high spatial resolution data comes from our recent programs with IDs 2017.1.01301.S and 2018.1.00908.S.
We have excluded cycle 0 data from our selection to avoid possibly lower quality data and known difficulties\footnote{E.g., fewer antennas available, sometimes missing calibration data, no pipeline support.} of combining these observations with subsequent cycles.
We have also excluded higher resolution (sub-kpc) observations available for two of the quasar host galaxies, because there is insufficient overlap of available baseline lengths compared to our main data sample. 
Finally, we have excluded quasar host galaxies without  $\sim1$\,kpc observations from the analysis, in order to have a consistent sample across the paper series.
These selection criteria yield a main sample of 27 quasar host galaxies, where all objects, by construction, have $\sim1$\,kpc data available.

Although the high-resolution data resolve the gas and dust structure and the kinematics of the host galaxy, they may not be sensitive to large-scale emission, which may be either below the detection limit, or out-resolved due to the lack of shorter baselines.
In order to quantify this effect and mitigate the issue as much as possible, we make use of all available archival data targeting the \cii\ emission line in the galaxies of our sample. 
Datasets with lower spatial resolution observations, obtained across multiple ALMA observational cycles, are available for 20 objects in our sample.

In total, this work uses 54 different observations targeting 27 objects, as listed in Table~\ref{tab:data}, where all datasets correspond to 12\,m array observations, unless noted otherwise.
The redshift, \cii\ integrated line flux density, and the full width at half maximum (FWHM), reported in Table~\ref{tab:data}, are all measured from the aperture-extracted \cii\ spectra, whereas coordinates are measured from the \cii\ intensity maps, as reported in \cite{venemans20}. For each object we  list used observation runs in various ALMA cycles, including their time on source (TOS), the synthesized beam size,  and the maximum recoverable scale (MRS). The last two values are computed based on the baseline statistics following the ALMA Technical Handbook (Eqs. 7.4 and 7.7), i.e.\ $\theta_{\mathrm{beam}} [\mathrm{rad}]=0.574\lambda/L_{80}$ and $\theta_{\mathrm{MRS}} [\mathrm{rad}]=0.983\lambda/L_{5}$, where $\lambda$ is the observing wavelength, and $L_{5}$ and $L_{80}$ are the 5\textsuperscript{th} and the 80\textsuperscript{th} percentile of the available $uv$-distances, respectively. The MRS is defined as the largest angular size at which at least 10\% of the total flux density of a uniform disk is recovered and should only be used to describe the data limitations to first order.

\begin{table*}
	\centering
	\scriptsize
	\caption{ALMA observations of \cii\ emission in $z\sim6$ quasar host galaxies used in our study (27 objects, 54 datasets).}
	\label{tab:data}
	\begin{tabular}{cccccccccccc}
		\hline
		Quasar host & z$_{\text{[C\textsc{ii}]}}$ & RA (ICRS )& Dec (ICRS) & $F_{\cii}$ & FWHM & Cycle & TOS & Beam\footnote{Synthesized beam obtained from baseline statistics (see main text), given in arcsec. Nominal high-resolution ($\sim$1\,kpc) datasets are  marked with boldface.} & MRS & Project code & Member ObsUnitSet ID \\
		galaxy &  & $\mathrm{{}^{\circ}}$ & $\mathrm{{}^{\circ}}$ & $\mathrm{Jy\,km\,s^{-1}}$& $\mathrm{km\,s^{-1}}$ &  & $\mathrm{min}$ & $\mathrm{{}^{\prime\prime}}$ & $\mathrm{{}^{\prime\prime}}$ &  &  \\
		\hline
		P007+04 & 6.0015 & 7.02736 & 4.95706 &  1.7  $\pm$ 0.1  &  370  $\pm$ 22  & 3\textsuperscript{\ref{fn:recal}}  & 7.4 & 0.44 & 5.0 & 2015.1.01115.S & uid://A001/X2fb/X3b8 \\
		&  &  &  &  & & 5 & 29.1 & {\bf 0.23} & 3.2 & 2017.1.01301.S & uid://A001/X1273/X360 \\
		\hline
		P009--10 & 6.0040 & 9.73553 & -10.43168 &  9.6  $\pm$ 0.7  &  437  $\pm$ 33  & 3\textsuperscript{\ref{fn:recal}, \ref{fn:baduv}}  & 8.4 & 0.4 & 4.5 & 2015.1.01115.S & uid://A001/X2fb/X3bc \\
		&  &  &  &  & & 5 & 25.7 & {\bf 0.23} & 3.3 & 2017.1.01301.S & uid://A001/X1273/X364 \\
		\hline
		J0100+2802 & 6.3268 & 15.05426 & 28.04051 &  3.7  $\pm$ 0.2  &  405  $\pm$ 20  & 3\footnote{\label{fn:baduv}Removed from  uv  stacking due to  data weights outliers or partial line frequency coverage (in case of P009--10, cycle 3).} & 72.4 & {\bf 0.18} & 3.5 & 2015.1.00692.S & uid://A001/X2d6/X1a8 \\
		\hline
		J0109--3047 & 6.7904 & 17.47135 & -30.79065 &  1.7  $\pm$ 0.1  &  354  $\pm$ 34  & 1 & 15.4 & 0.27 & 4.0 & 2012.1.00882.S & uid://A002/X5a9a13/X53b \\
		&  &  &  &  & & 2\footnote{\label{fn:recal}Recalibrated due to wrong Ceres or Pallas flux calibrator models.} & 35.8 & 0.2 & 2.1 & 2013.1.00273.S & uid://A001/X148/X6b \\
		&  &  &  &  & & 3 & 32.4 & {\bf 0.16} & 2.9 & 2015.1.00399.S & uid://A001/X5a3/X52 \\
		\hline
		J0129--0035\footnote{\label{fn:c0}Cycle 0 observations are also available, but are excluded from our study.} & 5.7788 & 22.49381 & -0.59440 &  2.1  $\pm$ 0.1  &  206  $\pm$ 9  & 1\textsuperscript{\ref{fn:recal}, \ref{fn:baduv}} & 76.5 & {\bf 0.18} & 2.0 & 2012.1.00240.S & uid://A002/X7fb989/X1c \\
		&  &  &  &  & & 3  & 60.7 & 0.34 & 4.4 & 2015.1.00997.S & uid://A001/X2fb/X5de \\
		&  &  &  &  & & 3 - 7m & 209.4 & 4.0 & 25.0 & 2015.1.00997.S & uid://A001/X2fb/X5e0 \\
		\hline
		J025--33 & 6.3373 & 25.68218 & -33.46264 &  5.5  $\pm$ 0.2  &  370  $\pm$ 16  & 3 & 7.9 & 0.61 & 5.8 & 2015.1.01115.S & uid://A001/X2fb/X3c4 \\
		&  &  &  &  & & 5 & 24.2 & {\bf 0.23} & 3.2 & 2017.1.01301.S & uid://A001/X1273/X368 \\
		\hline
		P036+03 & 6.5405 & 36.50782 & 3.04979 &  3.2  $\pm$ 0.1  &  237  $\pm$ 7  & 3 & 75.5 & {\bf 0.13} & 1.6 & 2015.1.00399.S & uid://A001/X5a3/X4a \\
		\hline
		J0305--3150\footnote{\label{fn:comp}Merger or nearby companion identified at  $<$ 10 \,kpc.}\textsuperscript{,}\footnote{\label{fn:uberres}Higher resolution ($<$40 \,mas) also available, but discarded here (see main text).} & 6.6139 & 46.32052 & -31.84888 &  5.4  $\pm$ 0.3  &  225  $\pm$ 15  & 1 & 15.9 & 0.26 & 3.7 & 2012.1.00882.S & uid://A002/X5a9a13/X543 \\
		&  &  &  &  & & 2\textsuperscript{\ref{fn:recal}}  & 15.5 & 0.2 & 2.1 & 2013.1.00273.S & uid://A001/X148/X6f \\
		&  &  &  &  & & 3 & 37.7 & {\bf 0.17} & 3.0 & 2015.1.00399.S & uid://A001/X5a3/X4e \\
		\hline
		P065--26 & 6.1871 & 65.40851 & -26.95432 &  1.7  $\pm$ 0.2  &  289  $\pm$ 31  & 3 & 15.8 & 0.74 & 7.3 & 2015.1.01115.S & uid://A001/X2fb/X3e4 \\
		&  &  &  &  & & 5 & 24.7 & {\bf 0.23} & 3.3 & 2017.1.01301.S & uid://A001/X1273/X36c \\
		\hline	
		J0842+1218 & 6.0754 & 130.62266 & 12.31402 &  0.8  $\pm$ 0.1  &  378  $\pm$ 52  & 3 & 7.4 & 0.98 & 9.2 & 2015.1.01115.S & uid://A001/X2fb/X3ec \\
		&  &  &  &  & & 4 & 53.6 & {\bf 0.25} & 2.7 & 2016.1.00544.S & uid://A001/X885/X36b \\
		\hline
		J1044--0125\textsuperscript{\ref{fn:c0}} & 5.7846 & 161.13767 & -1.41724 &  1.8  $\pm$ 0.2  &  454  $\pm$ 60  & 1\textsuperscript{\ref{fn:baduv}} & 76.5 & {\bf 0.17} & 1.9 & 2012.1.00240.S & uid://A002/X7fb989/X20 \\
		&  &  &  &  & & 3  & 60.7 & 0.57 & 5.0 & 2015.1.00997.S & uid://A001/X2fb/X5e4 \\
		&  &  &  &  & & 3 - 7m & 179.4 & 3.9 & 24.0 & 2015.1.00997.S & uid://A001/X2fb/X5e6 \\
		\hline
		J1048--0109 & 6.6759 & 162.07949 & -1.16123 &  1.9  $\pm$ 0.1  &  299  $\pm$ 24  & 3 & 11.9 & 0.95 & 9.1 & 2015.1.01115.S & uid://A001/X2fb/X3f4 \\
		&  &  &  &  & & 5 & 25.7 & {\bf 0.23} & 2.8 & 2017.1.01301.S & uid://A001/X1273/X370 \\
		\hline
		P167--13\textsuperscript{\ref{fn:comp}}  & 6.5144 & 167.64160 & -13.49607 &  5.3  $\pm$ 0.3  &  519  $\pm$ 25  & 3 - CW\textsuperscript{\ref{fn:baduv}, }\footnote{\label{fn:initials}The letters refer to the PI initials distinguishing between same cycle observations.} & 42.8 & 0.61 & 5.4 & 2015.1.00606.S & uid://A001/X2d6/X7d \\
		&  &  &  &  & & 3 - FW\textsuperscript{\ref{fn:initials}}  & 7.9 & 0.88 & 8.5 & 2015.1.01115.S & uid://A001/X2fb/X3f8 \\
		&  &  &  &  & & 4 & 42.7 & {\bf 0.25} & 3.7 & 2016.1.00544.S & uid://A001/X885/X367 \\
		\hline	
		J1120+0641 & 7.0848 & 170.00611 & 6.68996 &  1.0  $\pm$ 0.1  &  416  $\pm$ 39  & 1 & 160.7 & {\bf 0.22} & 2.6 & 2012.1.00882.S & uid://A002/X5a9a13/X537 \\
		\hline
		P183+05 & 6.4386 & 183.11240 & 5.09266 &  6.8  $\pm$ 0.3  &  397  $\pm$ 19  & 3 & 8.4 & 0.92 & 8.9 & 2015.1.01115.S & uid://A001/X2fb/X408 \\
		&  &  &  &  & & 4 & 47.1 & {\bf 0.25} & 2.7 & 2016.1.00544.S & uid://A001/X885/X363 \\
		\hline
		J1306+0356\textsuperscript{\ref{fn:comp}} & 6.0330 & 196.53441 & 3.94061 &  1.2  $\pm$ 0.1  &  246  $\pm$ 26  & 3 & 8.4 & 0.85 & 8.2 & 2015.1.01115.S & uid://A001/X2fb/X40c \\
		&  &  &  &  & & 5 & 27.7 & {\bf 0.23} & 3.3 & 2017.1.01301.S & uid://A001/X1273/X374 \\
		\hline
		J1319+0950\textsuperscript{\ref{fn:c0}, \ref{fn:comp}}  & 6.1347 & 199.79701 & 9.84763 &  4.1  $\pm$ 0.4  &  532  $\pm$ 57  & 1\textsuperscript{\ref{fn:baduv}} & 50.4 & {\bf 0.22} & 2.0 & 2012.1.00240.S & uid://A002/X7fb989/X18 \\
		&  &  &  &  & & 3  & 30.3 & 0.93 & 9.0 & 2015.1.00997.S & uid://A001/X2fb/X5d2 \\
		&  &  &  &  & & 3 - 7m & 119.6 & 3.9 & 26.0 & 2015.1.00997.S & uid://A001/X2fb/X5d4 \\
		\hline
		J1342+0928\textsuperscript{\ref{fn:comp}}  & 7.5400 & 205.53375 & 9.47736 &  1.0  $\pm$ 0.1  &  353  $\pm$ 27  & 5 & 114.1 & {\bf 0.16} & 3.0 & 2017.1.00396.S & uid://A001/X1296/X976 \\
		\hline
		P231--20\textsuperscript{\ref{fn:comp}} & 6.5869 & 231.65765 & -20.83354 &  3.3  $\pm$ 0.3  &  393  $\pm$ 35  & 3 & 7.4 & 0.87 & 8.5 & 2015.1.01115.S & uid://A001/X2fb/X440 \\
		&  &  &  &  & & 4 & 43.1 & {\bf 0.17} & 2.1 & 2016.1.00544.S & uid://A001/X885/X35f \\
		\hline
		P308--21\textsuperscript{\ref{fn:comp}} & 6.2355 & 308.04167 & -21.23399 &  3.4  $\pm$ 0.2  &  541  $\pm$ 32  & 3 & 12.4 & 0.63 & 5.9 & 2015.1.01115.S & uid://A001/X2fb/X418 \\
		&  &  &  &  & & 4 & 55.6 & {\bf 0.24} & 3.6 & 2016.A.00018.S & uid://A001/X11a4/Xf \\
		\hline	
		J2054--0005 & 6.0389 & 313.52708 & -0.08735 &  3.2  $\pm$ 0.1  &  236  $\pm$ 12  & 6 & 84.9 & {\bf 0.11} & 2.1 & 2018.1.00908.S & uid://A001/X133d/X261d \\
		\hline
		J2100--1715 & 6.0807 & 315.22792 & -17.25610 &  1.4  $\pm$ 0.1  &  361  $\pm$ 41  & 3\textsuperscript{\ref{fn:recal}}  & 7.9 & 0.57 & 5.2 & 2015.1.01115.S & uid://A001/X2fb/X41c \\
		&  &  &  &  & & 5 & 25.2 & {\bf 0.22} & 3.2 & 2017.1.01301.S & uid://A001/X1273/X378 \\
		\hline
		P323+12 & 6.5872 & 323.13826 & 12.29865 &  1.3  $\pm$ 0.2  &  271  $\pm$ 38  & 6 & 42.9 & {\bf 0.1} & 1.7 & 2018.1.00908.S & uid://A001/X133d/X2621 \\
		\hline
		J2318--3113 & 6.4429 & 349.57651 & -31.22955 &  1.5  $\pm$ 0.1  &  344  $\pm$ 34  & 3\textsuperscript{\ref{fn:recal}}  & 7.9 & 0.66 & 5.7 & 2015.1.01115.S & uid://A001/X2fb/X428 \\
		&  &  &  &  & & 5 & 23.7 & {\bf 0.24} & 3.5 & 2017.1.01301.S & uid://A001/X1273/X37c \\
		\hline
		J2318--3029 & 6.1456 & 349.63792 & -30.49266 &  2.3  $\pm$ 0.1  &  293  $\pm$ 17  & 6 & 39.9 & {\bf 0.1} & 1.7 & 2018.1.00908.S & uid://A001/X133d/X2619 \\
		\hline
		J2348--3054\textsuperscript{\ref{fn:uberres}} & 6.9007 & 357.13895 & -30.90285 &  1.5  $\pm$ 0.2  &  457  $\pm$ 49  & 1 & 16.7 & 0.41 & 5.3 & 2012.1.00882.S & uid://A002/X5a9a13/X53f \\
		&  &  &  &  & & 2\textsuperscript{\ref{fn:recal}}  & 36.8 & 0.19 & 2.0 & 2013.1.00273.S & uid://A001/X148/X73 \\
		&  &  &  &  & & 3 & 51.8 & {\bf 0.17} & 3.1 & 2015.1.00399.S & uid://A001/X5a3/X56 \\
		\hline
		P359--06 & 6.1719 & 359.13517 & -6.38313 &  2.7  $\pm$ 0.1  &  341  $\pm$ 18  & 3\textsuperscript{\ref{fn:recal}}  & 7.9 & 0.66 & 6.3 & 2015.1.01115.S & uid://A001/X2fb/X430 \\
		&  &  &  &  & & 5 & 25.7 & {\bf 0.22} & 3.1 & 2017.1.01301.S & uid://A001/X1273/X380 \\
		\hline
	\end{tabular}

\end{table*}

\subsection{Data reduction}
\label{sec:datareduc}

To obtain the calibrated $uv$-visibilities for imaging, we reduced the raw data using the default pipeline restoration scripts executed in the appropriate version of the Common Astronomy Software Applications (CASA) package \citep{mcmullin07}, i.e.\ the same version applied during the original calibration (CASA  {4.1} to {5.4.0-68}). Additional data quality checks and data manipulation were performed as described in the following section.

\subsubsection{Flux calibration accuracy}
\label{sec:fluxcalib}

We combine data from multiple objects and ALMA observation cycles. Because the addition of visibilities at incorrect flux scales could translate into erroneous spatial structure in the image plane,  we performed several quality assurance checks on the flux calibration.
As noted by \cite{stanley19}, several datasets required recalibration in a newer CASA version due to incorrect flux calibrator models present for Pallas and Ceres in CASA version 4.6 or lower. There are a total of nine datasets affected by this issue, as indicated in Table~\ref{tab:data}. Flux scales after recalibration were $60 - 80\%$ of the original values, with the mean flux level reduction of 15\% for these nine datasets. 

As a precaution, we reran the full pipeline calibration process on each of the 54 datasets, using the latest CASA version at the time of writing (version {5.6.1-8}), as described in the ALMA Science Pipeline User's Guide for CASA 5.6.1 (Section 6.4). Additionally, we manually enabled the query to the online ALMA flux calibrator catalog. 
The goal of this approach was to check for the existence of additional significant flux calibration changes introduced in the CASA development and debugging process, as well as to get the most up-to-date flux calibrator model values. For most of the datasets, the process finished successfully\footnote{Due to data format changes, CASA task definition changes, and backward incompatibility, this process failed for some datasets.} and yielded no flux level changes beyond 10\%, which is the usually reported calibration accuracy value. Because we found no further anomalies, we continued to use the calibrated visibilities provided by the default restoration scripts (except in the case of Pallas and Ceres calibrators as mentioned above).

\subsubsection{Data products per galaxy }
\label{sec:dataproduct}

We have  created several data products (measurement sets, maps and cubes)  in CASA for every observational cycle of each of the studied quasar (see Table~\ref{tab:data}), as described in \cite{venemans20}. We briefly summarize these here. 
We imaged the calibrated visibilities, using the CASA task {\sc tclean}, with the H\"{o}gbom deconvolver, and a channel spacing of 30~MHz (corresponding to 35~\kms\ on average). We weighted the data with the Briggs\footnote{We note that we recover the same aperture flux densities in both natural and Briggs weighted maps using the residual scaling method explained in a later section.} algorithm ({\sc robust = 0.5}) and cleaned (deconvolved) the maps down to $2\sigma$ level, within a cleaning mask radius of $2\arcsec$. We use these clean cubes to extract the \cii\ spectrum including the continuum emission. With the knowledge of the line position, we performed the continuum subtraction using the {\sc uvcontsub} task on the two spectral windows around the \cii\ line, excluding channels in total width of $2.5\times\mathrm{FWHM}$ around the \cii\ line (i.e.\ the total excluded bandwidth is on average 900\,\kms), while allowing for a continuum slope ({\sc fitorder = 1}). Using these datasets we imaged the continuum-free cube (same imaging parameters as above), as well as the \cii\ intensity map by averaging channels across $1.2\times\mathrm{FWHM}$ of the line, chosen to optimize the signal-to-noise ratio (S/N). We justify this integration width in Appendix~\ref{sec:app_optimalsn}.

\subsubsection{Stacking the data in the $uv$-plane}
\label{sec:uvstackmethod}

Stacking can be performed in the image plane by either averaging the  pixel/voxel values of maps or cubes, or by averaging extracted properties such as the spectra. Both methods are susceptible to several issues pertaining to the nature of interferometric data. I.e., the synthesized beam is not the same between datasets, and the faint stacked emission is usually uncleaned, which will yield ill-defined hybrid units in the stacked map (i.e., every pixel value will be a non-trivial combination of various cleaned and dirty beams). Another side-effect of image based stacking is that different spatial scales might be probed in different sources. In order to circumvent these issues,
we aim to add all visibilities in the $uv$-plane and thus obtain the best estimate of the mean emission of the population. To this effect we have adapted the method described in e.g.\ \cite{fujimoto19}, and extended it to be applicable to the whole spectrum. This method requires the data to be further reduced as follows.

For every individual cycle, starting from the continuum subtracted visibilities, the required channels were split out. Depending on the science goal, we either split out the equivalent of the \cii\ intensity map ($1.2\times\mathrm{FWHM}$), or, a smaller chunk corresponding to a specific velocity bin (i.e.\ a channel map, and/or the resolution of our spectral extraction). 
We used the function {\sc im.advisechansel} from the CASA toolkit to select the channels of interest in the kinematic local standard of rest frame (LSRK). Individual channels are selected based on the nearest neighbour algorithm. The data that we consider for $uv$-stacking were all observed at velocity resolutions of $\lesssim10\,\kms$, which is high enough to render any additional velocity-based data regridding or interpolation unnecessary. 
The data was further time-averaged inside 30\,s bins to reduce the data volume\footnote{The value is large enough to accomodate all datasets, but small enough not to introduce any time smearing effects.}. The coordinate reference system in the header was changed to the International Celestial Reference System (ICRS) as it was  sometimes incorrectly written as J2000. The visibilities were phase-shifted to the centroid of the \cii\ emission using the task {\sc fixvis}\footnote{The task uses small angle approximation, which is valid in our case as shifts are at most few arcseconds large, and less than an arcsecond in most of the cases.}. The header was modified further to artificially place the object at the $(0,0)$ coordinate\footnote{This was achieved by putting zeros into {\sc PHASE\_DIR, DELAY\_DIR, REFERENCE\_DIR} columns of the measurement set.}. Finally, the data weights were recomputed with the {\sc statwt} task using the same time and channels bin for all of the datasets. 
The calculation of the default weights is implemented differently across various CASA versions and can depend on the system temperature or gain factors. We manually checked that the ranges of recomputed weights are similar between all datasets, allowing the data to be co-added in a way that ensures the weighted average is not erroneously dominated by any single source. 
Several anomalous datasets were identified as having (at least) an order of magnitude larger data weights (they also have much lower observed velocity resolution of  $\gtrsim30\,\kms$), and were excluded from any further $uv$-stacking analysis. These exclusions are noted in Table~\ref{tab:data}.

We imaged the split, renormalized and re-centered data using the CASA task {\sc tclean}, considering the full list of all sources and cycles. 
We imaged the data in the continuum mode ({\sc specmode = mfs}) and applied the multi-scale deconvolver, with scales corresponding to 0 (delta-function), 1, 3, 5 and 10 synthesized beam sizes. Cleaning was performed within a 2$\arcsec$ radius mask down to a $2\sigma$ threshold.
To obtain the $uv$-stacked cube we imaged every velocity bin individually. The velocity bin was either of fixed size, i.e.\ 30\,\kms, or scaled with the width of the \cii\ line, i.e.\ $0.2\sigma_{\mathrm{line}}$. We used both approaches in our analysis. These separately imaged velocity bin slices were further joined into a single cube using the CASA toolkit task {\sc ia.imageconcat}.

The preparation of the data for $uv$-plane continuum stacking was performed in the same way as in the \cii\ case, but instead of  selecting the line emission channels, we flagged them. A total of width of $2.5\times\mathrm{FWHM}$  centered at \cii\ was removed, and the remaining channels inside individual spectral windows were averaged together. All available continuum data were used (i.e.\ data from all four science spectral windows).
We validated that our choice of excluded frequencies/velocities did not bias the dust continuum results by re-imaging the data, this time excluding a much broader bandwidth of $8\times\mathrm{FWHM}$ around the line ($\sim$2800\,\kms\ on average). 

Spectral stacking will smooth any spatial and velocity structure of individual sources. The stack itself is only meaningful if the mean property of interest (e.g. extent of the faint emission, or the line shape) is well-defined for the sample, which we cannot know a priori, and deeper observations are necessary to confirm any potential structure in individual objects.

\subsection{Mergers and companions}
\label{sec:comp}

Our aim is to measure  extended structure and high-velocity outflows in quasar host galaxies. Thus, we  take special care of objects exhibiting ongoing mergers, and of those that have a nearby companion galaxy. In both cases the accompanying source might bias our results, by either providing additional flux at larger distances from the quasar host, or at larger velocity separations. 
Where possible, we visually identified the presence of accompanying sources,  distinct in terms of either their spatial, or velocity position (\cii\ intensity maps are available in Appendix~\ref{sec:app_sources}; see also accompanying papers by \citealt{venemans20} and \citealt{neeleman20b}).
Morphological details on sub-kpc scales cannot be recovered with these observations. What we consider as a single component source in this work might break into multiple sources upon a higher resolution follow-up. 
Thus, definitions such as a merger or a galaxy pair are, by construction, arbitrary, especially in the low S/N regime. With these limitations in mind, we classify these special cases based on visual inspection into four groups as follows
(see also \citealt{decarli17}, \citealt{decarli19}, \citealt{neeleman19} and \citealt{banados19}):

I) Merging system blended inside the aperture:
These systems exhibit at least two clearly separated (spatially or in velocity) peaks at distances of less than $\sim 5$\,kpc. In our data it is not possible to draw a clear line between the two sources or easily measure their individual fluxes due to a connecting bridge of \cii\ emission or blending. We classify five of such systems: J1319+0950, J1342+0928, P167--13, P308--21 and J0305--3150. 

II) Nearby companion:
These are defined as sources that have a spatial offset of $\sim 5 - 10$\,kpc between the quasar host and its galaxy pair, such that an aperture flux can be obtained for both objects individually. However, the system will show increased flux on the smallest baselines that cover both sources simultaneously.
We classify two such systems: J1306+0356 and P231--20.

III) Distant companions: These are cases where there is a galaxy pair in the field, but at separations larger than 10\,kpc. No significant amplitude increase can be seen on the smallest baselines due to this large separation (the primary beam response at the companion source position is also significantly lower).
We classify two such systems: J0842+1218 and J2100--1715.

IV) Complex morphologies: These systems show a surface brightness morphology that is more complex than a single component source, but its origin is unclear. It is possible that these objects have companions detectable only at higher resolution, or that they contain extended and more patchy gas emission. In the absence of conclusive companion or merger evidence, we do not exclude these objects from the stacking analysis. We classify three such systems: J0100+2802, J025--33 and P009--10.

Groups I and II  are removed from some of our samples, as explained in the next section. Groups III and IV are always included in our stacking samples, with the latter containing potential candidates for follow-up observations.

\subsection{Data samples}
\label{sec:samples}

Throughout the paper we refer to four different data samples, all drawn from Table~\ref{tab:data}, as follows.

1) High-resolution data sample:  It contains all 27 quasar host galaxies, where each one is represented by a single high-resolution ($\sim1$\,kpc) dataset. This corresponds to the data used in the accompanying papers \citep{venemans20, neeleman20b}.

2) High-resolution clean data sample: It is a subset of the previous sample, obtained by removing five objects classified as merging systems blended inside the aperture (group I from Section~\ref{sec:comp}). This sample is used to stack aperture spectra extracted from individually imaged maps. Nearby companions (group II) will not contribute significantly to the aperture fluxes, and are therefore included in this sample.

3) Full $uv$-stacking data sample: It contains all of the datasets (including low resolution data), except five observing runs that have outlying data weights, and one without a full line coverage (see Table~\ref{tab:data} footnote and Section~\ref{sec:uvstackmethod}). This selection results in 26 quasar host galaxies (out of 27) and a total of 48 datasets (out of 54). The cumulative on source time is approximately 34\,h (out of which 25.5\,h were taken with the 12\,m array). This sample is used to obtain the highest S/N in the $uv$-stack, while attempting to recover broad-line emission, which would be indicative of outflows.

4) Clean $uv$-stacking data sample: It is a subset of the previous sample, obtained by further removing systems classified as mergers inside the aperture and near companions (groups I and II from Section~\ref{sec:comp}), as well as any 7\,m dataset (for details see Appendix~\ref{sec:app_subsamples}). This selection results in 19 quasar host galaxies and a total of 32 datasets. The cumulative time on source is approximately 18\,h. We consider this sample to be the least biased, and use it to derive our main stacking results.

\section{Methods}
\label{sec:methods}

In the following sections we present different methods employed in our search for extended \cii\ emission and outflows. We discuss flux measurements taken from the image and the $uv$-plane, and the applicability and shortcomings of each method. We also discuss the details of the stacking procedures.
By applying various techniques we aim to reach a consensus on the measured galaxy properties (e.g. total flux and spatial extent). This is particularly relevant for detections of low S/N.
Throughout this whole section we demonstrate various diagnostics on the source J1306+0356, which was chosen as a good example, due to it being a complex system (presence of a companion, see also \citealt{neeleman19}), with data available from multiple cycles at different resolutions.
Results for all individual quasar host galaxies are shown in Appendix~\ref{sec:app_sources}.

\subsection{Measuring the spatial extent of emission}

\subsubsection{Analysis in the image plane}
\label{sec:methods_aperture}

\begin{figure*}
	\centering

	\includegraphics[width=\linewidth]{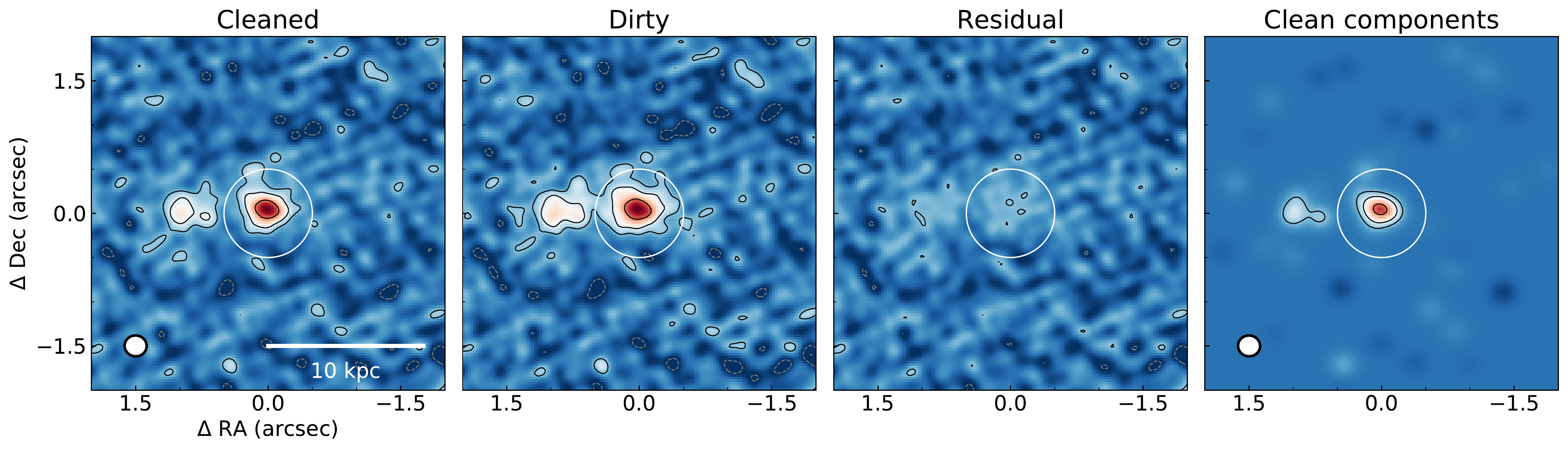}
	\caption{An example of the \cii\ intensity map, imaged from the  high-resolution dataset using $1.2\times\mathrm{FWHM}$ of the line width (the system shown is J1306+0356). Contours are logarithmic starting from $2\sigma$, continuing in powers of 2, with dashed ones indicating negative values. The four panels show maps from various deconvolution stages. The clean beam is shown in the bottom left corner. The white circle is the manually chosen aperture (radius of $0\farcs5$), that captures the emission of the quasar host galaxy only, which intensionally excludes the companion galaxy (encompassing some fraction of emission in the connecting bridge is unavoidable). The cleaned map is the sum of the residual map (units: Jy / dirty beam) and the cleaned components map (units: Jy / clean beam), and is therefore affected by ill-defined units. Excess flux in the residual map can lead to a non-negligible bias in the total flux measurement.}
	\vspace{0.3cm}
	\label{fig:example_map}
\end{figure*}

\begin{figure}
	\centering
	\includegraphics[width=\linewidth]{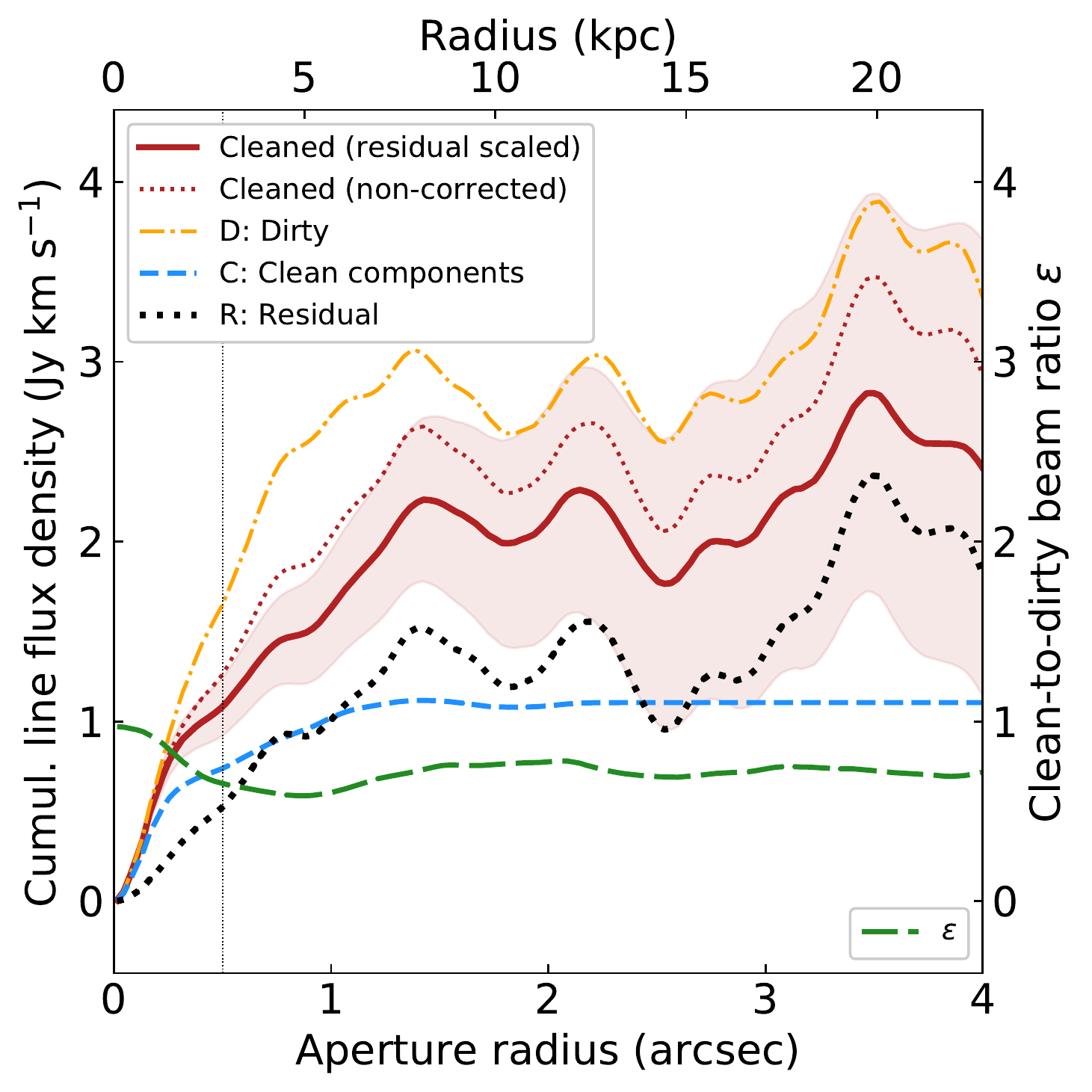}
	\caption{
		An example of the curves of growth used in the residual scaling method. As in Figure~\ref{fig:example_map}, we show the results for J1306+0356.
		The corrected flux density is obtained by scaling the uncleaned flux with the proper beam size, and its $1\sigma$ uncertainty is shown with the shaded region. The vertical line shows the manually chosen aperture (radius of $0\farcs5$), used to derive the total flux of the quasar host galaxy. Additional flux at larger radii is due to the presence of a companion galaxy.
	The clean-to-dirty beam ratio is roughly constant at larger radii.
	Within the chosen aperture, more than a third of the measured flux is contributed by the residual.
	The residual-scaled cleaned flux density is 85\% of the one obtained from the standard cleaned map.
	}
	\label{fig:example_aper_dirty}
\end{figure}

\begin{figure}
	\centering
	\includegraphics[width=0.95\linewidth]{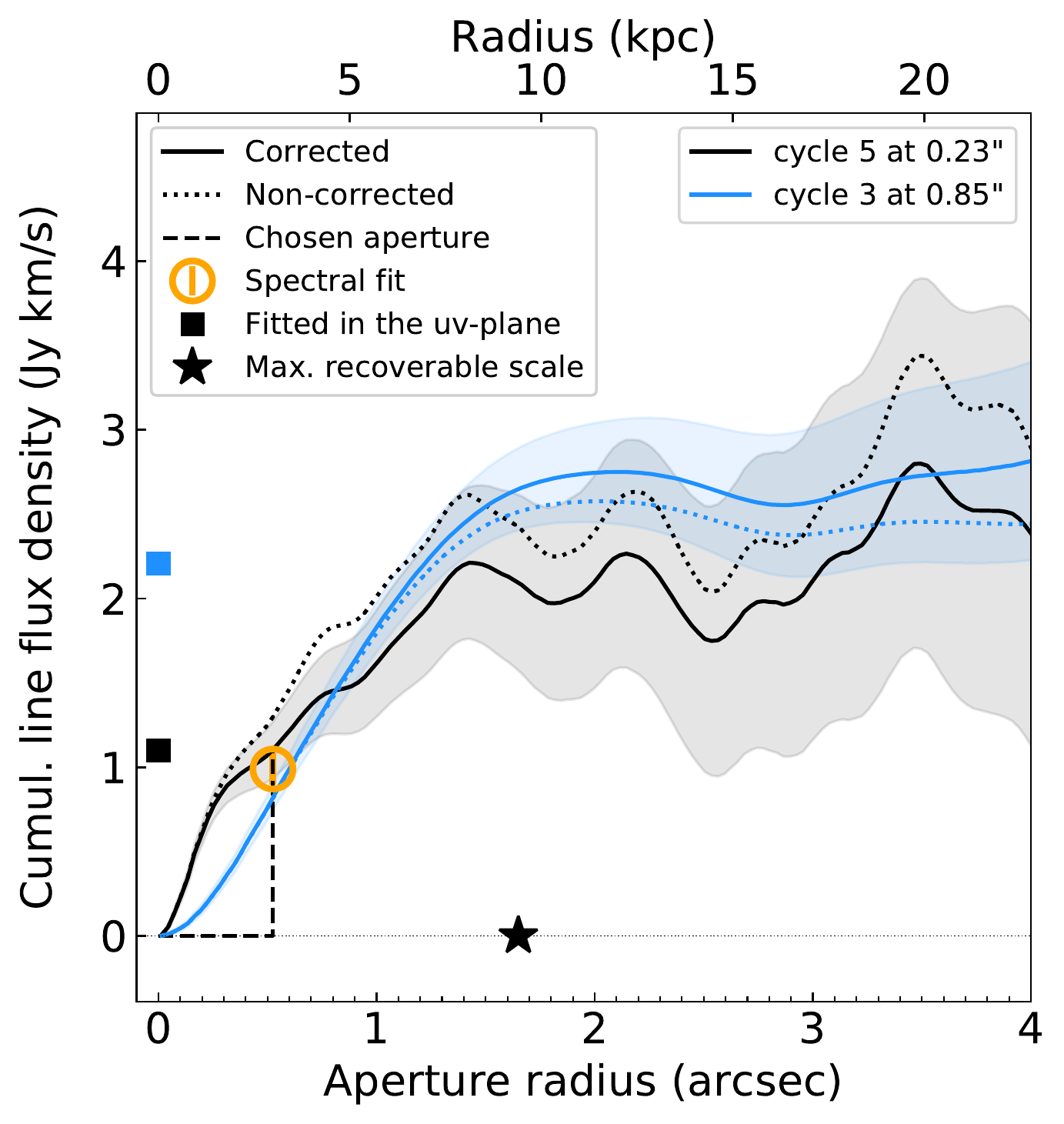}
	\caption{ 
		An example of the curves of growth measured from \cii\ intensity maps of two different observation cycles (3 and 5, as shown in the legend) targeting J1306+0356.  
	The shaded area indicates the $1\sigma$ uncertainty of the residual-scaled flux density. 
	The dashed line marks the aperture size chosen to encompass the quasar host galaxy only (see Figure~\ref{fig:example_map}).
	The squares on the zero radius line show total flux densities derived from single Gaussian fits in the $uv$-plane (colored by cycle, same as in Figure~\ref{fig:example_uv}). The star shows the MRS, which is larger than the plotted range in the case of cycle 3 observations.
	The orange circle shows the flux density obtained from a spectrum fit of the high-resolution cycle data (see Figure~\ref{fig:example_spec}), multiplied by 0.84 to remove flux beyond $1.2\times\mathrm{FWHM}$.
	This figures demonstrates that the companion is blended with the quasar host galaxy in the lower resolution data, and that residual scaling corrections differ based on the beam shape, and can go in either positive or negative direction.}
	\label{fig:example_aper}
\end{figure}

A common method of measuring the flux and source size from a map is by fitting a 2D Gaussian to the galaxy in the image plane. We do so by employing the {\sc imfit} task included in CASA. The second approach that requires no priors on the source morphology is the aperture integration, for which we choose to apply circular apertures. In the remainder of this section we explore only the aperture integration, but the discussed systematics are applicable to any image based analysis (i.e.\ fitting a 2D Gaussian).

One problem present in image plane analysis is the non-trivial unit definition (\jybeam). Due to the nature of the deconvolution (cleaning) process, the final cleaned map will contain hybrid units. Several maps from the cleaning process, shown in Figure~\ref{fig:example_map}, demonstrate this issue. The final cleaned map is obtained by summing the residual map (in units of Jy / dirty beam) and the clean components map (in units of Jy / clean beam), whose units differ due to the beam definition. The clean beam is obtained\footnote{This can be manually overridden by the {\sc restoringbeam} parameter in {\sc tclean}, but we made no such attempt.} by fitting a 2D Gaussian to the peak of the point spread function (PSF; the dirty beam). It has a well defined volume (integral) equal to $\Omega_{\mathrm{beam}}=\pi/ (4\ln 2)\times\theta_{\mathrm{maj}}\theta_{\mathrm{min}}$, where $\theta_{\mathrm{maj}}$ and $\theta_{\mathrm{maj}}$ are major and minor axis FWHMs of the elliptical 2D Gaussian. On the other hand, the dirty beam is a sum of many sine/cosine waves, whose integral oscillates around zero. Therefore, the volume of the dirty beam is ill-defined and will depend on the integration limits (i.e.\ the aperture size).

The clean components map is defined as the cleaned sky model (i.e.\ point sources / delta functions, and, in case of the multi-scale algorithm, additional Gaussians of manually chosen sizes) convolved with the Gaussian clean beam. During the sky reconstruction, the clean algorithm does not distinguish between positive and negative peaks. The true emission is positive, therefore the cleaning should not progress too deep, in order to avoid cleaning both positive and negative noise peaks. After the sky model is subtracted from the visibilities, the residual remains.
Any flux measured from the residual (i.e.\ flux below the cleaning threshold) will be incorrectly assigned the clean beam size (defined to be a Gaussian in the restoration process, and written in the image header of the final map). 

To mitigate this issue we  scale the flux collected from the residual map \citep[for details see][]{jorsater95, walter08, novak19}. In practice, this is performed by measuring aperture fluxes from three maps: the dirty (D), the residual (R), and  the clean components\footnote{The clean components map can be obtained by either convolving the sky model map (units of Jy/pixel) with the clean beam, or by subtracting the residual map from the final map output by CASA.} (C) map (see Figure~\ref{fig:example_map}). The flux density measurement is obtained by summing the values of pixels contained within the aperture, and dividing the sum with the number of pixels in the clean beam: $F_\mathrm{aper} [\mathrm{Jy}]=\sum_{\mathrm{pixels}}\mathrm{map [\jybeam]} / (\Omega_{\mathrm{beam}}/\mathrm{pixelsize}^2$), where the beam axes sizes and the pixel size are all expressed in the same units (e.g. arcseconds).
Corrected flux inside an aperture can be obtained as $F_{\mathrm{correct}}=C +\epsilon R=\epsilon D$, where $ \epsilon= C/(D-R)$ is the clean-to-dirty beam area ratio defined inside a specific aperture (see  Appendix~\ref{sec:app_residualscale}).
The drawback of this method is that it requires some cleaned flux, and becomes numerically unstable when $R$ approaches $D$. 
The error on the aperture flux can be approximated by $ \mathrm{rms}\times \sqrt{N}$, where $N$ is the number of beams contained inside the aperture, and the root-mean-square (rms) is the noise variation measured in the entire map.

The amount of the residual scaling depends on the exact shape of the dirty beam (defined by the $uv$-coverage, applied imaging weights, and any additional $uv$-tapering) and the aperture size, and can either boost or decrease the measured flux density.
If the scaling is significant, not applying the correction is likely to produce larger systematic uncertainties in the case of extended sources that are resolved over multiple beams, sources with low surface brightness (where significant amount of  emission remains in the residual), and in the case of a non-Gaussian dirty beam (more likely to happen with naturally weighted imaging). Image-based stacking is also likely to be affected, especially if  maps with different dirty beams are being stacked, because the knowledge of the beam shape and the fraction of cleaned flux is lost in the stacking process. Furthermore, the units become increasingly non-trivial after 2D image convolution, often employed for beam matching\footnote{The convolution kernel is computed from the clean beam only, therefore making its effect on the dirty beam unknown, requiring further quantification.}.

In the first panel of Figure~\ref{fig:example_map}, we show an example of a high-resolution \cii\ intensity map, produced by imaging averaged channels in total width of $1.2\times\mathrm{FWHM}$ (range of $\pm1.4\sigma_{\mathrm{line}}$) from the continuum subtracted measurement set. 
The measured aperture flux density can be corrected for the missing tails beyond the $1.2\times\mathrm{FWHM}$, assuming a Gaussian profile, by multiplying by 1.19. 
In Figure~\ref{fig:example_map}, a second $4\sigma$ peak is apparent 5\,kpc away \citep[see also][]{neeleman19}. 
Due to the large separation we are able to define an aperture containing the full source emission, with no significant contribution from the companion or the connecting bridge.

In Figure~\ref{fig:example_aper_dirty} we show flux densities measured inside a growing aperture (hereafter referred to as the curves of growth) for all of the maps from Figure~\ref{fig:example_map}. The clean beam size can only be used to measure the emission in the clean components map. If used on other maps, it will introduce systematic errors if the clean-to-dirty beam ratio defined inside the aperture is different from one. In this example, the ratio is 0.65 at the chosen aperture of $0\farcs5$, therefore any uncleaned flux is being over-counted by a factor of $1/0.65\approx1.5$. Because of this, the corrected flux density is 85\% of the one measured in the cleaned map without residual scaling.

In Figure~\ref{fig:example_aper} we show measurements from two observation cycles of the same source, taken at different resolutions. The quasar host galaxy and its companion are blended together in the low resolution data (beam $\approx0\farcs85$). The total flux density of the system, i.e.\ the value at which the curves of growth flatten, are consistent within the errors. The residual scaling correction is smaller for the low resolution data.

\subsubsection{$uv$-plane analysis}
\label{sec:methods_uv}

\begin{figure}
	\centering
	\includegraphics[width=\linewidth]{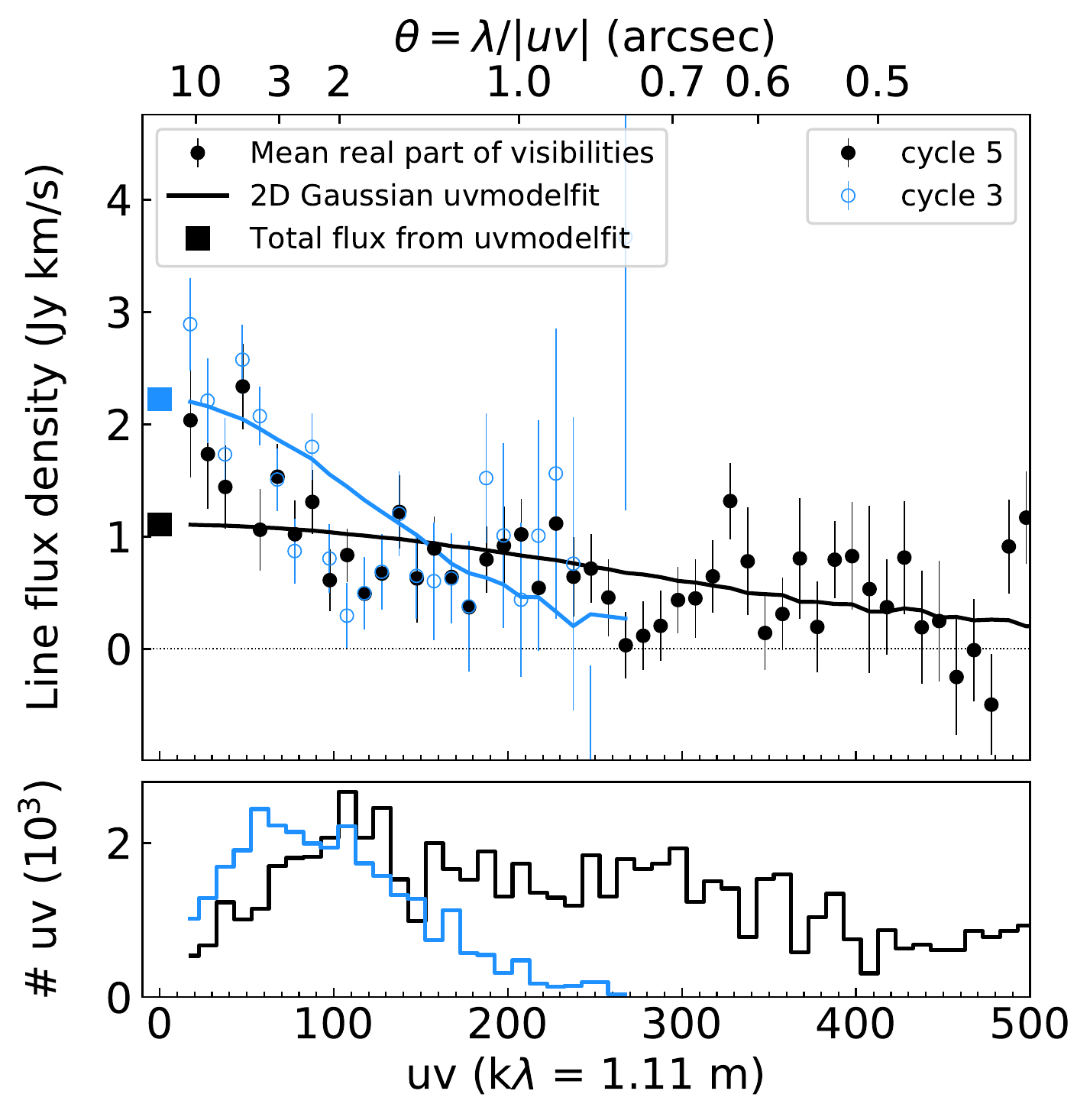}
	\caption{Example flux density  measured in annuli of $uv$-distances of two observation cycles of J1306+0356. The error bars represent the standard error on the mean of visibilities (real part) in a given annulus. The solid lines are single Gaussian fits performed in the $uv$-plane.
	Histograms in the lower panel show the number of visibilities available in a given annulus after data averaging. 
	The values on the $x$-axis are measured in kilowavelengths, where we also provide the value of 1\,k$\lambda$ in meters.
	The upper $x$-axis shows the fringe resolution of a given projected baseline length (defined as the observing wavelength divided by the  $uv$-distance). A single Gaussian is a poor model choice in this example, exacerbated by the presence of the companion, resulting in significantly different flux estimates depending on available baseline lengths (i.e.\ cycle 5 data is dominated by longer baselines and fitting a compact source is preferred). }
	\label{fig:example_uv}
\end{figure}

Due to discrete and sparse $uv$-coverage, the map produced by the clean algorithm is not uniquely defined, i.e.\ missing Fourier components must be extrapolated. To avoid such issues altogether, the  measurement can be directly performed in the $uv$-plane.
Approaches vary from simpler ones, such as examining 1D amplitudes vs. $uv$-distance plots \citep[e.g.,][]{hodge16}, to more complex ones, such as using specialized multi-component fitting software, e.g.\ {\sc uvmultifit} \citep{uvmultifit}, as demonstrated in e.g. \cite{rujopakarn19}, or fitting morphology models based on Bayesian statistics \citep[see e.g.,][]{pavesi16}.
We employ two techniques: 1) we attempt to  fit a 2D Gaussian in the $uv$-plane using the {\sc uvmodelfit} task in CASA, and 2) we analyze amplitudes in radially-averaged annuli of $uv$-distances (units of k$\lambda$).
For high S/N data, a poor fit to the visibilities indicates a more complex source morphology (i.e.\ non-Gaussian), or it can imply additional sources/companions in the field (i.e.\ the second source is blended together with the main one on the shortest baselines only). One downside of this simple $uv$-plane analysis method is that it does not account for the complex morphology encoded in the visibilities' phases, so it is best applied in parallel with the map analysis.

In  Figure~\ref{fig:example_uv} we show the flux densities measured at specific $uv$-distances (i.e.\ projected baseline lengths) of two different observation cycles. 
Only the real part of the complex visibilities is shown as a proxy for the visibility amplitude\footnote{Using only the real part helps to see oscillations around zero flux densities at larger $uv$-distances, whereas amplitudes are by definition always positive.}. For a source in the phase center, the imaginary part of the visibilities will always have a mean value of zero. We have confirmed that this to be the case for all of our datasets.  Histograms in the lower panel show the number of visibilities from different cycles that depend on the number of antennas and the total integration time (all data was averaged to the same interval of 30\,s, see Section~\ref{sec:uvstackmethod}). 

It is expected that a single Gaussian would be a poor representation of this system, which contains two galaxies at close ($\sim5$\,kpc) separation. Therefore, an attempt to fit a single 2D Gaussian to the visibilities, using the  {\sc uvmodelfit} task in CASA, yields significantly different models and total fluxes (shown as squares in Figure~\ref{fig:example_uv}) for the two cycles. The low-resolution data confirms the upturn in visibilities that is only hinted at in the high-resolution data  below 100\,k$\lambda$. In this example it is clear that there is a second galaxy that is causing this upturn. In a different scenario, where no such companion is obvious from the image plane analysis, the interpretation becomes more difficult, as the faint spatially extended emission can have the same signature.

\subsubsection{Comparing flux measurements}
\label{sec:methods_compare}

\begin{figure}
	\centering
	\includegraphics[width=\linewidth]{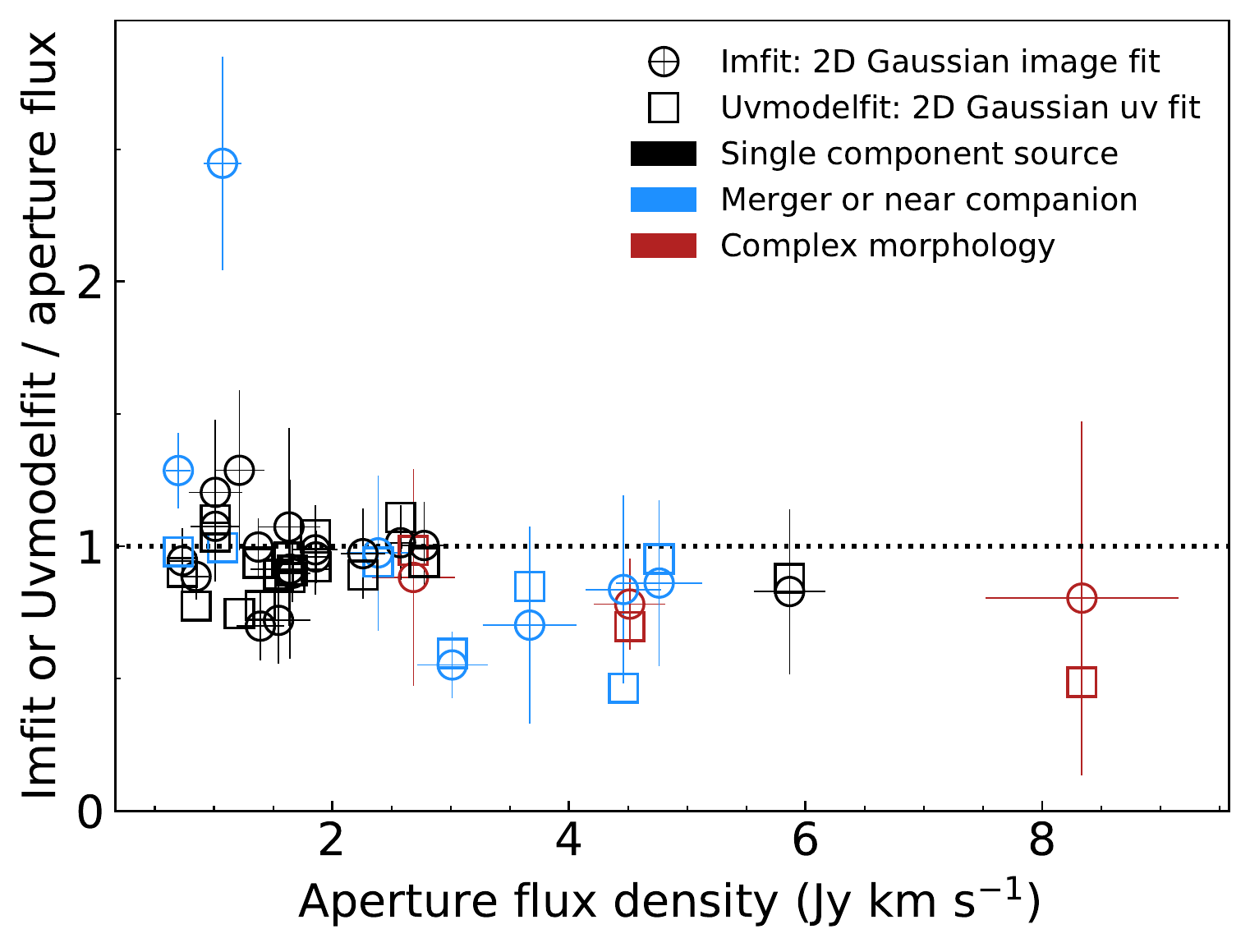}
	\caption{Systematic differences between three different flux measurements ({\sc imfit}, {\sc uvmodelfit} and the aperture integration) for all sources. The horizontal dashed line corresponds to our best estimate using the residual scaled aperture measurements. The colored points indicate sources with complex morphologies.}
	\label{fig:fluxcompare}
\end{figure}

We apply the different methods of  measuring the flux density, outlined above, to our high-resolution sample of quasars (sample 1 from Section~\ref{sec:samples}), and  compare the results in Figure~\ref{fig:fluxcompare}.
Aperture sizes for individual sources were chosen manually (after visual inspection) to be at a radius where the curve of growth begins to flatten, approaching values consistent with short baseline $uv$-amplitudes.
 We limit the 2D Gaussian fitting in the image plane  to a circle of a 2$\arcsec$ radius, and apply no additional constraints to $uv$-plane fits. 

Disregarding the outlier with a nearby companion at $\sim1\,\jykms$ (in example source J1306+0356), all of the measurements are consistent within a factor of two, and generally consistent within the errors. Two tentative trends can be observed in Figure~\ref{fig:fluxcompare}.
Compared to the aperture measurements, image plane fitting using {\sc imfit} (circles in Figure~\ref{fig:fluxcompare}) produces larger values at the fainter end (i.e.\ Gaussian tails apparently account for positive noise peaks), while both image plane based {\sc imfit} fits (circles), and $uv$-plane based {\sc uvmodelfit} (squares) single Gaussian fits, yield generally lower fluxes at the brighter end, possibly indicating the presence of more extended faint emission. In some of the cases, fitting a single 2D Gaussian is obviously a poor description of the observations, nevertheless we show this simple approach to demonstrate the resulting scatter.

Measurements of both the aperture flux density, and amplitudes at the shortest baselines, can be sensitive to large scale structure, whereas fitting a 2D Gaussian in the image plane will unlikely provide good estimates for the extended  faint emission.  For this reason we refrain from fitting 2D Gaussians in the image plane in the remainder of the paper.

\subsection{Measuring the spectrum}

\subsubsection{Extracting aperture spectra}
\label{sec:methods_spec}

\begin{figure}
	\centering
	\includegraphics[width=\linewidth]{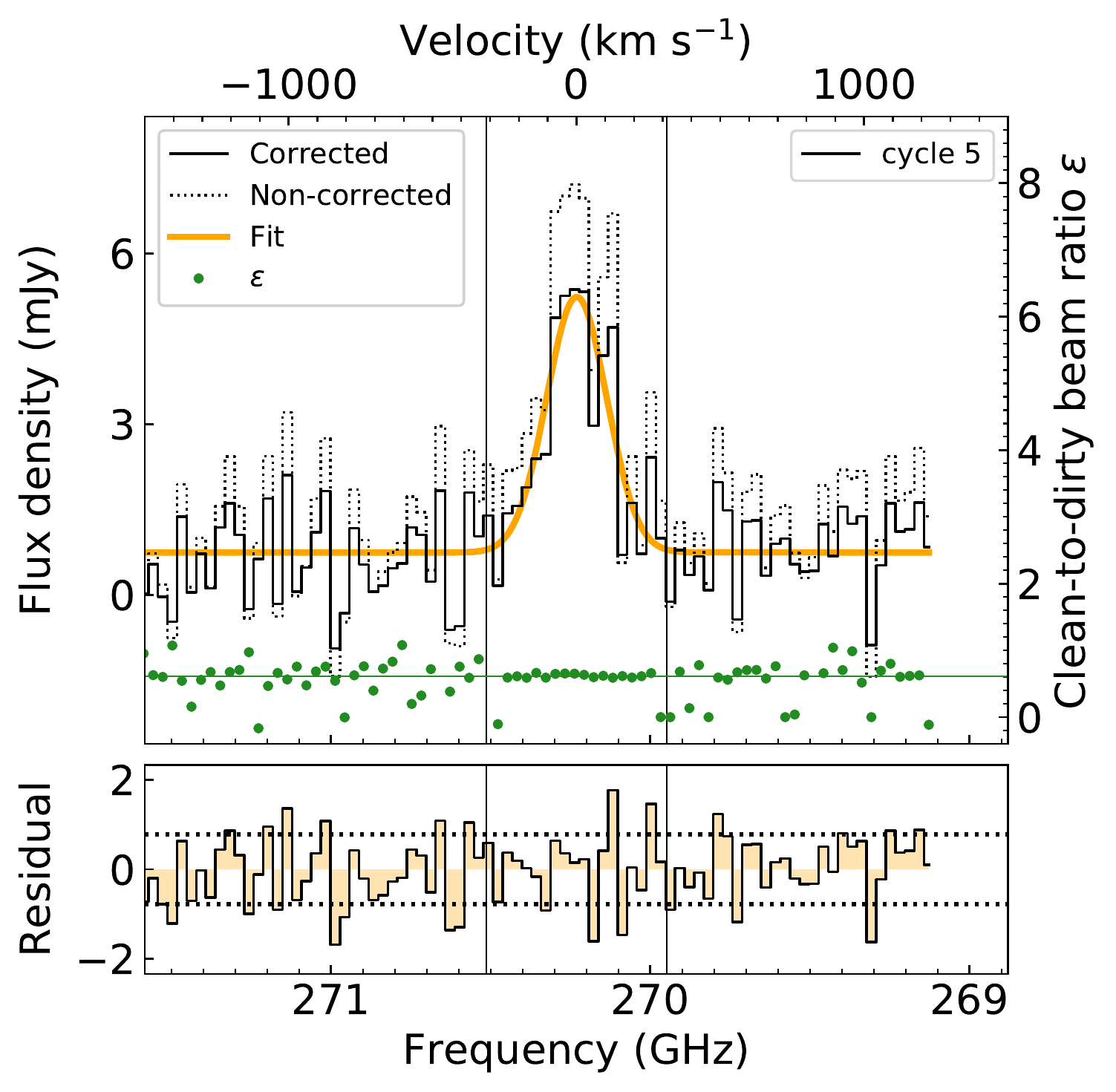}
	\caption{
		Example of an integrated spectrum, for J1306+0356, extracted from individual channels, using a manually chosen aperture with a radius of  $0\farcs5$. The orange line is a single Gaussian plus a constant fit to the corrected spectrum. Green points indicate the clean-to-dirty beam ratios measured in individual channels. The green line shows the best estimate of $\epsilon=0.61$ (see Section~\ref{sec:methods_aperture}), used to obtain the final corrected spectrum.
	Residuals from the fit are shown in the lower panel. The dotted horizontal lines outline $\pm1\sigma$ variations in the residual. Vertical lines mark $\pm3\sigma_{\mathrm{line}}$. With smaller S/N of emission in individual channels, compared to the \cii\ intensity map, the effect of the residual scaling is more pronounced (the difference between dotted line and full line spectra), and is larger than 20\% in this specific case.
}
	\label{fig:example_spec}
\end{figure}

To  measure the spectrum, we apply residual scaling on the aperture flux densities obtained from individual channels before a spectrum is extracted. The clean-to-dirty beam ratio, $\epsilon$, can be estimated from the channels with the highest S/N, and applied on the remaining channels, as the beam shape is not expected to vary significantly between neighboring channels. This is demonstrated in Figure~\ref{fig:example_spec}, where the factor $\epsilon$ exhibits the least scatter when measured across the line. If the continuum is absent, $\epsilon$ can  be measured exclusively over the line (as it requires some clean flux $C$ to calculate). The value of $\epsilon=0.61$, obtained here, is slightly smaller compared to $\epsilon=0.65$, derived from the \cii\ intensity map (see Figure~\ref{fig:example_aper_dirty}), because the shape of the dirty beam will vary slightly due to different visibilities being averaged in both cases (varying frequency ranges are being averaged).
The S/N of the emission is lower inside individual channels compared to a broader intensity map collapse, and with less cleaned emission in smaller velocity bins, the contribution of the residual (uncleaned) flux is larger.
This results in a more significant residual-scaling effect, as the corrected value now corresponds to 78\% of the non-corrected one (compared to the previous 85\% in the intensity map). 
In this example, a single Gaussian plus a constant fit yields the \cii\ line FWHM of 246  $\pm$ 26\,\kms. The residual spectrum after the subtraction of the fit  (the lower panel of Figure~\ref{fig:example_spec}) is consistent with pure noise. There is no indication of an additional broad spectral component.

\subsubsection{Stacking spectra}

To increase the S/N of the spectra and recover possible fainter emission we also perform spectral stacking. Due to the range of \cii\  linewidths in our sample (see Table~\ref{tab:data}) it is necessary to rescale the spectra by their respective widths before stacking.
This step ensures that we do not misinterpret the contribution from broader lines as potential outflows in the mean spectrum.
Therefore, we resample all spectra before stacking, and present the stacked spectra in the units of velocity divided by $\sigma_{\mathrm{line}}$, which has an average value of 150\,\kms. 

We perform spectral stacking in two ways. First, we stack individually measured and residual-scaled aperture spectra from the data cube (aperture sizes were chosen  manually), where we consider the underlying continuum to be flat, within the bandwidth of interest, and estimated from the fitting of a Gaussian plus constant profile. This constant value\footnote{The dust continuum spectral energy distribution is not flat. However, in our case this slope would result in only a few percent difference in flux between the two ends of the observed bandwidth. In a single spectral setup, ALMA provides up to 3.75\,GHz of a contiguous bandwidth coverage, which corresponds to $\sim$4000\,\kms\ at  the observed frequency of $\sim$270\,GHz, where the \cii\ line of $z\sim6$ galaxies is redshifted.}
for the continuum is subtracted before the addition of the spectra.  No additional weights to individual spectra are applied. 
Second, we perform spectral $uv$-stacking (as described in  Section~\ref{sec:uvstackmethod}).
To obtain the full spectral cube, we selected velocity slices of $0.2\sigma_{\mathrm{line}}$ from individual datasets and imaged them together. This velocity corresponds to a width of 30\,\kms\ on average.
The $uv$-data was continuum subtracted prior to stacking and, in contrast to the first method, this continuum subtraction allowed for a spectral slope in the fit. Having two different methods of continuum estimation largely mitigates any potential bias of subtracting a  broad component as continuum.

\section{Results}
\label{sec:res}

In the following sections we present the recovered spatial extent of both \cii\ and the dust continuum, as well as evidence for/against outflows.

\subsection{Spatial extent of the  \cii\ emission}
\label{sec:results_ciiextent}

\begin{figure*}
	\centering
	\includegraphics[width=\linewidth]{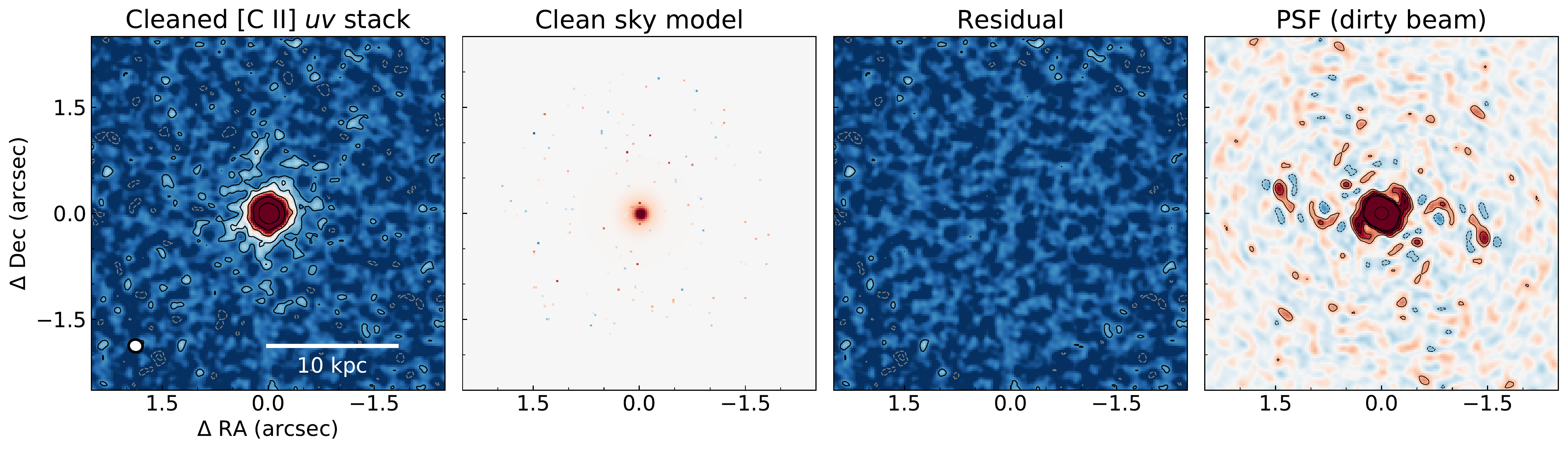}\\
	\includegraphics[height=6.8cm]{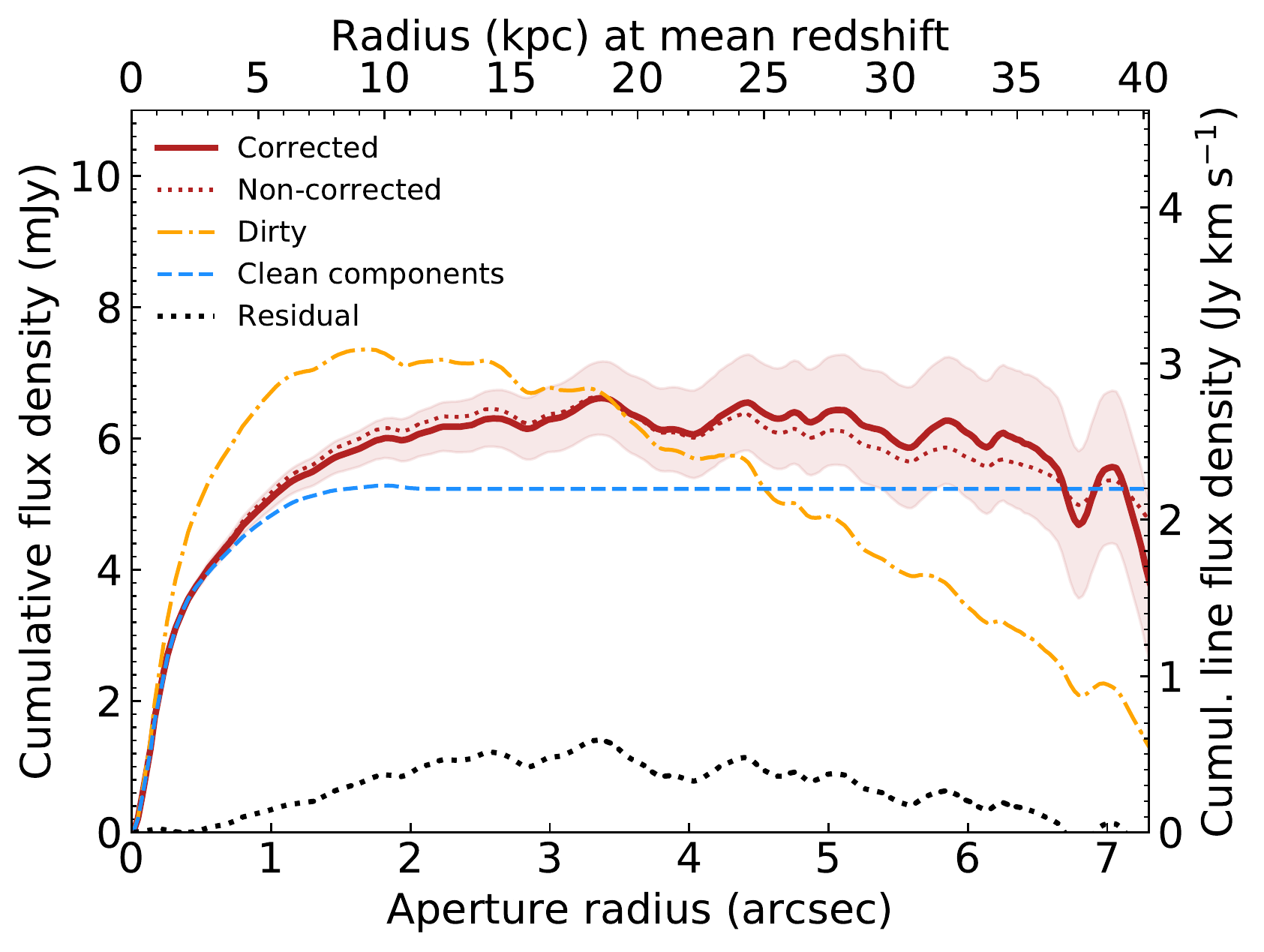} 
	\includegraphics[height=6.8cm]{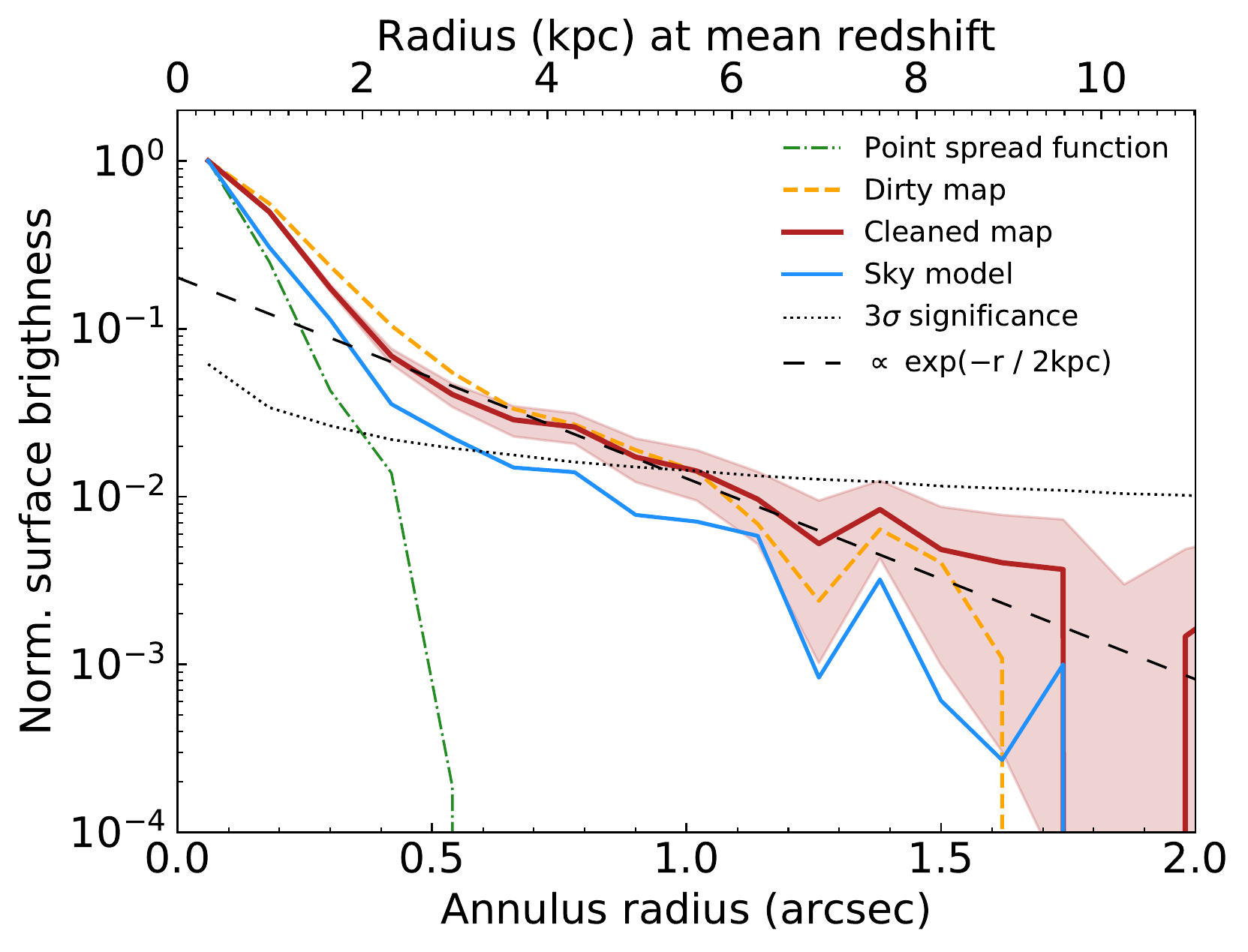}\\

	\caption{The emission of the \cii\ line measured in the $uv$-stack of the clean sample of $z\gtrsim6$ quasar host galaxies. 
		{\it The upper four panels} display the final cleaned map, as well as the multi-resolution clean model, the associated residuals, and the dirty beam (a different deconvolution method is discussed in Appendix~\ref{sec:app_ms}). Contours for the cleaned map and the residual are logarithmic in powers of 2 starting with $2\sigma$, where $1\sigma=18.8\,\mujybeam$, and dashed lines represent negative values. The peak of the emission is detected at  $\mathrm{S/N} = 60$.
		Contours for the PSF indicate 1, 2, 3, 4, 5 and 50\% levels (the last one defining the beam FWHM of $0\farcs20\times0\farcs18$, also shown in the corner of the first panel). 
		{\it The lower left panel} shows the curves of growth measured in different maps as specified in the legend.  The cleaning was performed within a circular mask of 2$\arcsec$ radius. The right-hand $y$-axis is scaled assuming an average bandwidth of 420\,\kms. The flux values are not corrected for the missing flux outside the  bandwidth used (i.e. total fluxes need to be multiplied by 1.19, see Section~\ref{sec:methods_aperture}).  The shaded area indicates the $1\sigma$ uncertainty on the cumulative flux.
		{\it The lower right panel} shows surface brightness in annuli measured in different maps, as specified in the legend, and normalized to the maximum value. The shaded area indicates $1\sigma$ error on the mean, computed as the  $\mathrm{rms}/\sqrt{N}$, where $N$ is the number of beams contained in the annulus. An exponential profile with a scale length of 2\,kpc is shown with the long-dashed line for comparison.}
	
	\label{fig:uvstackcii}
\end{figure*}

Using the analysis template presented in the previous section, we performed  a visual inspection for each of the 27 quasar host galaxies in our sample (for details see Appendix~\ref{sec:app_sources}). 
We find that, in most cases, the full extent of the source is difficult to quantify.  Our chosen aperture sizes have a mean radius of 4\,kpc, but the curves of growth usually hint at additional larger scale emission. Most sources for which the amplitudes are well-fit by a single Gaussian exhibit slightly higher amplitudes at the lowest baselines that probe the largest spatial scales. 
Fluxes measured between different cycles of individual sources are usually consistent, although discrepancies can reach up to 50\% in some cases. This may be attributed to the low S/N of the particular observation, but some remaining calibration issues may also be present.
Unfortunately, the S/N available at short baselines is generally insufficient to make any significant claims regarding extended emission in individual sources.
Three sources with additional 7\,m data show at least 10\% larger amplitudes on short baselines compared to the 12\,m data.
High-resolution imaging ($\sim$1\,kpc), can still be successfully used to recover emission on significantly larger scales because the MRS (see Section~\ref{sec:datalist}) is approximately an order of magnitude larger than the synthesized beam.  
In summary,  visual inspection of individual sources show tentative evidence for extended emission, with apertures larger than the usual $2\sigma$ outlining contours  necessary to collect this emission.

We investigate the mean extent of this \cii\ emission via $uv$-stacking.
The stack was created as described in Section~\ref{sec:uvstackmethod} using the 19 quasar host galaxies in the clean sample (sample 4 from  Section~\ref{sec:samples}, totaling 32 observation runs). In short, from each dataset, which was first continuum subtracted, we selected frequency channels that encompass a total of $1.2\times\mathrm{FWHM}$ of the \cii\ line, and then imaged them all combined, thus obtaining a single $uv$-stacked \cii\ intensity map.
The results are presented in Figure~\ref{fig:uvstackcii}, where the maps highlight the quality of the combined data and the success of the deconvolution process. 
The anisotropic 2D structure is the result of stacking galaxies with random inclinations, that might also have faint undetected companions, and is therefore not meaningful. 

From the curve of growth in Figure~\ref{fig:uvstackcii}, we  conclude that the emission extends up to 10\,kpc ($\sim2\arcsec$), beyond which we do not recover any significant additional flux.
The value at which the cumulative flux saturates ($2.5\pm0.1$\,\jykms), when corrected for missing tails, is consistent with the mean of the line fluxes measured individually from the spectra. 
The corrected and the non-corrected flux densities are consistent within the errors, because most of the flux was cleaned,  and the clean-to-dirty beam ratio is 0.75--1, up to a radius of $3\farcs5$. The dirty beam sidelobes levels are 4\% at most.
We report a simple one parameter size estimate, the half-light radius, defined as the radius where the curve of growth equals to 50\% of the value at 10\,kpc (where the curve reaches saturation).
The \cii\ emission half-light radius is $1.6 \pm 0.1$\,kpc, and the region within 2, 3, 4, 5 and 10\,kpc contains 55\%, 65\%, 75\%, 80\% and 100\% of the flux, respectively.

We also examine the surface brightness profile of the \cii\ emission,  shown in Figure~\ref{fig:uvstackcii}.
The azimuthally-averaged profile measured in the cleaned map (red line) indicates the presence of two components of emission; a steep core, and a broad fainter component. 
The extended component is detected at $>3\sigma$ significance up to a radius of 5\,kpc (and $>2\sigma$ up to 8\,kpc), and is well described by an exponential function with a scale length of 2\,kpc (also shown in Figure~\ref{fig:uvstackcii}). The emission drops to 1\% of the peak beyond 6\,kpc.
Here, we did not perform any residual scaling, but instead calculated the mean value of pixels within some annulus, as the corrections are negligible.

%

We ascertained to what extent the multi-scale cleaning algorithm may bias the shape of the profile by repeating the imaging process with a different and simpler algorithm. The results are presented in Appendix~\ref{sec:app_ms}, and show that, regardless of the clean procedure, we recover the same shape of the surface brightness profile.
Furthermore, in Appendix~\ref{sec:app_uvfit} we fit the brightness profile models directly to the $uv$-data. The results obtained with this method are in agreement with our conclusions drawn from the image plane analysis.
We investigate further possible selection biases by repeating this analysis for several different subsamples in Appendix~\ref{sec:app_subsamples} and find no systematics.

\subsection{Spatial extent of the dust continuum emission}
\label{sec:results_dustextent}

\begin{figure*}
	\centering
	\includegraphics[width=\linewidth]{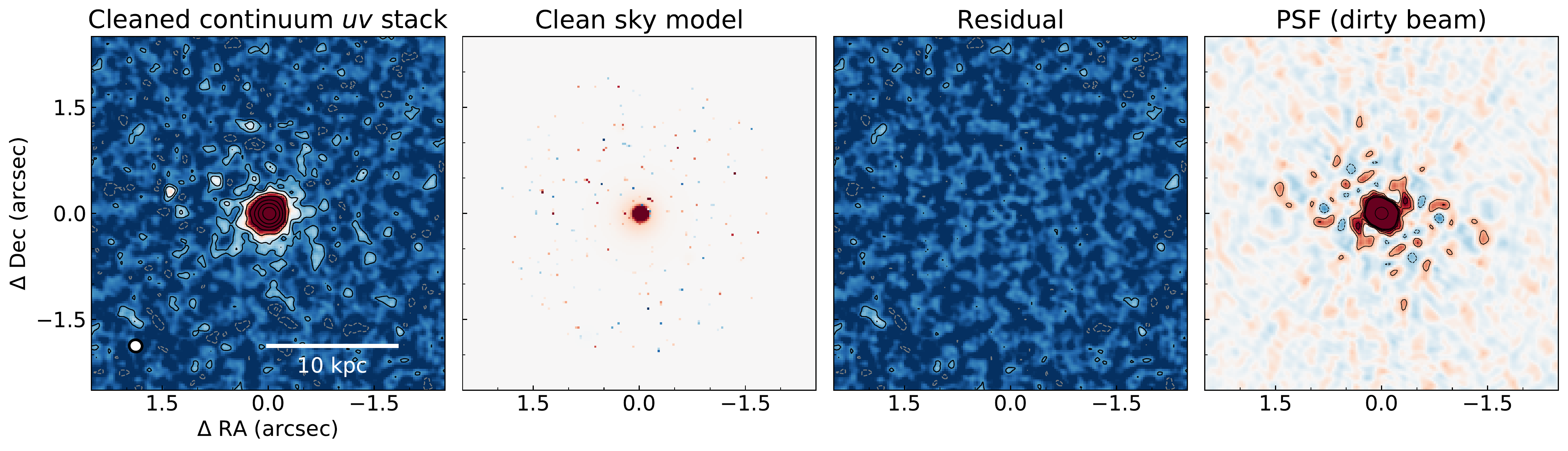}\\
	\includegraphics[height=6.8cm]{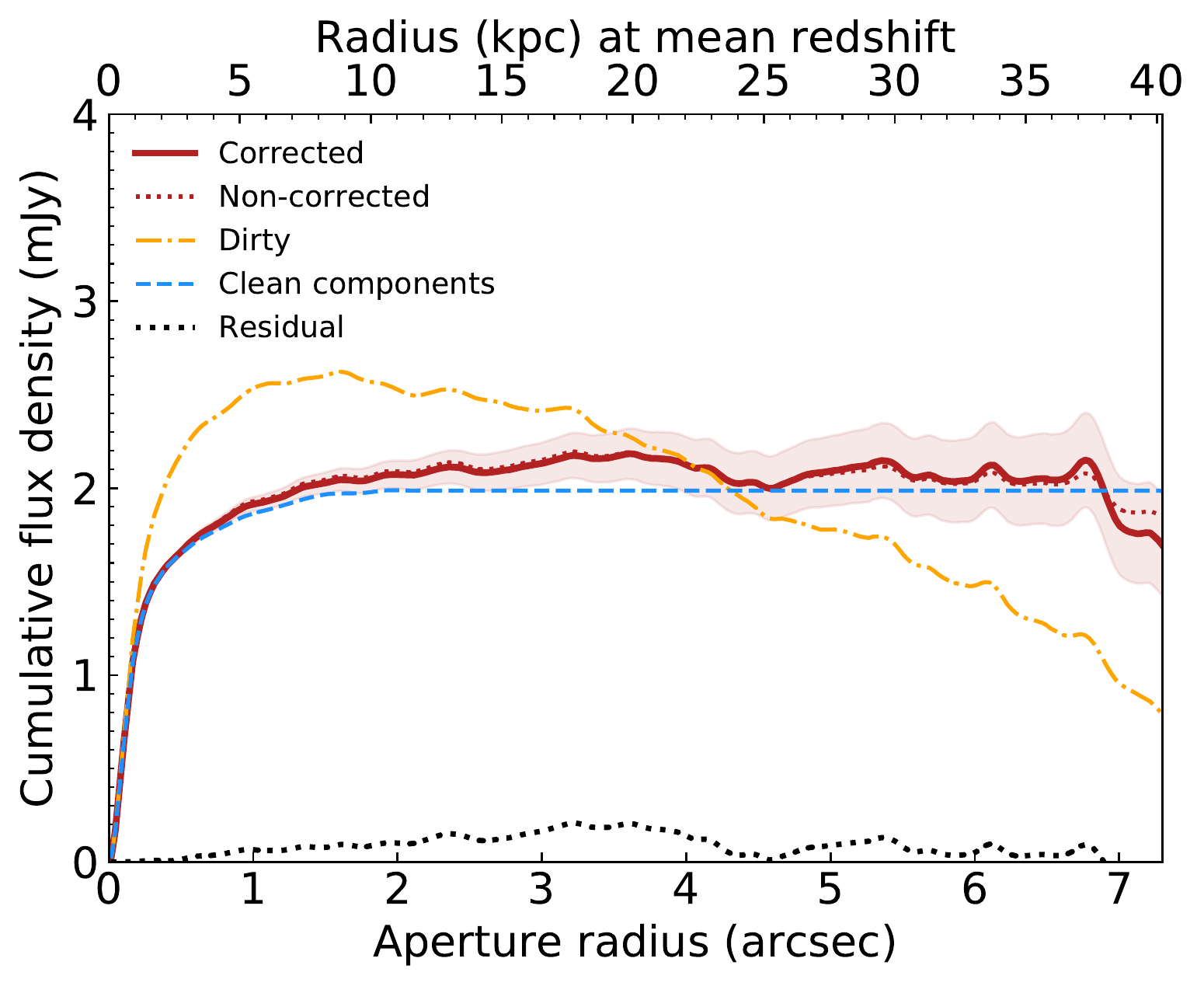}
	\includegraphics[height=6.8cm]{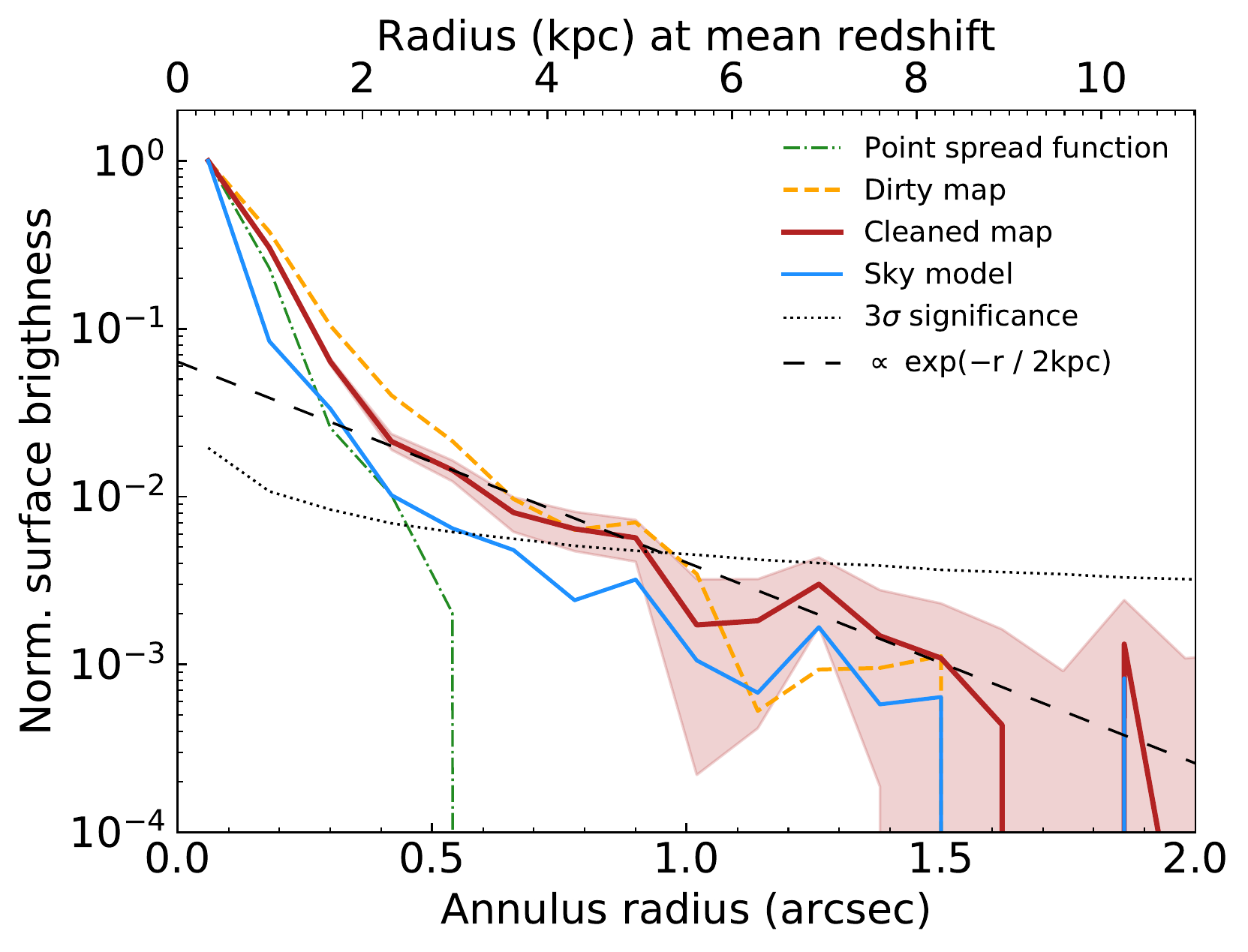}\\
	\caption{Same as Figure~\ref{fig:uvstackcii}, but for the dust continuum emission. The mean observed continuum frequency is 256\,GHz ($\approx1.2\,$mm), the rms in the cleaned map is $1\sigma=4\,\mujybeam$, the peak is observed at the S/N of 200, and the resolution is $0\farcs18\times0\farcs17$.
	}
	\label{fig:uvstackcont}
\end{figure*}

We investigate whether the extended emission is specific to the \cii\ line by replicating the analysis for the dust-continuum emission. 
The stacking method is fully described in Section~\ref{sec:uvstackmethod}. In short, from each full dataset (no continuum subtraction performed) of the clean sample, we removed a total bandwidth of $2.5\times\mathrm{FWHM}$ around the \cii\ line, and imaged the remaining channels together, thereby creating the $uv$-stacked dust continuum map.
We show the results in Figure~\ref{fig:uvstackcont}.
Based on the cumulative aperture flux, the total flux density saturates at 10\,kpc, as was the case with the \cii\ emission. The final value of $2.05 \pm 0.07$\,mJy is consistent with the mean of individual continuum measurements \citep[see][]{venemans20} indicating that no object is biasing our stacked results at a significant level.  The dirty beam sidelobes levels are 3\% at most.
The curve of growth is steeper in the core for the continuum, compared to the \cii. The dust continuum half-light radius, defined as in the previous section, is $0.86 \pm 0.03$\,kpc, and the emission contained within 2, 3, 4, 5 and 10\,kpc corresponds to 75\%, 80\%, 85\%, 90\% and 100\% of the total, respectively.

The compactness of the dust continuum emission is evident in the surface brightness profile, shown in Figure~\ref{fig:uvstackcont}. 
The emission drops to 1\% of the peak beyond 4\,kpc (compared to 6\,kpc in case of \cii).
However, at radii beyond 2\,kpc we also find an extended  component with a significance of  $>3\sigma$ up to a radius of 5\,kpc (and $>2\sigma$ up to 7\,kpc). The exponential function has a scale length consistent with the one we find for the \cii\ emission. 

In summary, both the \cii\ and the dust continuum emission show two-component profiles, where the scale lengths of the extended components are consistent between the two. The central component, which dominates the half-light radius measurement, is more compact in the case of the dust continuum emission.

\subsection{High-velocity outflows}
\label{sec:results_outflow}

\begin{figure}
	\centering
	\includegraphics[width=\linewidth]{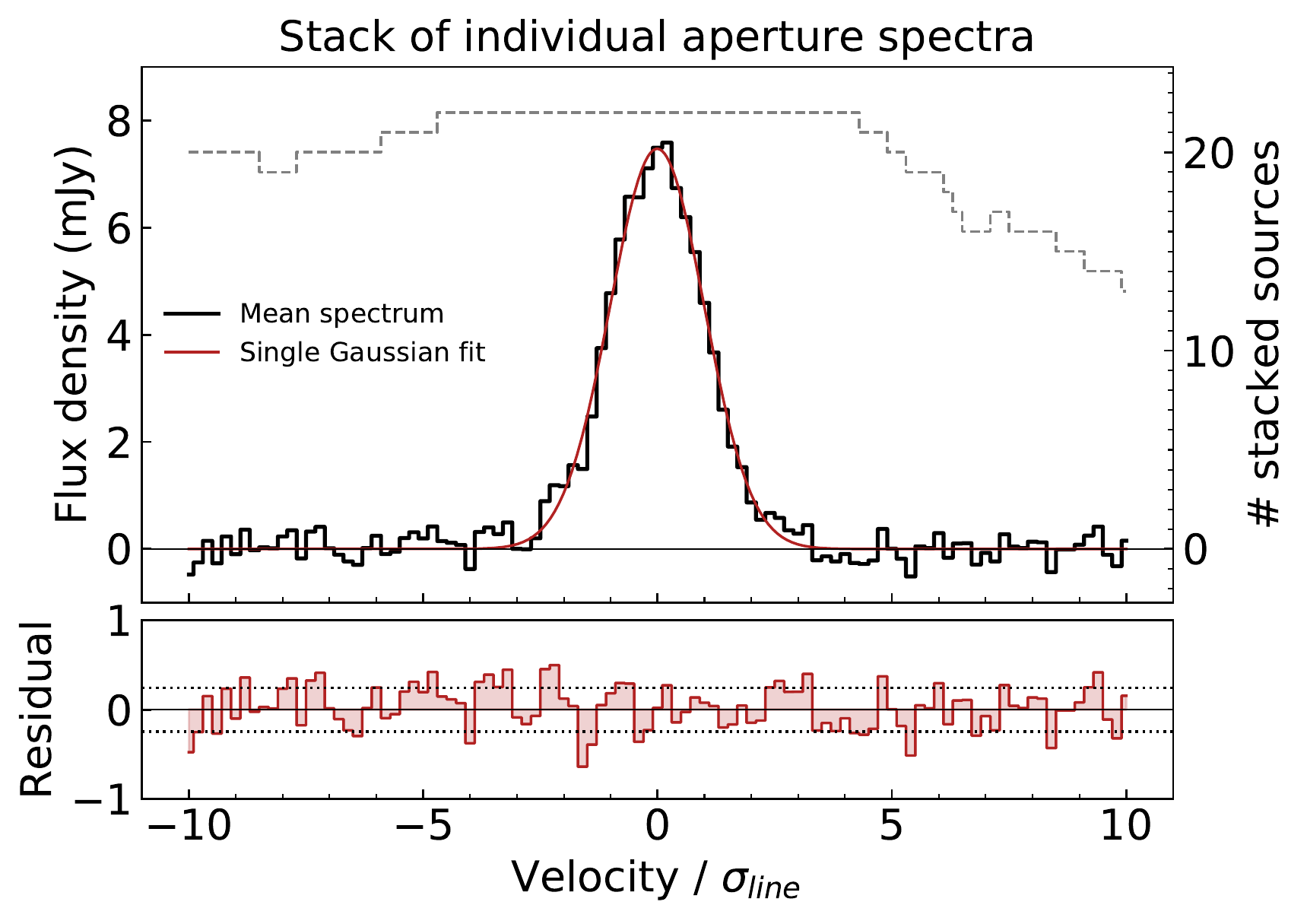}
	\caption{{\it The upper panel} shows the stack of individual \cii\ spectra of the quasar host galaxies, drawn from the clean high-resolution sample (sample 2 from Section~\ref{sec:samples}). Prior to stacking, all spectra were rescaled in velocities by their respective linewidths (mean $\sigma_{\mathrm{line}}=150\,\kms$), and corrected for a constant continuum. The dashed gray line indicates the number of sources available per channel bin.
		{\it The lower panel} shows the residual (in units of mJy) after the subtraction of a single Gaussian. Horizontal dotted lines indicate the $1\sigma$ standard deviation in the residuals. A single Gaussian is a good fit to the stacked spectrum.}
	\label{fig:aperspecstack}
\end{figure}

\begin{figure*}
	\centering
	\includegraphics[width=0.48\linewidth]{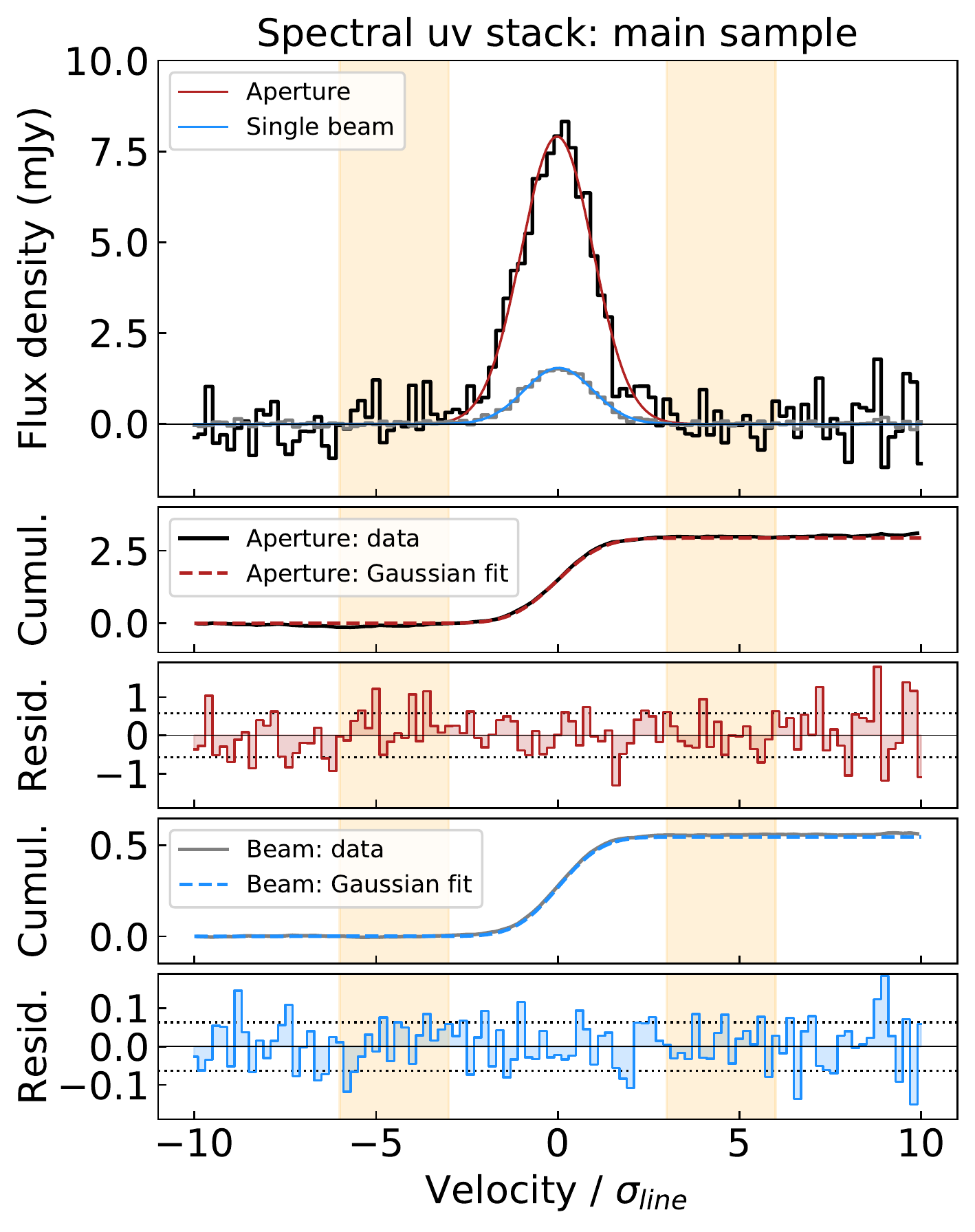}
	\includegraphics[width=0.48\linewidth]{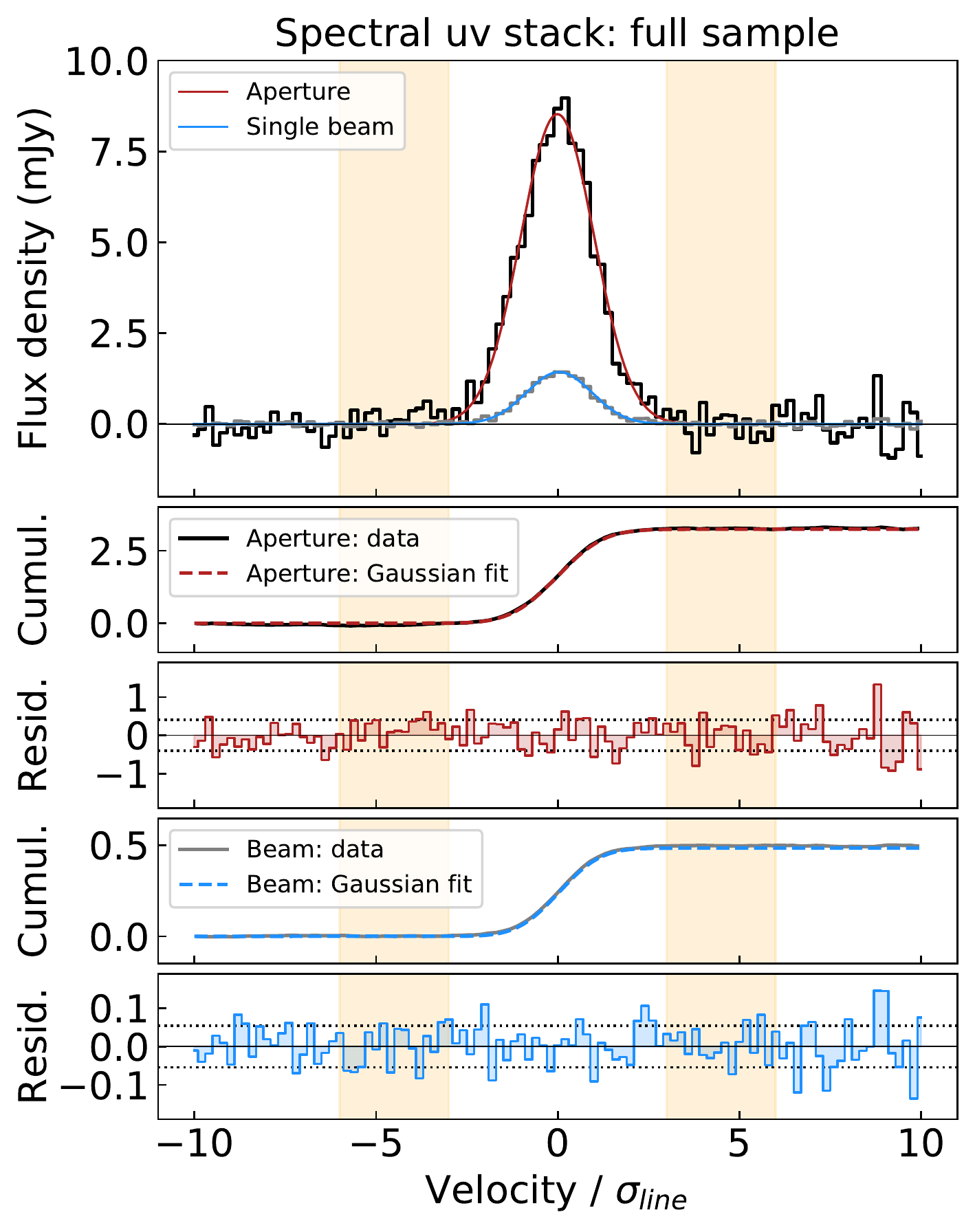}\\
	\hspace{3mm}\includegraphics[width=0.4\linewidth]{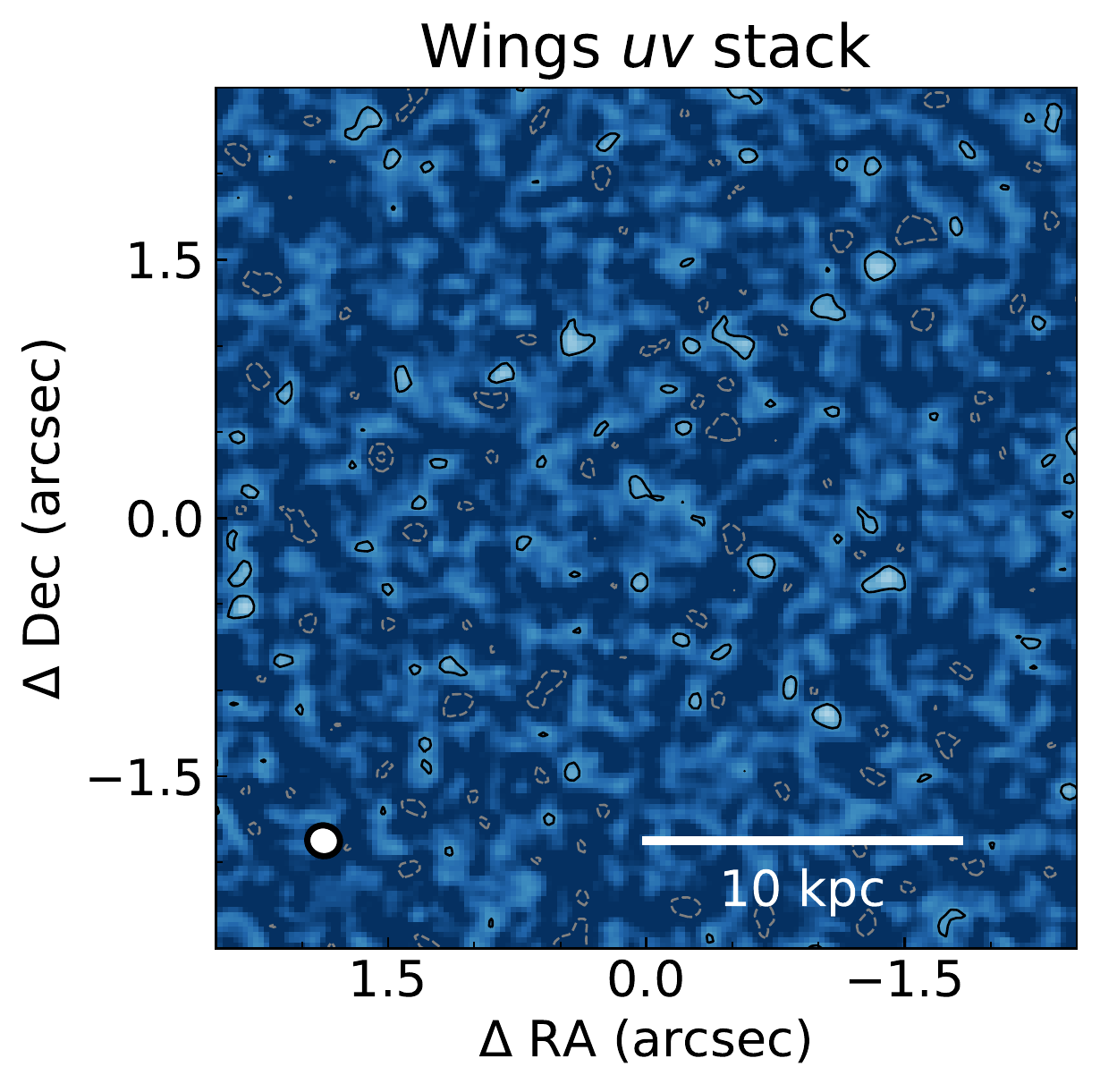}
	\hspace{1cm}
	\hspace{3mm}\includegraphics[width=0.4\linewidth]{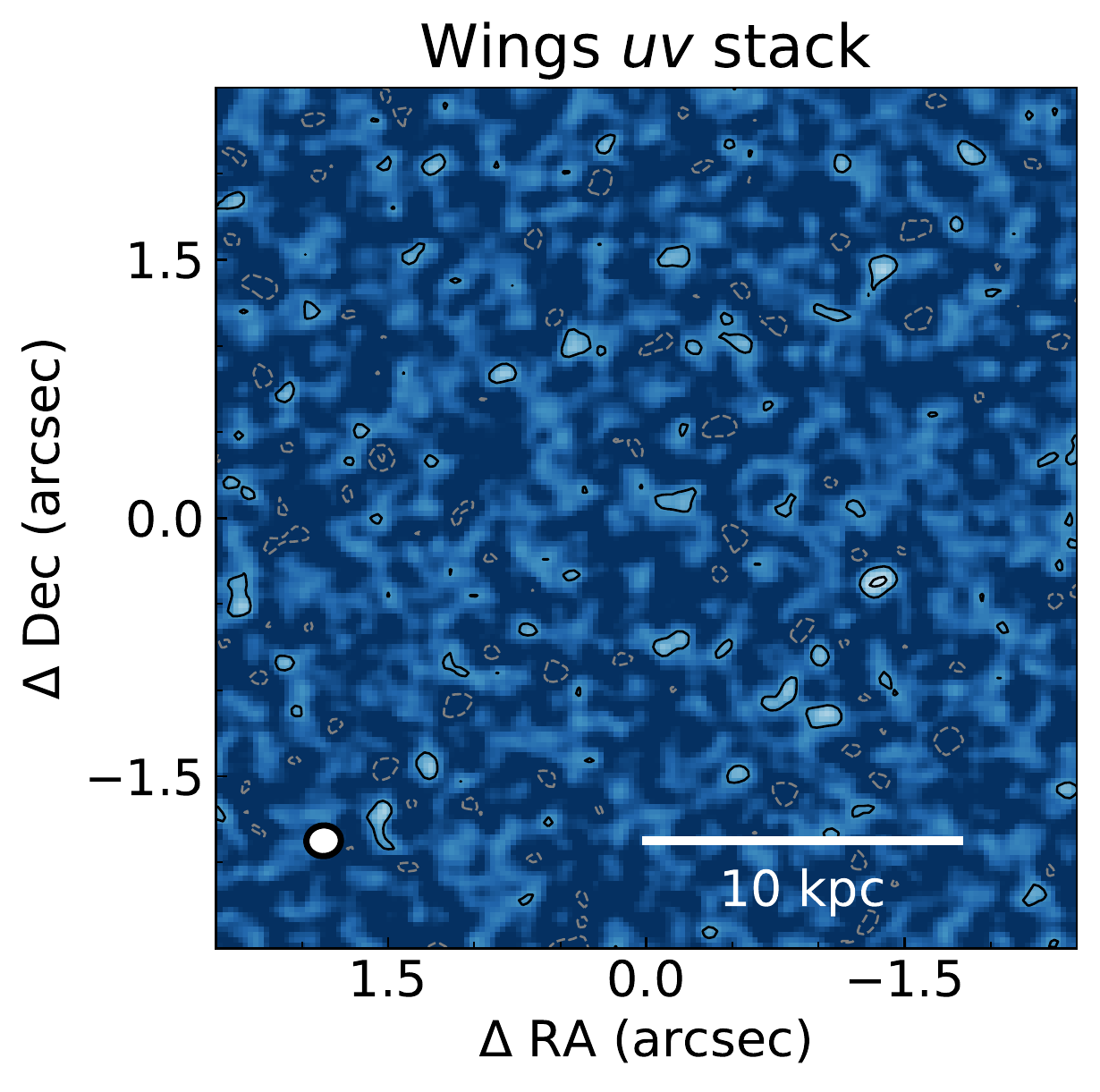}
	\caption{
		The mean \cii\ spectrum of a sample of $z\gtrsim6$ quasar host galaxies, measured from the $uv$-stacked and imaged data.
		 {\it The top panels} show the \cii\ spectra in the main clean sample (left) and the full sample (right).
	The mean linewidth in the sample is $\sigma_{\mathrm{line}}=150\,\kms$.
		The residual-corrected $2\arcsec$ radius aperture spectrum is shown with a black line, with the corresponding single Gaussian fit in red. The gray line shows the single beam (i.e.\ single pixel, the central $\sim$1\,kpc) spectrum, with the corresponding single Gaussian fit in blue.
		The cumulative distribution (in \jykms) and the fitting residual (in mJy) can be seen in the subsequent panels for the aperture and the single beam spectrum, respectively. Horizontal dotted lines in the residual panels show $1\sigma$ standard deviation.
		The orange shaded regions outline the velocity ranges imaged to produce the wings map shown below. We observe no evidence for a broad-line component.
		{\it The bottom panels} show the maps obtained by imaging velocity range between 3$\sigma_{\mathrm{line}}$ and 6$\sigma_{\mathrm{line}}$ (both positive and negative velocity components combined), to maximize the sensitivity to detect a broad emission line feature. The rms level in the map is 10.3\mujybeam\ for the main sample (left) and 8.3\mujybeam\ for the full sample (right). Solid (dashed) contours outline $\pm2\sigma$ and $\pm4\sigma$ significant positive (negative) emission.
		The beam size is shown in the lower-left corner. Again, we find no evidence for a broad emission, that could manifest as, e.g.,  a peak in the center of the map, or an excess of positive emission at larger annuli.
	}
	\label{fig:uvspecstack}
\end{figure*}

We turn our attention to the \cii\ spectra to investigate the possible presence of broad components that may be indicative of outflows.

A single Gaussian describes the \cii\ spectrum well in all of the individual cases (see  Appendix~\ref{sec:app_sources}), except in a few that are known for hosting a merger or a companion system (see Section~\ref{sec:comp}).
The stacked spectrum is also well fit by a single Gaussian, as shown in Figure~\ref{fig:aperspecstack}.
Here, the peak of the stacked \cii\ line is detected at a S/N of 32 in velocity bins of $0.2\sigma_{\mathrm{line}}$. There is no evidence of any broad spectral component.
Fitting two Gaussians would collect marginally more emission, only at velocities between $-6\sigma_{\mathrm{line}}$ and $-2\sigma_{\mathrm{line}}$, yielding 3\% larger integrated line flux density.
One caveat of this stacking  approach is that it relies on manually-chosen aperture sizes, which means that all stacked objects do not contribute equally on all spatial scales. 
We further improve on the result by performing a $uv$-plane spectral stack next.

We also find no evidence for a broad spectral component, indicative of a high-velocity outflow, in the \cii\ spectrum of the $uv$-stacked data.
The \cii\ spectrum drawn from the on the $uv$-stacked and imaged data cube of the clean sample (sample 4 from the Section~\ref{sec:samples}) is presented in the left panels of Figure~\ref{fig:uvspecstack}. The spectral $uv$-stacking procedure is described in detail in Section~\ref{sec:uvstackmethod}.
In short, from each continuum subtracted dataset we selected channels that accumulate to slices of $0.2\sigma_{\mathrm{line}}$. We then imaged together slices that correspond to the same offset from the line peak. There are 100 such slices, covering velocities between $\pm10\sigma_{\mathrm{line}}$.

The rms noise in the velocity range between  $\pm6\sigma_{\mathrm{line}}$ is 60\,\mujybeam\ per channel of $0.2\sigma_{\mathrm{line}}$ (which on average corresponds to 30\,\kms).
We measured the $2\arcsec$ radius aperture spectrum (residual corrected) from the $uv$-stack imaged cube, which should encompass the entire \cii\ emission, according to our previous spatial scale analysis. Additionally, we performed a single beam (i.e.\ single pixel) measurement at the central \cii\ peak position following \cite{decarli18}. These two measurements are both necessary as we do not know the spatial scale of the potential broad spectral component.  Outflows are more likely to be dominant at small angular offsets from the quasar, however this is not the case if the high velocity component is due to companion galaxies.
The residuals that remain after subtracting a single Gaussian from the spectrum show no evidence of a broad spectral component, either in the core, or on large scales (see Figure~\ref{fig:uvspecstack}). 
The total \cii\ flux contained within the single Gaussian fit, taking the $\sigma_{\mathrm{line}}=150\,\kms$, is equal to $2.9 \pm 0.1\,\jykms$, consistent with the value measured from the $uv$-stacked intensity map (see Section~\ref{sec:results_ciiextent}). 
The difference in the cumulative flux density across the spectrum between the observed data and the fit is less than 1\%. 
If we measure the \cii\ spectrum in an annulus between 1$\arcsec$ and 2$\arcsec$, the spectral shape is equivalent to the single beam and the full aperture measurement. 

Even when maximising the available signal, we recover no broad component of emission.
We therefore selected and stacked velocity ranges between $\pm (3~\mathrm{to}~6) \sigma_{\mathrm{line}}$ in the $uv$-plane. This velocity range corresponds to 450 -- 900\,\kms\ on average.  If a broad spectral component were to exist, its S/N would be maximized in such a map, which is shown in Figure~\ref{fig:uvspecstack}. This map has an rms noise level of $10.8\,\mujybeam$, and can be considered as a collapse over 900\,\kms. No significant emission is detected, i.e.\ the map is consistent with pure random noise.
To be certain that we have not missed any fainter, more extended emission, we repeated the analysis on the full $uv$-stacking data sample (sample 3 from the Section~\ref{sec:samples}).
The spectrum and map of these wings are  shown in the right panels of Figure~\ref{fig:uvspecstack}. The rms noise in the cube is 48\,\mujybeam\ per channel, which is 20\% lower compared to the more restricted clean sample. The final result remains the same. 

We conclude that, on average, $z\gtrsim6$ quasar host galaxies do not show evidence of broad-linewidth \cii\ outflow, in either the aperture spectrum, single beam spectrum, or the extended wings map.
However, we cannot rule out the possibility of a very faint outflow, below our stack detection limit. For that purpose we define an upper limit on the outflow emission as follows. By taking the
rms level of the wings map, and its bandwidth, we estimate a $3\sigma$ upper limit on the outflow in the very core (single $\sim1$\,kpc beam) of 0.03\,\jykms, which corresponds to less than 5\% of the core peak emission.
Alternatively, the spatially integrated \cii\ line of average $\mathrm{FWHM}\sim 350\,\kms$ detected at peak $\mathrm{S/N}\sim30$ in velocity bins of $\sim30\,\kms$  shows no outflow signatures.
If outflows are ubiquitous in high-redshift quasar host galaxies, the flux contained in such a broad component is negligible compared to the main narrow component. Although individual objects may exhibit significant outflows, these galaxies should then be considered as being outliers (we found no such object in our studied sample). 
We note that a secondary broad spectral feature can be obtained if one does not employ linewidth normalization, as discussed in Appendix~\ref{sec:app_velstack}. 

\section{Discussion} \label{sec:disc}

In the following sections we interpret and discuss our results in the context of published work.

\subsection{Spatial extent of the \cii\ and dust emission}

\begin{figure}
	\centering
	\includegraphics[width=\linewidth]{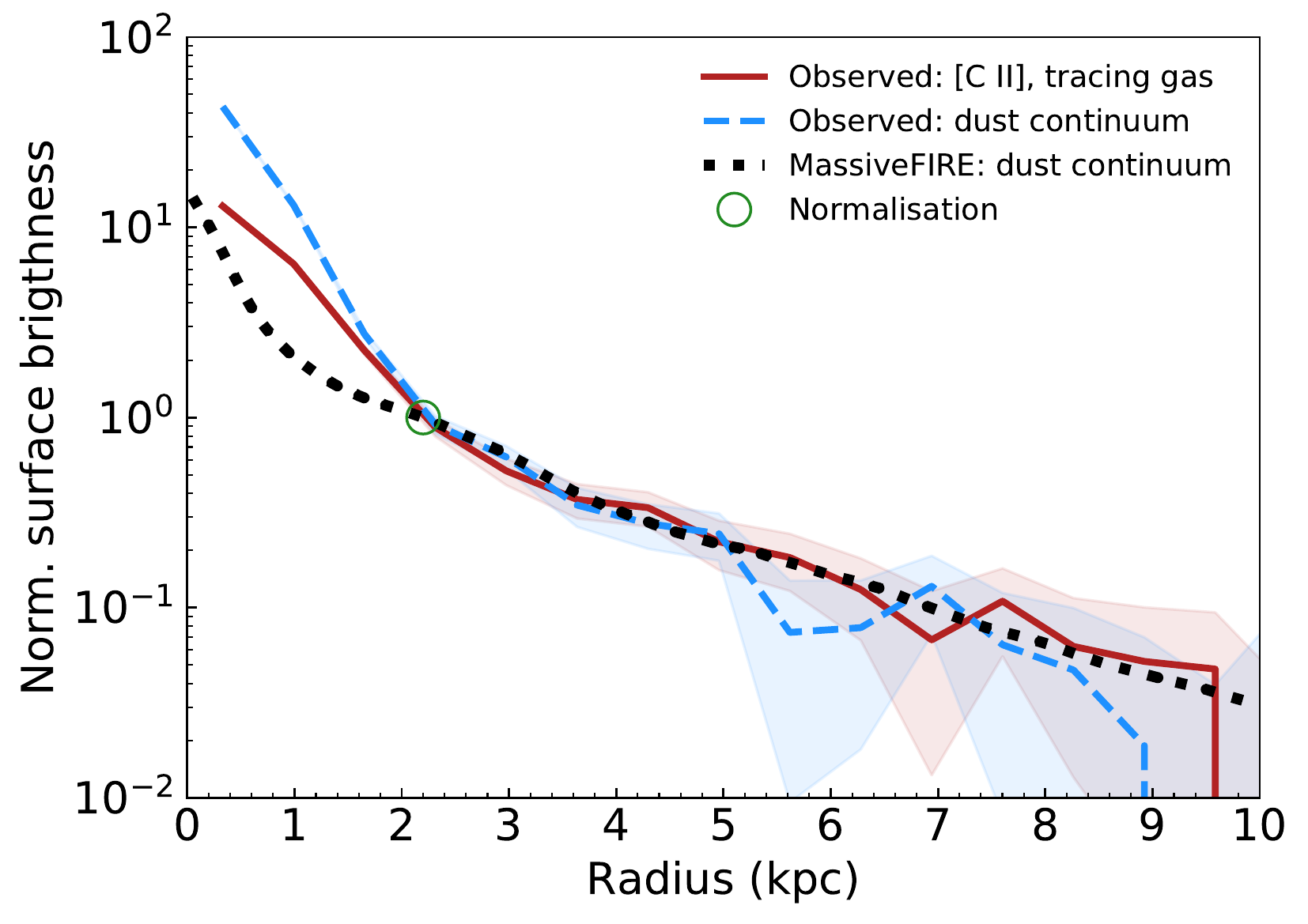}
	\caption{Comparison of average surface brightness profiles, normalized in the region of the extended faint emission. 
		Observed profiles are measured from the $uv$-stacked sample of  $z\gtrsim6$ quasar host galaxies (clean sample, same profiles as in Figures~\ref{fig:uvstackcii} and \ref{fig:uvstackcont}). 
	The simulated profile is measured in the stack of a sample of massive galaxies at $z\sim6$, where no SMBH modeling was performed \citep[see][]{feldmann16,feldmann17}. The tails of the profiles have comparable exponential scale lengths in all three cases.}
	\label{fig:profiles_compare}
\end{figure}

\begin{figure}
	\centering
	\includegraphics[width=\linewidth]{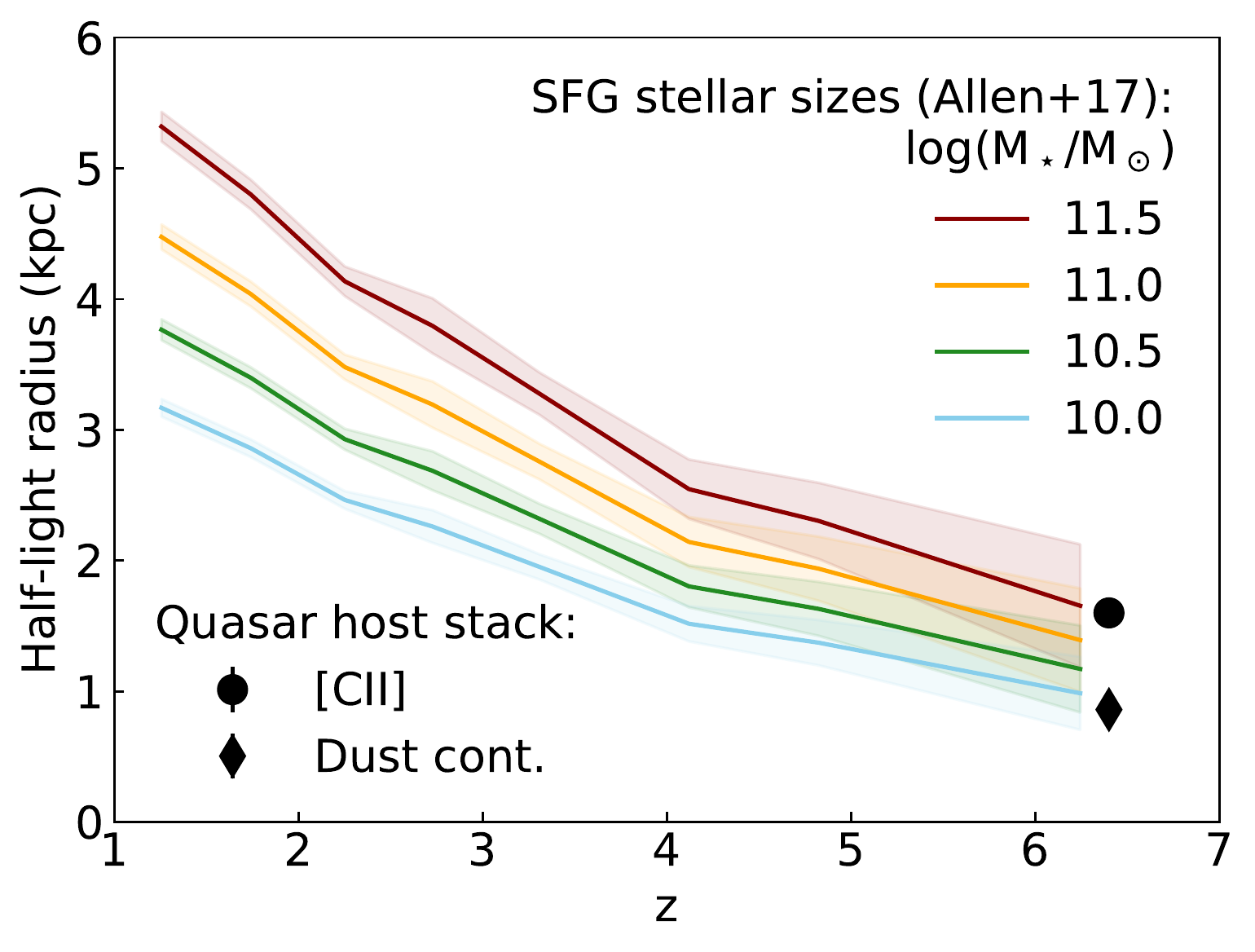}
	\caption{Comparison between the stellar half-light radii of star-forming galaxies (SFGs) from a mass-complete sample  \citep{allen17}, with different colors referring to specific stellar mass bins, and the two measurements from our stacks of $z\gtrsim6$ quasar host galaxies. The uncertainties of our measurements correspond only to the errors on the mean half-light radii, and are smaller than the symbols.}
	\label{fig:size_evol}
\end{figure}

We find that the dust continuum and the interstellar gas, traced by the \cii\ emission, follow the same exponential surface brightness profiles extending between 2 and 10\,kpc, as shown in Figure~\ref{fig:profiles_compare}, implying that the \cii-to-FIR ratio\footnote{This refers to the surface-brightness ratio only. Computing the luminosity ratio would require assuming a dust SED, and a spatial dust temperature distribution, which is unknown for these sources \citep[see accompanying paper,][]{venemans20}.} is constant at large radii. These similar extents imply that the \cii\ emission is tracing the ISM of the galaxies, rather than halos or outflows.

The central ISM component dominates the dust and \cii\ emission. The emission within 2\,kpc contains 55\% (75\%) of the total \cii\ (dust continuum) flux density with a 20\% (10\%) contribution from beyond 5\,kpc.
Such observations are rare for high-redshift quasar host galaxies. For the $z\sim6.4$ quasar SDSS J1148+5251 \cite{walter03, walter09} found a compact \cii\ emission embedded in a more extended gas reservoir (traced by the CO line). However, these observations could not recover the faint component probed here via stacking due to sensitivity constraints. Recently, more high-resolution observations were conducted targeting additional quasar host galaxies. 
These showed that the dust emission is also centrally concentrated, even more so than the \cii\ emission \citep[e.g.,][]{wang19, venemans19}.

Although their properties differ from those of our $z\sim6$ quasar host galaxies, the majority of previously observed and simulated star-forming galaxies also appear to exhibit a dominant, compact component of \cii\ and/or dust emission \citep[e.g.,][]{olsen15,hodge16, chen17, calistro-rivera18, gullberg18,gullberg19,cochrane19, rybak19}. Some high-redshift observations also recover extended faint components of \cii\ or dust emission \citep[e.g.,][]{rybak19, gullberg19, fujimoto19,fujimoto20}.

The extent of the ISM in our quasar host galaxies, traced by \cii\ and dust emission, appears to be consistent with the expected extent of the stellar component. For high-redshift quasar host galaxies, such as our sample, no stellar emission has been detected likely due to the presence of a strong central point source and dust obscuration \citep[see e.g.,][]{mechtley12}.
We therefore compare the \cii\ and dust sizes measured here to the predictions from the ZFOURGE survey, see Figure~\ref{fig:size_evol}. Based on the simulations of \cite{marshall19b}, which probe the properties of quasar host galaxies, we assume that galaxies in our sample have stellar masses of $10^{10.5 - 11}$\,\msol. For galaxies of these stellar masses at $z\sim7$, the stellar mass - size evolution fit by \cite{allen17} indicates that our sample should have stellar half-light radii of 1 -- 2\,kpc, consistent with the \cii\ and dust half-light radii. Note that for the $z\sim4.3$ star-forming galaxy, for which the \cii\ was probed at high sensitivity, \cite{neeleman20} measure consistent \cii\ and stellar half-light radii. Moreover, studies of local galaxies typically find consistent stellar and dust half-light radii, albeit with some scatter depending on the wavelength of the two tracers \citep[e.g.,][]{leroy09,hunt15, casasola17}.

\subsection{Dust emission from the MassiveFIRE simulation}

\begin{figure*}
	\centering
		\includegraphics[width=0.24\linewidth]{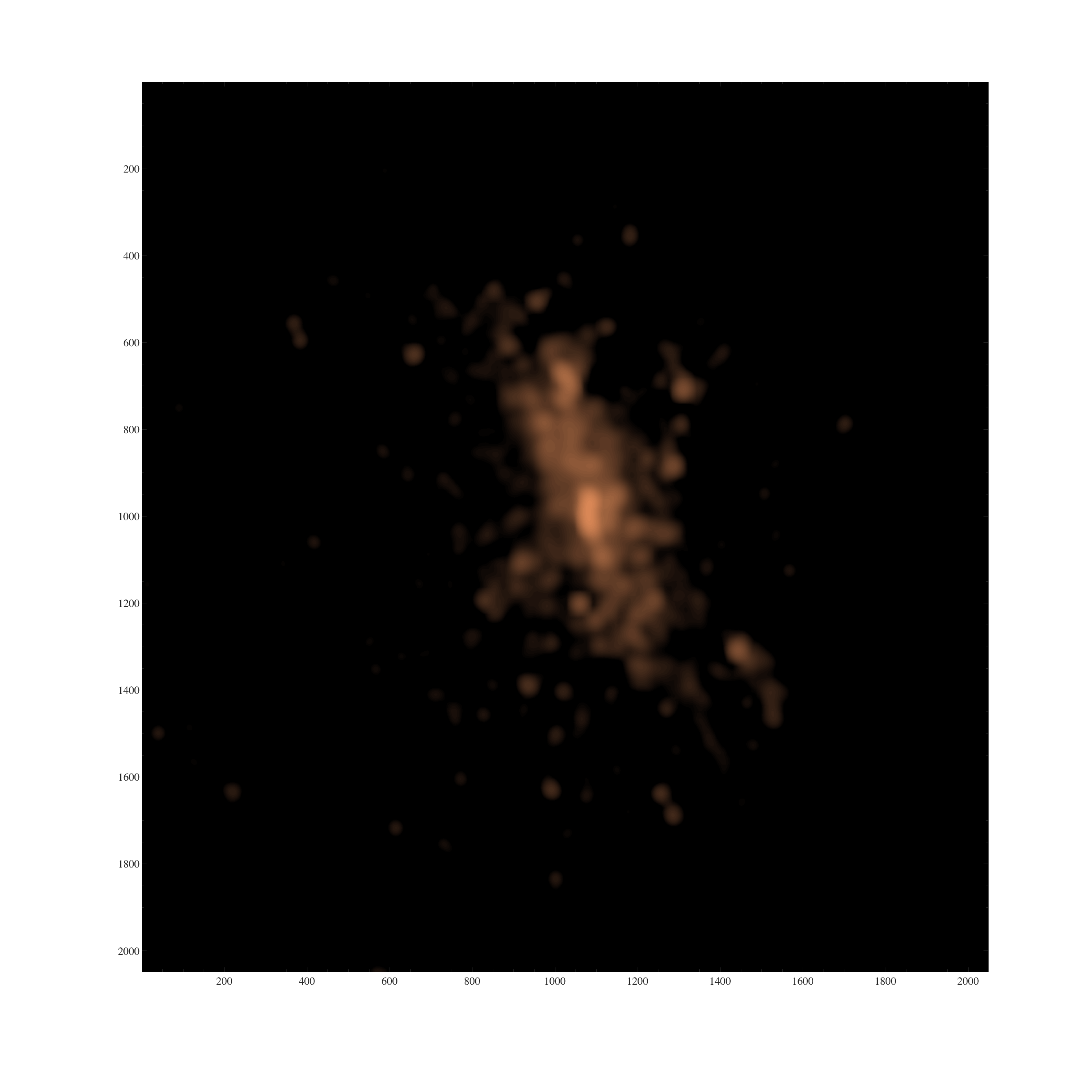}
		\includegraphics[width=0.24\linewidth]{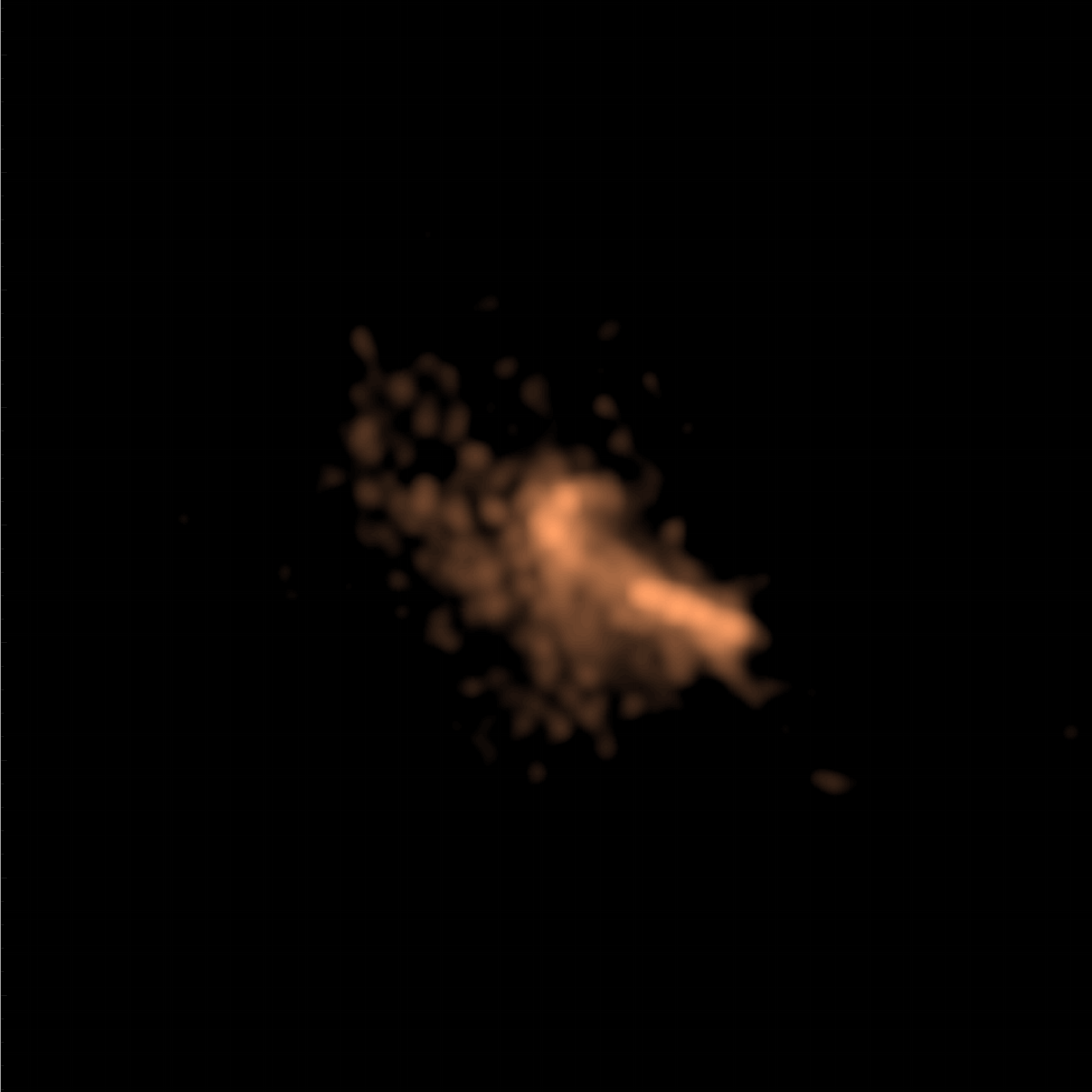}
		\includegraphics[width=0.24\linewidth]{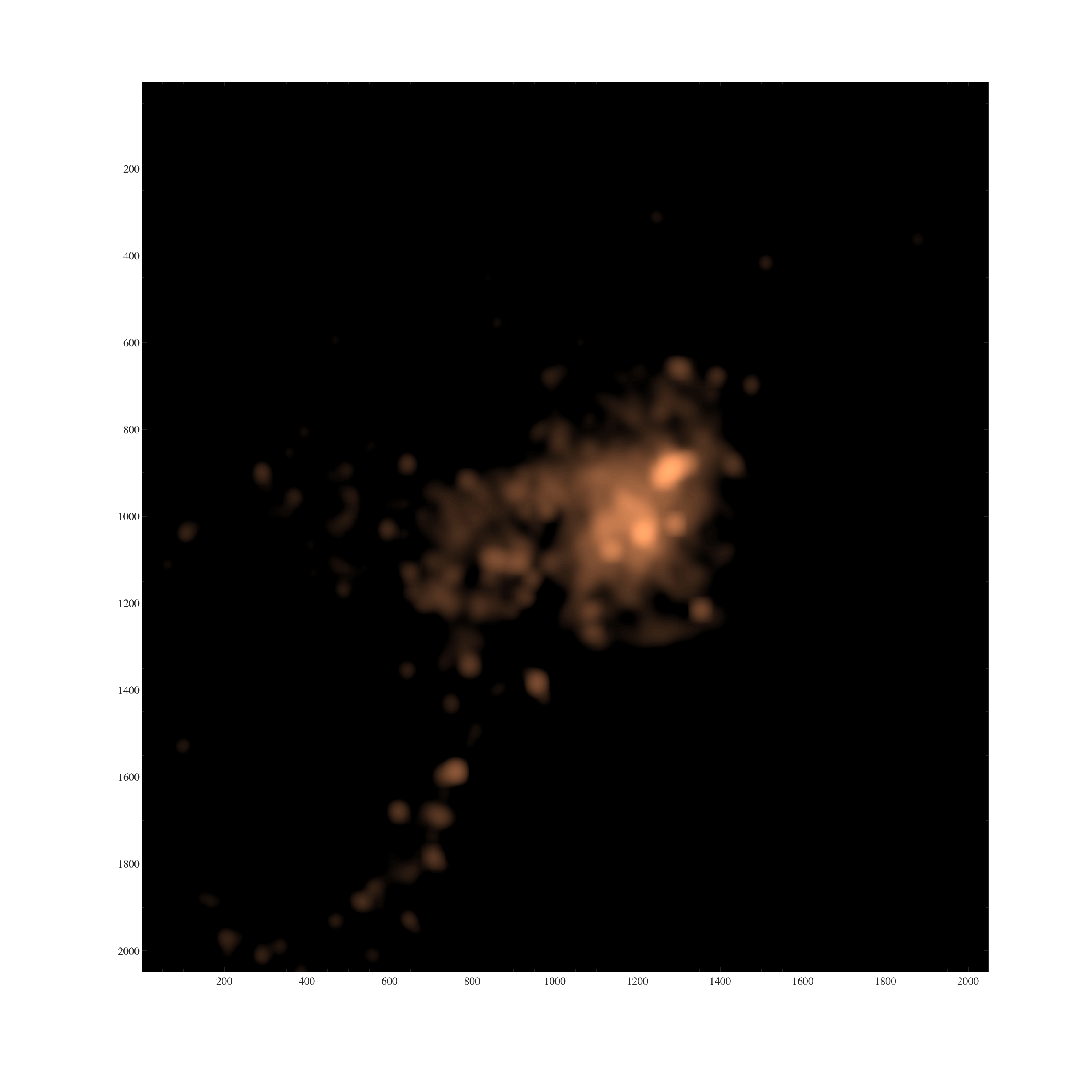}
		\includegraphics[width=0.24\linewidth]{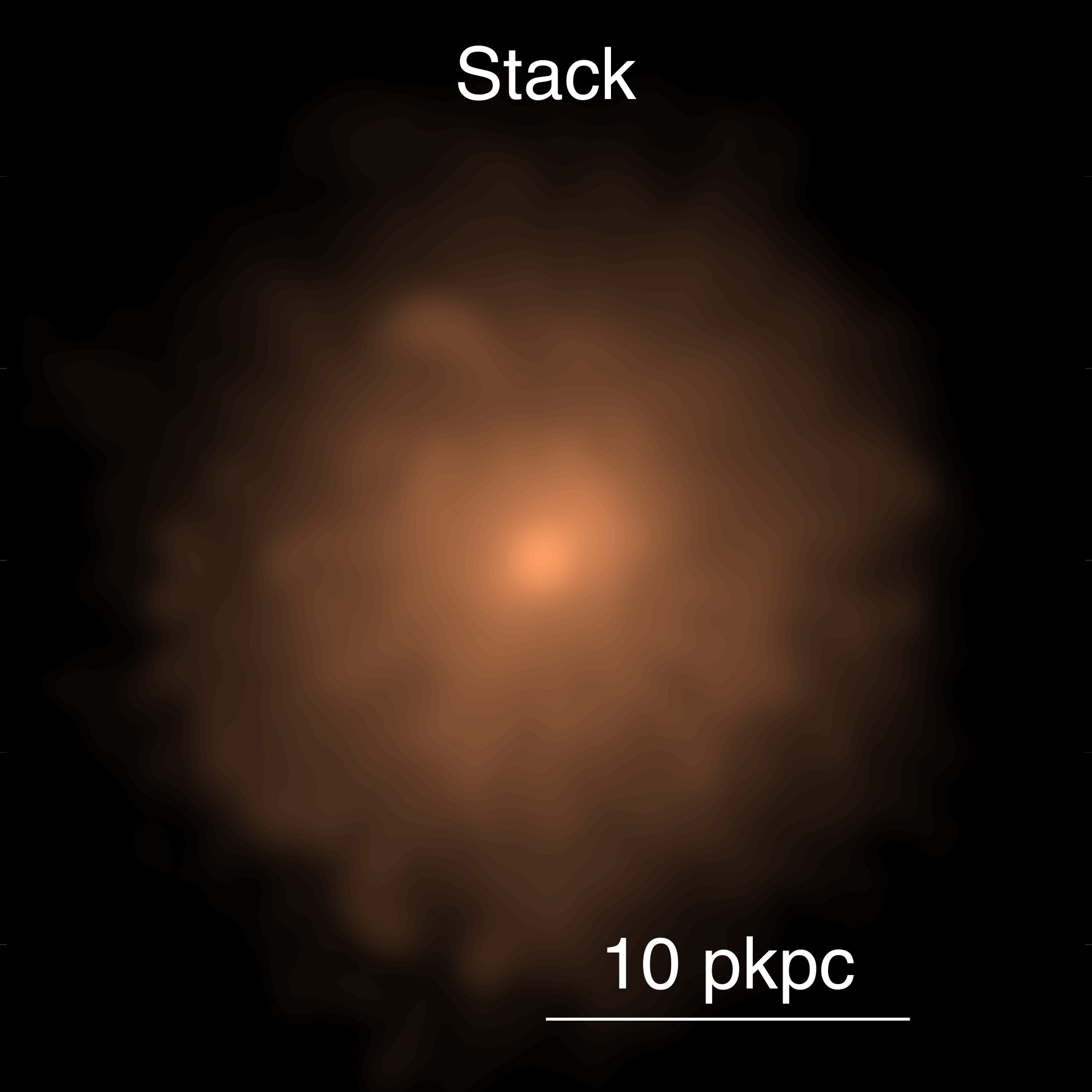}
	\caption{
			{\it The first three panels} show the observed FIR emission in ALMA Band 6 (1.2\,mm) of three examples galaxies at $z\sim 6$ from the MassiveFIRE simulation. These galaxies have IR luminosities comparable to our quasar host galaxy sample. The maps are smoothed with a Gaussian kernel with FWHM of 1\,kpc, and are 30\,kpc on the side.
		{\it The last panel} shows the stack of 5 galaxies at 18 random viewing angles.
	}
	\label{fig:simul}
\end{figure*}

We compare our observational data with predictions from the MassiveFIRE suite of cosmological zoom-in simulations \citep{feldmann16,feldmann17}, which is a part of the Feedback in Realistic Environments (FIRE) project \citep{hopkins14}. 
Galaxy formation simulations offer the prospect of studying the properties of massive, high-redshift galaxies with high spatial resolution and from multiple viewing angles.

All simulations start from cosmological initial conditions and account for star formation and various stellar feedback channels such as momentum injection from supernovae, stellar winds, photo-heating, and radiation pressure. Feedback from SMBHs is not included. The physical model employed by these FIRE simulations is described in detail in \cite{hopkins14}. The spatially resolved UV-to-mm spectral energy distributions (SEDs) are calculated in post-processing via self-consistent dust radiative transfer calculations with the help of SKIRT \citep{baes11,baes15}.
A cosmic microwave background temperature floor is included in the calculation. ALMA broadband fluxes are computed by convolving the redshifted SEDs with the ALMA transmission functions. Further details are described in \cite{liang19}.

In Figure~\ref{fig:simul}, we show several examples of galaxies from the D and E series of the MassiveFIRE suite at $z = 6$. These simulated galaxies reside in halos with virial masses of $(2-6)\times10^{12}\,\msol$, have star-formation rates (SFRs) of 70 -- 400\,\msolyr, and total IR luminosities (8 -- 1000\,$\mu$m) of about $10^{12}$\,\lsol, comparable to our sample of quasar hosts.
We calculate the radial profile of the sub-mm emission by stacking the predicted 1.2\,mm ALMA flux density maps of five simulated galaxies, each viewed from 18 random directions, and by averaging the flux densities in radial annuli. 
As for the observational data, all intrinsic structure (different disk orientations etc.) are smoothed out by stacking the simulated galaxies' emission.
The stacked 1.2 mm flux density profile of MassiveFIRE galaxies closely follows both the SFR surface density as well as the dust column density profiles. 

The radial dust profile of the stacked MassiveFIRE galaxies closely follows the radial profile of the dust and \cii\ emission of our quasar host galaxies at radii beyond 2\,kpc  (shown in Figure~\ref{fig:profiles_compare}), while the core emission differs, most likely due to selection bias.
The observed sources host an actively accreting SMBH at their centers, which may be associated with enhanced levels of star formation activity, whereas the simulated galaxies do not, and are simply matched to our sample in FIR luminosity.
Although the simulations and observations do not pertain to the same population of galaxies, we believe that the comparison of the extended profiles is valid, assuming that the feedback from the central SMBH does not severely affect the host (gas and dust) at radii beyond 2 kpc.
Our findings suggest that the observation of extended sub-mm emission around IR-luminous, high-redshift galaxies may be a consequence of an extended distribution of dust, gas, and active star formation, partially due to the large but irregular morphology of some of these luminous objects.
The agreement between simulations and observations suggests that the extended emission seen in the ALMA data are part of the main galaxy's ISM. Furthermore, the lack of correlation between the quasar's bolometric luminosity and the host's FIR emission \citep[see accompanying paper by][]{venemans20} suggests that the feedback from the active galactic nucleus is not dominating the FIR emission.

\subsection{Outflows}

We find no evidence for a broad \cii\ emission line component in the mean spectrum of either the clean or the full quasar host galaxy sample, using multiple analysis approaches, including a novel technique of spectral $uv$-stacking. 
Individual objects also show no evidence for broad-line outflows. 
The same spectral shape, consistent with a single Gaussian, is measured in the central beam ($<1$\,kpc), the larger aperture spectrum ($<10$\,kpc), and the extended annulus (5 -- 10\,kpc) \citep[see also][]{neeleman20b}. A single Gaussian collects 99\% of the observed emission, based on the cumulative flux integral.  Imaging a wider velocity range, where the broad feature may be expected, also yielded a non-detection. 
Our results therefore suggest that high-velocity outflows are atypical for $z>6$ quasar host galaxies.
Objects in our sample are the brightest emitters at $z\gtrsim6$, and we may expect strong quasar feedback based on their bolometric luminosities. However, the lack of outflow signatures in our results suggest that the ISM is not significantly affected by the wind-like feedback in these systems.

The observational evidence for \cii\ outflows in $z\sim6$ quasar host galaxies remains contested.
\cite{decarli18} found no broad spectral feature in their spectral stack of 23 quasar host galaxy population, although only a relatively short $\sim 8$\,min integration time was available per source.
Another stacking analysis, utilizing  26 quasar host galaxies at $z\sim6$, was performed by \cite{stanley19}.
They normalized all line widths prior to stacking and report a broad component of $>700\,$\kms, albeit at less than $1.5\sigma$ significance, which reaches a maximum of $2.5\sigma$ from a favourably-selected subsample. Their result is therefore formally consistent with a non-detection, which is also supported by the deeper data used in our analysis.

\cite{maiolino12} and \cite{cicone15} performed a study of the quasar host galaxy SDSS J1148+5251 and presented evidence for a significant outflow identified as a broad spectral component of $\sim 900\,\kms$, extending out to spatial scales of 30\,kpc. 
To date, these studies represent the only cases where such strong features are reported for a single high-redshift object, which makes SDSS J1148+5251 an outlier in the $z\gtrsim6$ quasar host galaxy population.
A spectral stack of 48 quasar host galaxies beyond $z>4.5$, performed by \cite{bischetti19}, also appears to reveal a very broad $\sim1730\,$\kms\ mean spectral feature (ranging from 700 -- 2500\,\kms, depending on the subsample), with an integrated flux ratio between the broad and the narrow component of $\sim0.2$ (and 0.05 for the peak ratio). 

Our findings appear to be in tension with the results of \cite{bischetti19}, although the studies are of comparable depths.
Given the high significance of their detection, and the large percentage of the total flux being in the broad component, we would also expect to see  a broad line.
This discrepancy could be explained by one or a combination of three factors.
The first explanation is that stacking spectra of different linewidths in velocity space without  witdth normalization may change the spectral shape. 
This effect could be exacerbated in the sample studied in \cite{bischetti19}, as it contains additional objects at $4.5 < z < 6$ with much larger linewidths of up to a FWHM of 800\,\kms\ (other aforementioned studies limit themselves to $z\gtrsim6$ objects).
In Appendix~\ref{sec:app_velstack}, we demonstrate that the broad wings are identified in the spectra stacked without line width renormalization. However, from the shape of the residual, we argue that this is, at least in part, an artifact resulting from the linear addition of Gaussians with various widths.
The second explanation for the discrepancy is that faint or blended companion galaxies that are unresolved at the available resolution could contribute to the broadening of the \cii\ line (with the recent addition of new high resolution data, three additional quasar host galaxies in their sample now have confirmed nearby companions, see Section~\ref{sec:comp}, besides those already considered by the authors).
The third possible issue is that when computing the continuum for its subtraction, \cite{bischetti19} use a zeroth order polynomial on a quarter of the available bandwidth, which ensures that the \cii\ line is properly excluded from the fit, but reduces the S/N of the fit.

The quasar host galaxy samples used in \cite{decarli18}, \cite{bischetti19}, and \cite{stanley19}, largely overlap with our work (our studied sources constitute  15/27, 22/48, and 18/26 of the sample in the three mentioned papers, respectively), which makes the range of different reported results puzzling.
We emphasize that our study is building upon the above mentioned studies by analyzing previously unavailable $\sim$1\,kpc resolution data, which allows us to better identify merging or companion systems, as well as to resolve extended features, if present.
However, a weak link of the analysis in all of the present studies lies in securely separating the continuum from the potential broad spectral component. The trade-off between selecting velocities far enough from the \cii\ line, and the resulting S/N of the continuum fit is likely the greatest culprit causing inconsistent results. In this paper we employed two methods of continuum subtraction. The first one depends on extracting the spectrum from the full (line plus continuum) cube, and fitting it with a Gaussian plus a constant line. The second one relies on the {\sc uvcontsub} task in CASA to fit the continuum component in the visibilities using a slope and disregarding on average $\sim900$\,\kms\ worth of data centered on the \cii\ line.
Both methods yielded the same result, i.e.\ a non-detection.
The only way to properly address this issue would be to observe much wider bandwidths around the line.
In addition, there are multiple effects inherent to interferometric data that could bias the measurements if not properly accounted for. Image plane based stacking will result in a  beam that is difficult to interpret, and the final map will be a complex combination of dirty residuals, each with a different PSF, and cleaned components. We attempted to mitigate these issues by performing stacking in the $uv$-plane, and applying the residual scaling method to account for the subtleties (see Section~\ref{sec:methods_aperture}) inherent to any cleaned map.

\section{Conclusions} \label{sec:conc}

We have investigated the extent of the interstellar gas and dust in $z\gtrsim6$ quasar host galaxies using ALMA observations of the \cii\ emission line  and the underlying continuum. Our sample consists of 27 objects observed at the high resolution of $\sim1$\,kpc, with lower resolution archival data available for 20 objects. 
We apply the novel approach of $uv$-plane spectral stacking in order to mitigate the issue of beam matching and uncleaned flux, which can become a significant hindrance in image-based stacking methods. This improved the S/N and the fidelity of our analysis. 
Depending on the subsample under investigation, we can accumulate up to 34\,h (with 25.5\,h on the 12\,m array) of on-source time.
From the individual and stacked analysis we conclude the following.

\begin{enumerate}
	\item 	Quasar host galaxies at $z\gtrsim 6$ are characterized by a central ($<2\,$kpc) concentration of \cii\ and dust emission, with mean half-light radii of $1.6 \pm 0.1$\,kpc and $0.86 \pm 0.03$\,kpc, respectively. 
	By stacking we recover a region of fainter \cii\ and dust emission that extends out to 10\,kpc, which is well described by an exponential function with a scale length of 2\,kpc. 
	Beyond 5\,kpc, the mean surface brightness is observed at values less than 1 -- 2\% of the peak, and contributes 10 -- 20\% to the total measured flux density.

	\item The large spatial extent of the dust continuum matches that of the \cii, suggesting that we are observing the gas and dust of the interstellar medium of the quasar host galaxies extending out to 10\,kpc. The extended dust surface brightness profile follows the same trend as obtained from the simulations of massive galaxies (MassiveFIRE) at similar far-infrared luminosities as our sample, further strengthening our interpretation. 
	
	\item We find no evidence of a high-velocity ($\gtrsim 500\,\kms$) \cii\ component, that may be indicative of outflows,  in our $z\gtrsim6$ quasar host galaxy sample. The mean velocity profile of the \cii\ line is consistent with a single Gaussian, with less than 1\% flux excess on both small ($<2\,$kpc), and large ($<10\,$kpc) spatial scales. This finding suggests that the ISM in these extremely active objects is not significantly affected by wind-like feedback.
\end{enumerate}

Our findings imply that the extended emission is due to the extended ISM in the host galaxies, and that no additional component (e.g. a galaxy halo, or a broad-line outflow) needs to be invoked to interpret the measurements.

\acknowledgments
We thank the anonymous referee for providing helpful suggestions that improved the clarity of the paper.
Ml.N., B.P.V., F.W., Ma.N., and A.B.D. acknowledge support from the ERC Advanced Grant 740246 (Cosmic Gas).
M.K. acknowledges support from the International Max Planck Research School for Astronomy and Cosmic Physics at Heidelberg University (IMPRS-HD).
R.F. acknowledges financial support from the Swiss National Science Foundation (grant no 157591). MassiveFIRE simulations were run with resources provided by the NASA High-End Computing (HEC) Program through the NASA Advanced Supercomputing (NAS) Division at Ames Research Center, proposal SMD-14-5492. Additional computing support was provided by HEC allocations SMD-14-5189, SMD-15-5950, and NSF XSEDE allocations AST120025, AST150045.

This paper makes use of the following ALMA data:
ADS/JAO.ALMA\#2012.1.00240.S, \\
ADS/JAO.ALMA\#2012.1.00882.S, \\
ADS/JAO.ALMA\#2013.1.00273.S, \\
ADS/JAO.ALMA\#2015.1.00399.S, \\
ADS/JAO.ALMA\#2015.1.00606.S, \\
ADS/JAO.ALMA\#2015.1.00692.S, \\
ADS/JAO.ALMA\#2015.1.00997.S, \\
ADS/JAO.ALMA\#2015.1.01115.S, \\
ADS/JAO.ALMA\#2016.1.00544.S, \\
ADS/JAO.ALMA\#2016.A.00018.S, \\
ADS/JAO.ALMA\#2017.1.00396.S, \\
ADS/JAO.ALMA\#2017.1.01301.S, \\
ADS/JAO.ALMA\#2018.1.00908.S.\\
ALMA is a partnership of ESO (representing its member states), NSF (USA) and NINS (Japan), together with NRC (Canada), NSC and ASIAA (Taiwan), and KASI (Republic of Korea), in cooperation with the Republic of Chile. The Joint ALMA Observatory is operated by ESO, AUI/NRAO and NAOJ.

This research makes use of CASA  \citep{mcmullin07}, Astropy,\footnote{http://www.astropy.org} a community-developed core Python package for Astronomy \citep{astropy:2013, astropy:2018}, Matplotlib \citep{Hunter:2007}, and Topcat \citep{topcat} .

\bibliographystyle{aasjournal}
\bibliography{refs}

\begin{thebibliography}{}
\expandafter\ifx\csname natexlab\endcsname\relax\def\natexlab#1{#1}\fi
\providecommand{\url}[1]{\href{#1}{#1}}
\providecommand{\dodoi}[1]{doi:~\href{http://doi.org/#1}{\nolinkurl{#1}}}
\providecommand{\doeprint}[1]{\href{http://ascl.net/#1}{\nolinkurl{http://ascl.net/#1}}}
\providecommand{\doarXiv}[1]{\href{https://arxiv.org/abs/#1}{\nolinkurl{https://arxiv.org/abs/#1}}}

\bibitem[{{Allen} {et~al.}(2017){Allen}, {Kacprzak}, {Glazebrook}, {Labb{\'e}},
  {Tran}, {Spitler}, {Cowley}, {Nanayakkara}, {Papovich}, {Quadri},
  {Straatman}, {Tilvi}, \& {van Dokkum}}]{allen17}
{Allen}, R.~J., {Kacprzak}, G.~G., {Glazebrook}, K., {et~al.} 2017, \apjl, 834,
  L11, \dodoi{10.3847/2041-8213/834/2/L11}

\bibitem[{{Arnaboldi} {et~al.}(2007){Arnaboldi}, {Neeser}, {Parker}, {Rosati},
  {Lombardi}, {Dietrich}, \& {Hummel}}]{arnaboldi07}
{Arnaboldi}, M., {Neeser}, M.~J., {Parker}, L.~C., {et~al.} 2007, The
  Messenger, 127, 28

\bibitem[{{Astropy Collaboration} {et~al.}(2013){Astropy Collaboration},
  {Robitaille}, {Tollerud}, {Greenfield}, {Droettboom}, {Bray}, {Aldcroft},
  {Davis}, {Ginsburg}, {Price-Whelan}, {Kerzendorf}, {Conley}, {Crighton},
  {Barbary}, {Muna}, {Ferguson}, {Grollier}, {Parikh}, {Nair}, {Unther},
  {Deil}, {Woillez}, {Conseil}, {Kramer}, {Turner}, {Singer}, {Fox}, {Weaver},
  {Zabalza}, {Edwards}, {Azalee Bostroem}, {Burke}, {Casey}, {Crawford},
  {Dencheva}, {Ely}, {Jenness}, {Labrie}, {Lim}, {Pierfederici}, {Pontzen},
  {Ptak}, {Refsdal}, {Servillat}, \& {Streicher}}]{astropy:2013}
{Astropy Collaboration}, {Robitaille}, T.~P., {Tollerud}, E.~J., {et~al.} 2013,
  \aap, 558, A33, \dodoi{10.1051/0004-6361/201322068}

\bibitem[{{Ba{\~n}ados} {et~al.}(2016){Ba{\~n}ados}, {Venemans}, {Decarli},
  {Farina}, {Mazzucchelli}, {Walter}, {Fan}, {Stern}, {Schlafly}, {Chambers},
  {Rix}, {Jiang}, {McGreer}, {Simcoe}, {Wang}, {Yang}, {Morganson}, {De Rosa},
  {Greiner}, {Balokovi{\'c}}, {Burgett}, {Cooper}, {Draper}, {Flewelling},
  {Hodapp}, {Jun}, {Kaiser}, {Kudritzki}, {Magnier}, {Metcalfe}, {Miller},
  {Schindler}, {Tonry}, {Wainscoat}, {Waters}, \& {Yang}}]{banados16}
{Ba{\~n}ados}, E., {Venemans}, B.~P., {Decarli}, R., {et~al.} 2016, \apjs, 227,
  11, \dodoi{10.3847/0067-0049/227/1/11}

\bibitem[{{Ba{\~n}ados} {et~al.}(2018){Ba{\~n}ados}, {Venemans},
  {Mazzucchelli}, {Farina}, {Walter}, {Wang}, {Decarli}, {Stern}, {Fan},
  {Davies}, {Hennawi}, {Simcoe}, {Turner}, {Rix}, {Yang}, {Kelson}, {Rudie}, \&
  {Winters}}]{banados18}
{Ba{\~n}ados}, E., {Venemans}, B.~P., {Mazzucchelli}, C., {et~al.} 2018, \nat,
  553, 473, \dodoi{10.1038/nature25180}

\bibitem[{{Ba{\~n}ados} {et~al.}(2019{\natexlab{a}}){Ba{\~n}ados}, {Novak},
  {Neeleman}, {Walter}, {Decarli}, {Venemans}, {Mazzucchelli}, {Carilli},
  {Wang}, {Fan}, {Farina}, \& {Rix}}]{banados19}
{Ba{\~n}ados}, E., {Novak}, M., {Neeleman}, M., {et~al.} 2019{\natexlab{a}},
  \apjl, 881, L23, \dodoi{10.3847/2041-8213/ab3659}

\bibitem[{{Ba{\~n}ados} {et~al.}(2019{\natexlab{b}}){Ba{\~n}ados}, {Rauch},
  {Decarli}, {Farina}, {Hennawi}, {Mazzucchelli}, {Venemans}, {Walter},
  {Simcoe}, {Prochaska}, {Cooper}, {Davies}, \& {Chen}}]{banados19b}
{Ba{\~n}ados}, E., {Rauch}, M., {Decarli}, R., {et~al.} 2019{\natexlab{b}},
  \apj, 885, 59, \dodoi{10.3847/1538-4357/ab4129}

\bibitem[{{Baes} \& {Camps}(2015)}]{baes15}
{Baes}, M., \& {Camps}, P. 2015, Astronomy and Computing, 12, 33,
  \dodoi{10.1016/j.ascom.2015.05.006}

\bibitem[{{Baes} {et~al.}(2011){Baes}, {Verstappen}, {De Looze}, {Fritz},
  {Saftly}, {Vidal P{\'e}rez}, {Stalevski}, \& {Valcke}}]{baes11}
{Baes}, M., {Verstappen}, J., {De Looze}, I., {et~al.} 2011, \apjs, 196, 22,
  \dodoi{10.1088/0067-0049/196/2/22}

\bibitem[{{Bischetti} {et~al.}(2019){Bischetti}, {Maiolino}, {Carniani},
  {Fiore}, {Piconcelli}, \& {Fluetsch}}]{bischetti19}
{Bischetti}, M., {Maiolino}, R., {Carniani}, S., {et~al.} 2019, \aap, 630, A59,
  \dodoi{10.1051/0004-6361/201833557}

\bibitem[{{Calistro Rivera} {et~al.}(2018){Calistro Rivera}, {Hodge}, {Smail},
  {Swinbank}, {Weiss}, {Wardlow}, {Walter}, {Rybak}, {Chen}, {Brand t},
  {Coppin}, {da Cunha}, {Dannerbauer}, {Greve}, {Karim}, {Knudsen},
  {Schinnerer}, {Simpson}, {Venemans}, \& {van der Werf}}]{calistro-rivera18}
{Calistro Rivera}, G., {Hodge}, J.~A., {Smail}, I., {et~al.} 2018, \apj, 863,
  56, \dodoi{10.3847/1538-4357/aacffa}

\bibitem[{{Carilli} \& {Walter}(2013)}]{carilliwalter13}
{Carilli}, C.~L., \& {Walter}, F. 2013, Annual Review of Astronomy and
  Astrophysics, 51, 105, \dodoi{10.1146/annurev-astro-082812-140953}

\bibitem[{{Casasola} {et~al.}(2017){Casasola}, {Cassar{\`a}}, {Bianchi},
  {Verstocken}, {Xilouris}, {Magrini}, {Smith}, {De Looze}, {Galametz},
  {Madden}, {Baes}, {Clark}, {Davies}, {De Vis}, {Evans}, {Fritz}, {Galliano},
  {Jones}, {Mosenkov}, {Viaene}, \& {Ysard}}]{casasola17}
{Casasola}, V., {Cassar{\`a}}, L.~P., {Bianchi}, S., {et~al.} 2017, \aap, 605,
  A18, \dodoi{10.1051/0004-6361/201731020}

\bibitem[{{Chambers} {et~al.}(2016){Chambers}, {Magnier}, {Metcalfe},
  {Flewelling}, {Huber}, {Waters}, {Denneau}, {Draper}, {Farrow}, {Finkbeiner},
  {Holmberg}, {Koppenhoefer}, {Price}, {Rest}, {Saglia}, {Schlafly}, {Smartt},
  {Sweeney}, {Wainscoat}, {Burgett}, {Chastel}, {Grav}, {Heasley}, {Hodapp},
  {Jedicke}, {Kaiser}, {Kudritzki}, {Luppino}, {Lupton}, {Monet}, {Morgan},
  {Onaka}, {Shiao}, {Stubbs}, {Tonry}, {White}, {Ba{\~n}ados}, {Bell},
  {Bender}, {Bernard}, {Boegner}, {Boffi}, {Botticella}, {Calamida},
  {Casertano}, {Chen}, {Chen}, {Cole}, {Deacon}, {Frenk}, {Fitzsimmons},
  {Gezari}, {Gibbs}, {Goessl}, {Goggia}, {Gourgue}, {Goldman}, {Grant},
  {Grebel}, {Hambly}, {Hasinger}, {Heavens}, {Heckman}, {Henderson}, {Henning},
  {Holman}, {Hopp}, {Ip}, {Isani}, {Jackson}, {Keyes}, {Koekemoer}, {Kotak},
  {Le}, {Liska}, {Long}, {Lucey}, {Liu}, {Martin}, {Masci}, {McLean}, {Mindel},
  {Misra}, {Morganson}, {Murphy}, {Obaika}, {Narayan}, {Nieto-Santisteban},
  {Norberg}, {Peacock}, {Pier}, {Postman}, {Primak}, {Rae}, {Rai}, {Riess},
  {Riffeser}, {Rix}, {R{\"o}ser}, {Russel}, {Rutz}, {Schilbach}, {Schultz},
  {Scolnic}, {Strolger}, {Szalay}, {Seitz}, {Small}, {Smith}, {Soderblom},
  {Taylor}, {Thomson}, {Taylor}, {Thakar}, {Thiel}, {Thilker}, {Unger},
  {Urata}, {Valenti}, {Wagner}, {Walder}, {Walter}, {Watters}, {Werner},
  {Wood-Vasey}, \& {Wyse}}]{chambers16}
{Chambers}, K.~C., {Magnier}, E.~A., {Metcalfe}, N., {et~al.} 2016, arXiv
  e-prints, arXiv:1612.05560.
\newblock \doarXiv{1612.05560}

\bibitem[{{Chen} {et~al.}(2017){Chen}, {Hodge}, {Smail}, {Swinbank}, {Walter},
  {Simpson}, {Calistro Rivera}, {Bertoldi}, {Brandt}, {Chapman}, {da Cunha},
  {Dannerbauer}, {De Breuck}, {Harrison}, {Ivison}, {Karim}, {Knudsen},
  {Wardlow}, {Wei{\ss}}, \& {van der Werf}}]{chen17}
{Chen}, C.-C., {Hodge}, J.~A., {Smail}, I., {et~al.} 2017, \apj, 846, 108,
  \dodoi{10.3847/1538-4357/aa863a}

\bibitem[{{Cicone} {et~al.}(2015){Cicone}, {Maiolino}, {Gallerani}, {Neri},
  {Ferrara}, {Sturm}, {Fiore}, {Piconcelli}, \& {Feruglio}}]{cicone15}
{Cicone}, C., {Maiolino}, R., {Gallerani}, S., {et~al.} 2015, \aap, 574, A14,
  \dodoi{10.1051/0004-6361/201424980}

\bibitem[{{Cochrane} {et~al.}(2019){Cochrane}, {Hayward},
  {Angl{\'e}s-Alc{\'a}zar}, {Lotz}, {Parsotan}, {Ma}, {Kere{\v{s}}},
  {Feldmann}, {Faucher-Gigu{\`e}re}, \& {Hopkins}}]{cochrane19}
{Cochrane}, R.~K., {Hayward}, C.~C., {Angl{\'e}s-Alc{\'a}zar}, D., {et~al.}
  2019, \mnras, 488, 1779, \dodoi{10.1093/mnras/stz1736}

\bibitem[{{De Rosa} {et~al.}(2011){De Rosa}, {Decarli}, {Walter}, {Fan},
  {Jiang}, {Kurk}, {Pasquali}, \& {Rix}}]{derosa11}
{De Rosa}, G., {Decarli}, R., {Walter}, F., {et~al.} 2011, \apj, 739, 56,
  \dodoi{10.1088/0004-637X/739/2/56}

\bibitem[{{De Rosa} {et~al.}(2014){De Rosa}, {Venemans}, {Decarli}, {Gennaro},
  {Simcoe}, {Dietrich}, {Peterson}, {Walter}, {Frank}, {McMahon}, {Hewett},
  {Mortlock}, \& {Simpson}}]{derosa14}
{De Rosa}, G., {Venemans}, B.~P., {Decarli}, R., {et~al.} 2014, \apj, 790, 145,
  \dodoi{10.1088/0004-637X/790/2/145}

\bibitem[{{Decarli} {et~al.}(2017){Decarli}, {Walter}, {Venemans},
  {Ba{\~n}ados}, {Bertoldi}, {Carilli}, {Fan}, {Farina}, {Mazzucchelli},
  {Riechers}, {Rix}, {Strauss}, {Wang}, \& {Yang}}]{decarli17}
{Decarli}, R., {Walter}, F., {Venemans}, B.~P., {et~al.} 2017, \nat, 545, 457,
  \dodoi{10.1038/nature22358}

\bibitem[{{Decarli} {et~al.}(2018){Decarli}, {Walter}, {Venemans},
  {Ba{\~n}ados}, {Bertoldi}, {Carilli}, {Fan}, {Farina}, {Mazzucchelli},
  {Riechers}, {Rix}, {Strauss}, {Wang}, \& {Yang}}]{decarli18}
---. 2018, \apj, 854, 97, \dodoi{10.3847/1538-4357/aaa5aa}

\bibitem[{{Decarli} {et~al.}(2019){Decarli}, {Dotti}, {Ba{\~n}ados}, {Farina},
  {Walter}, {Carilli}, {Fan}, {Mazzucchelli}, {Neeleman}, {Novak}, {Riechers},
  {Strauss}, {Venemans}, {Yang}, \& {Wang}}]{decarli19}
{Decarli}, R., {Dotti}, M., {Ba{\~n}ados}, E., {et~al.} 2019, \apj, 880, 157,
  \dodoi{10.3847/1538-4357/ab297f}

\bibitem[{{Fan} {et~al.}(2006){Fan}, {Strauss}, {Becker}, {White}, {Gunn},
  {Knapp}, {Richards}, {Schneider}, {Brinkmann}, \& {Fukugita}}]{fan06}
{Fan}, X., {Strauss}, M.~A., {Becker}, R.~H., {et~al.} 2006, \aj, 132, 117,
  \dodoi{10.1086/504836}

\bibitem[{{Farina} {et~al.}(2019){Farina}, {Arrigoni-Battaia}, {Costa},
  {Walter}, {Hennawi}, {Drake}, {Decarli}, {Gutcke}, {Mazzucchelli},
  {Neeleman}, {Georgiev}, {Eilers}, {Davies}, {Ba{\~n}ados}, {Fan}, {Onoue},
  {Schindler}, {Venemans}, {Wang}, {Yang}, {Rabien}, \& {Busoni}}]{farina19}
{Farina}, E.~P., {Arrigoni-Battaia}, F., {Costa}, T., {et~al.} 2019, \apj, 887,
  196, \dodoi{10.3847/1538-4357/ab5847}

\bibitem[{{Feldmann} {et~al.}(2016){Feldmann}, {Hopkins}, {Quataert},
  {Faucher-Gigu{\`e}re}, \& {Kere{\v{s}}}}]{feldmann16}
{Feldmann}, R., {Hopkins}, P.~F., {Quataert}, E., {Faucher-Gigu{\`e}re}, C.-A.,
  \& {Kere{\v{s}}}, D. 2016, \mnras, 458, L14, \dodoi{10.1093/mnrasl/slw014}

\bibitem[{{Feldmann} {et~al.}(2017){Feldmann}, {Quataert}, {Hopkins},
  {Faucher-Gigu{\`e}re}, \& {Kere{\v{s}}}}]{feldmann17}
{Feldmann}, R., {Quataert}, E., {Hopkins}, P.~F., {Faucher-Gigu{\`e}re}, C.-A.,
  \& {Kere{\v{s}}}, D. 2017, \mnras, 470, 1050, \dodoi{10.1093/mnras/stx1120}

\bibitem[{{Ferrara} {et~al.}(2019){Ferrara}, {Vallini}, {Pallottini},
  {Gallerani}, {Carniani}, {Kohandel}, {Decataldo}, \& {Behrens}}]{ferrara19}
{Ferrara}, A., {Vallini}, L., {Pallottini}, A., {et~al.} 2019, \mnras, 489, 1,
  \dodoi{10.1093/mnras/stz2031}

\bibitem[{{Fujimoto} {et~al.}(2020{\natexlab{a}}){Fujimoto}, {Oguri}, {Nagao},
  {Izumi}, \& {Ouchi}}]{fujimoto19b}
{Fujimoto}, S., {Oguri}, M., {Nagao}, T., {Izumi}, T., \& {Ouchi}, M.
  2020{\natexlab{a}}, \apj, 891, 64, \dodoi{10.3847/1538-4357/ab718c}

\bibitem[{{Fujimoto} {et~al.}(2019){Fujimoto}, {Ouchi}, {Ferrara},
  {Pallottini}, {Ivison}, {Behrens}, {Gallerani}, {Arata}, {Yajima}, \&
  {Nagamine}}]{fujimoto19}
{Fujimoto}, S., {Ouchi}, M., {Ferrara}, A., {et~al.} 2019, \apj, 887, 107,
  \dodoi{10.3847/1538-4357/ab480f}

\bibitem[{{Fujimoto} {et~al.}(2020{\natexlab{b}}){Fujimoto}, {Silverman},
  {Bethermin}, {Ginolfi}, {Jones}, {Le F{\`e}vre}, {Dessauges-Zavadsky},
  {Rujopakarn}, {Faisst}, {Fudamoto}, {Cassata}, {Morselli}, {Schaerer},
  {Capak}, {Yan}, {Vallini}, {Toft}, {Loiacono}, {Zamorani}, {Talia},
  {Narayanan}, {Hathi}, {Lemaux}, {Boquien}, {Amorin}, {Ibar}, {Koekemoer},
  {M{\'e}ndez-Hern{\'a}ndez}, {Bardelli}, {Vergani}, {Zucca}, {Romano}, \&
  {Cimatti}}]{fujimoto20}
{Fujimoto}, S., {Silverman}, J.~D., {Bethermin}, M., {et~al.}
  2020{\natexlab{b}}, arXiv e-prints, arXiv:2003.00013.
\newblock \doarXiv{2003.00013}

\bibitem[{{Gullberg} {et~al.}(2018){Gullberg}, {Swinbank}, {Smail}, {Biggs},
  {Bertoldi}, {De Breuck}, {Chapman}, {Chen}, {Cooke}, {Coppin}, {Cox},
  {Dannerbauer}, {Dunlop}, {Edge}, {Farrah}, {Geach}, {Greve}, {Hodge}, {Ibar},
  {Ivison}, {Karim}, {Schinnerer}, {Scott}, {Simpson}, {Stach}, {Thomson}, {van
  der Werf}, {Walter}, {Wardlow}, \& {Weiss}}]{gullberg18}
{Gullberg}, B., {Swinbank}, A.~M., {Smail}, I., {et~al.} 2018, \apj, 859, 12,
  \dodoi{10.3847/1538-4357/aabe8c}

\bibitem[{{Gullberg} {et~al.}(2019){Gullberg}, {Smail}, {Swinbank},
  {Dudzevi{\v{c}}i{\={u}}t{\.{e}}}, {Stach}, {Thomson}, {Almaini}, {Chen},
  {Conselice}, {Cooke}, {Farrah}, {Ivison}, {Maltby}, {Micha{\l}owski},
  {Simpson}, {Scott}, {Wardlow}, \& {Weiss}}]{gullberg19}
{Gullberg}, B., {Smail}, I., {Swinbank}, A.~M., {et~al.} 2019, \mnras, 490,
  4956, \dodoi{10.1093/mnras/stz2835}

\bibitem[{{Hodge} {et~al.}(2016){Hodge}, {Swinbank}, {Simpson}, {Smail},
  {Walter}, {Alexander}, {Bertoldi}, {Biggs}, {Brandt}, {Chapman}, {Chen},
  {Coppin}, {Cox}, {Dannerbauer}, {Edge}, {Greve}, {Ivison}, {Karim},
  {Knudsen}, {Menten}, {Rix}, {Schinnerer}, {Wardlow}, {Weiss}, \& {van der
  Werf}}]{hodge16}
{Hodge}, J.~A., {Swinbank}, A.~M., {Simpson}, J.~M., {et~al.} 2016, \apj, 833,
  103, \dodoi{10.3847/1538-4357/833/1/103}

\bibitem[{{Hollenbach} \& {Tielens}(1999)}]{hollenbach99}
{Hollenbach}, D.~J., \& {Tielens}, A.~G.~G.~M. 1999, Reviews of Modern Physics,
  71, 173, \dodoi{10.1103/RevModPhys.71.173}

\bibitem[{{Hopkins} {et~al.}(2014){Hopkins}, {Kere{\v{s}}}, {O{\~n}orbe},
  {Faucher-Gigu{\`e}re}, {Quataert}, {Murray}, \& {Bullock}}]{hopkins14}
{Hopkins}, P.~F., {Kere{\v{s}}}, D., {O{\~n}orbe}, J., {et~al.} 2014, \mnras,
  445, 581, \dodoi{10.1093/mnras/stu1738}

\bibitem[{{Hunt} {et~al.}(2015){Hunt}, {Draine}, {Bianchi}, {Gordon}, {Aniano},
  {Calzetti}, {Dale}, {Helou}, {Hinz}, {Kennicutt}, {Roussel}, {Wilson},
  {Bolatto}, {Boquien}, {Croxall}, {Galametz}, {Gil de Paz}, {Koda},
  {Mu{\~n}oz-Mateos}, {Sandstrom}, {Sauvage}, {Vigroux}, \& {Zibetti}}]{hunt15}
{Hunt}, L.~K., {Draine}, B.~T., {Bianchi}, S., {et~al.} 2015, \aap, 576, A33,
  \dodoi{10.1051/0004-6361/201424734}

\bibitem[{Hunter(2007)}]{Hunter:2007}
Hunter, J.~D. 2007, Computing In Science \& Engineering, 9, 90,
  \dodoi{10.1109/MCSE.2007.55}

\bibitem[{{Jiang} {et~al.}(2016){Jiang}, {McGreer}, {Fan}, {Strauss},
  {Ba{\~n}ados}, {Becker}, {Bian}, {Farnsworth}, {Shen}, {Wang}, {Wang},
  {Wang}, {White}, {Wu}, {Wu}, {Yang}, \& {Yang}}]{jiang16}
{Jiang}, L., {McGreer}, I.~D., {Fan}, X., {et~al.} 2016, \apj, 833, 222,
  \dodoi{10.3847/1538-4357/833/2/222}

\bibitem[{{Jorsater} \& {van Moorsel}(1995)}]{jorsater95}
{Jorsater}, S., \& {van Moorsel}, G.~A. 1995, \aj, 110, 2037,
  \dodoi{10.1086/117668}

\bibitem[{{Lawrence} {et~al.}(2007){Lawrence}, {Warren}, {Almaini}, {Edge},
  {Hambly}, {Jameson}, {Lucas}, {Casali}, {Adamson}, {Dye}, {Emerson},
  {Foucaud}, {Hewett}, {Hirst}, {Hodgkin}, {Irwin}, {Lodieu}, {McMahon},
  {Simpson}, {Smail}, {Mortlock}, \& {Folger}}]{lawrence07}
{Lawrence}, A., {Warren}, S.~J., {Almaini}, O., {et~al.} 2007, \mnras, 379,
  1599, \dodoi{10.1111/j.1365-2966.2007.12040.x}

\bibitem[{{Leroy} {et~al.}(2009){Leroy}, {Walter}, {Bigiel}, {Usero}, {Weiss},
  {Brinks}, {de Blok}, {Kennicutt}, {Schuster}, {Kramer}, {Wiesemeyer}, \&
  {Roussel}}]{leroy09}
{Leroy}, A.~K., {Walter}, F., {Bigiel}, F., {et~al.} 2009, \aj, 137, 4670,
  \dodoi{10.1088/0004-6256/137/6/4670}

\bibitem[{{Liang} {et~al.}(2019){Liang}, {Feldmann}, {Kere{\v{s}}}, {Scoville},
  {Hayward}, {Faucher-Gigu{\`e}re}, {Schreiber}, {Ma}, {Hopkins}, \&
  {Quataert}}]{liang19}
{Liang}, L., {Feldmann}, R., {Kere{\v{s}}}, D., {et~al.} 2019, \mnras, 489,
  1397, \dodoi{10.1093/mnras/stz2134}

\bibitem[{{Maiolino} {et~al.}(2012){Maiolino}, {Gallerani}, {Neri}, {Cicone},
  {Ferrara}, {Genzel}, {Lutz}, {Sturm}, {Tacconi}, {Walter}, {Feruglio},
  {Fiore}, \& {Piconcelli}}]{maiolino12}
{Maiolino}, R., {Gallerani}, S., {Neri}, R., {et~al.} 2012, \mnras, 425, L66,
  \dodoi{10.1111/j.1745-3933.2012.01303.x}

\bibitem[{{Marshall} {et~al.}(2019){Marshall}, {Ni}, {Di Matteo}, {Wyithe},
  {Wilkins}, \& {Croft}}]{marshall19b}
{Marshall}, M.~A., {Ni}, Y., {Di Matteo}, T., {et~al.} 2019, arXiv e-prints,
  arXiv:1912.03428.
\newblock \doarXiv{1912.03428}

\bibitem[{{Mart{\'\i}-Vidal} {et~al.}(2014){Mart{\'\i}-Vidal}, {Vlemmings},
  {Muller}, \& {Casey}}]{uvmultifit}
{Mart{\'\i}-Vidal}, I., {Vlemmings}, W.~H.~T., {Muller}, S., \& {Casey}, S.
  2014, \aap, 563, A136, \dodoi{10.1051/0004-6361/201322633}

\bibitem[{{Matsuoka} {et~al.}(2018){Matsuoka}, {Strauss}, {Kashikawa}, {Onoue},
  {Iwasawa}, {Tang}, {Lee}, {Imanishi}, {Nagao}, {Akiyama}, {Asami}, {Bosch},
  {Furusawa}, {Goto}, {Gunn}, {Harikane}, {Ikeda}, {Izumi}, {Kawaguchi},
  {Kato}, {Kikuta}, {Kohno}, {Komiyama}, {Lupton}, {Minezaki}, {Miyazaki},
  {Murayama}, {Niida}, {Nishizawa}, {Noboriguchi}, {Oguri}, {Ono}, {Ouchi},
  {Price}, {Sameshima}, {Schulze}, {Shirakata}, {Silverman}, {Sugiyama},
  {Tait}, {Takada}, {Takata}, {Tanaka}, {Toba}, {Utsumi}, {Wang}, \&
  {Yamashita}}]{matsuoka18}
{Matsuoka}, Y., {Strauss}, M.~A., {Kashikawa}, N., {et~al.} 2018, \apj, 869,
  150, \dodoi{10.3847/1538-4357/aaee7a}

\bibitem[{{Mazzucchelli} {et~al.}(2017){Mazzucchelli}, {Ba{\~n}ados},
  {Venemans}, {Decarli}, {Farina}, {Walter}, {Eilers}, {Rix}, {Simcoe},
  {Stern}, {Fan}, {Schlafly}, {De Rosa}, {Hennawi}, {Chambers}, {Greiner},
  {Burgett}, {Draper}, {Kaiser}, {Kudritzki}, {Magnier}, {Metcalfe}, {Waters},
  \& {Wainscoat}}]{mazzucchelli17}
{Mazzucchelli}, C., {Ba{\~n}ados}, E., {Venemans}, B.~P., {et~al.} 2017, \apj,
  849, 91, \dodoi{10.3847/1538-4357/aa9185}

\bibitem[{{McMullin} {et~al.}(2007){McMullin}, {Waters}, {Schiebel}, {Young},
  \& {Golap}}]{mcmullin07}
{McMullin}, J.~P., {Waters}, B., {Schiebel}, D., {Young}, W., \& {Golap}, K.
  2007, in Astronomical Data Analysis Software and Systems XVI, ed. R.~A.
  {Shaw}, F.~{Hill}, \& D.~J. {Bell}, Vol. 376, 127

\bibitem[{{Mechtley} {et~al.}(2012){Mechtley}, {Windhorst}, {Ryan},
  {Schneider}, {Cohen}, {Jansen}, {Fan}, {Hathi}, {Keel}, {Koekemoer},
  {R{\"o}ttgering}, {Scannapieco}, {Schneider}, {Strauss}, \&
  {Yan}}]{mechtley12}
{Mechtley}, M., {Windhorst}, R.~A., {Ryan}, R.~E., {et~al.} 2012, \apjl, 756,
  L38, \dodoi{10.1088/2041-8205/756/2/L38}

\bibitem[{{Morganson} {et~al.}(2012){Morganson}, {De Rosa}, {Decarli},
  {Walter}, {Chambers}, {McGreer}, {Fan}, {Burgett}, {Flewelling}, {Greiner},
  {Hodapp}, {Kaiser}, {Magnier}, {Price}, {Rix}, {Sweeney}, \&
  {Waters}}]{morganson12}
{Morganson}, E., {De Rosa}, G., {Decarli}, R., {et~al.} 2012, \aj, 143, 142,
  \dodoi{10.1088/0004-6256/143/6/142}

\bibitem[{{Mortlock} {et~al.}(2011){Mortlock}, {Warren}, {Venemans}, {Patel},
  {Hewett}, {McMahon}, {Simpson}, {Theuns}, {Gonz{\'a}les-Solares}, {Adamson},
  {Dye}, {Hambly}, {Hirst}, {Irwin}, {Kuiper}, {Lawrence}, \&
  {R{\"o}ttgering}}]{mortlock11}
{Mortlock}, D.~J., {Warren}, S.~J., {Venemans}, B.~P., {et~al.} 2011, \nat,
  474, 616, \dodoi{10.1038/nature10159}

\bibitem[{{Neeleman} {et~al.}(2020){Neeleman}, {Prochaska}, {Kanekar}, \&
  {Rafelski}}]{neeleman20}
{Neeleman}, M., {Prochaska}, J.~X., {Kanekar}, N., \& {Rafelski}, M. 2020,
  \nat, 581, 269, \dodoi{10.1038/s41586-020-2276-y}

\bibitem[{{Neeleman} {et~al.}(2020, in prep.){Neeleman}, {Venemans}, {Walter},
  \& {Novak}}]{neeleman20b}
{Neeleman}, M., {Venemans}, B., {Walter}, F., \& {Novak}, M. 2020, in prep.

\bibitem[{{Neeleman} {et~al.}(2019){Neeleman}, {Ba{\~n}ados}, {Walter},
  {Decarli}, {Venemans}, {Carilli}, {Fan}, {Farina}, {Mazzucchelli}, {Novak},
  {Riechers}, {Rix}, \& {Wang}}]{neeleman19}
{Neeleman}, M., {Ba{\~n}ados}, E., {Walter}, F., {et~al.} 2019, \apj, 882, 10,
  \dodoi{10.3847/1538-4357/ab2ed3}

\bibitem[{{Novak} {et~al.}(2019){Novak}, {Ba{\~n}ados}, {Decarli}, {Walter},
  {Venemans}, {Neeleman}, {Farina}, {Mazzucchelli}, {Carilli}, {Fan}, {Rix}, \&
  {Wang}}]{novak19}
{Novak}, M., {Ba{\~n}ados}, E., {Decarli}, R., {et~al.} 2019, \apj, 881, 63,
  \dodoi{10.3847/1538-4357/ab2beb}

\bibitem[{{Olsen} {et~al.}(2015){Olsen}, {Greve}, {Narayanan}, {Thompson},
  {Toft}, \& {Brinch}}]{olsen15}
{Olsen}, K.~P., {Greve}, T.~R., {Narayanan}, D., {et~al.} 2015, \apj, 814, 76,
  \dodoi{10.1088/0004-637X/814/1/76}

\bibitem[{{Pavesi} {et~al.}(2016){Pavesi}, {Riechers}, {Capak}, {Carilli},
  {Sharon}, {Stacey}, {Karim}, {Scoville}, \& {Smol{\v c}i{\'c}}}]{pavesi16}
{Pavesi}, R., {Riechers}, D.~A., {Capak}, P.~L., {et~al.} 2016, \apj, 832, 151,
  \dodoi{10.3847/0004-637X/832/2/151}

\bibitem[{{Price-Whelan} {et~al.}(2018){Price-Whelan}, {Sip{\H{o}}cz},
  {G{\"u}nther}, {Lim}, {Crawford}, {Conseil}, {Shupe}, {Craig}, {Dencheva},
  {Ginsburg}, {VanderPlas}, {Bradley}, {P{\'e}rez-Su{\'a}rez}, {de Val-Borro},
  {Paper Contributors}, {Aldcroft}, {Cruz}, {Robitaille}, {Tollerud},
  {Coordination Committee}, {Ardelean}, {Babej}, {Bach}, {Bachetti}, {Bakanov},
  {Bamford}, {Barentsen}, {Barmby}, {Baumbach}, {Berry}, {Biscani}, {Boquien},
  {Bostroem}, {Bouma}, {Brammer}, {Bray}, {Breytenbach}, {Buddelmeijer},
  {Burke}, {Calderone}, {Cano Rodr{\'\i}guez}, {Cara}, {Cardoso}, {Cheedella},
  {Copin}, {Corrales}, {Crichton}, {D{\textquoteright}Avella}, {Deil},
  {Depagne}, {Dietrich}, {Donath}, {Droettboom}, {Earl}, {Erben}, {Fabbro},
  {Ferreira}, {Finethy}, {Fox}, {Garrison}, {Gibbons}, {Goldstein}, {Gommers},
  {Greco}, {Greenfield}, {Groener}, {Grollier}, {Hagen}, {Hirst}, {Homeier},
  {Horton}, {Hosseinzadeh}, {Hu}, {Hunkeler}, {Ivezi{\'c}}, {Jain}, {Jenness},
  {Kanarek}, {Kendrew}, {Kern}, {Kerzendorf}, {Khvalko}, {King}, {Kirkby},
  {Kulkarni}, {Kumar}, {Lee}, {Lenz}, {Littlefair}, {Ma}, {Macleod},
  {Mastropietro}, {McCully}, {Montagnac}, {Morris}, {Mueller}, {Mumford},
  {Muna}, {Murphy}, {Nelson}, {Nguyen}, {Ninan}, {N{\"o}the}, {Ogaz}, {Oh},
  {Parejko}, {Parley}, {Pascual}, {Patil}, {Patil}, {Plunkett}, {Prochaska},
  {Rastogi}, {Reddy Janga}, {Sabater}, {Sakurikar}, {Seifert}, {Sherbert},
  {Sherwood-Taylor}, {Shih}, {Sick}, {Silbiger}, {Singanamalla}, {Singer},
  {Sladen}, {Sooley}, {Sornarajah}, {Streicher}, {Teuben}, {Thomas},
  {Tremblay}, {Turner}, {Terr{\'o}n}, {van Kerkwijk}, {de la Vega}, {Watkins},
  {Weaver}, {Whitmore}, {Woillez}, {Zabalza}, \& {Contributors}}]{astropy:2018}
{Price-Whelan}, A.~M., {Sip{\H{o}}cz}, B.~M., {G{\"u}nther}, H.~M., {et~al.}
  2018, \aj, 156, 123, \dodoi{10.3847/1538-3881/aabc4f}

\bibitem[{{Rujopakarn} {et~al.}(2019){Rujopakarn}, {Daddi}, {Rieke}, {Puglisi},
  {Schramm}, {P{\'e}rez-Gonz{\'a}lez}, {Magdis}, {Alberts}, {Bournaud},
  {Elbaz}, {Franco}, {Kawinwanichakij}, {Kohno}, {Narayanan}, {Silverman},
  {Wang}, \& {Williams}}]{rujopakarn19}
{Rujopakarn}, W., {Daddi}, E., {Rieke}, G.~H., {et~al.} 2019, \apj, 882, 107,
  \dodoi{10.3847/1538-4357/ab3791}

\bibitem[{{Rybak} {et~al.}(2019){Rybak}, {Calistro Rivera}, {Hodge}, {Smail},
  {Walter}, {van der Werf}, {da Cunha}, {Chen}, {Dannerbauer}, {Ivison},
  {Karim}, {Simpson}, {Swinbank}, \& {Wardlow}}]{rybak19}
{Rybak}, M., {Calistro Rivera}, G., {Hodge}, J.~A., {et~al.} 2019, \apj, 876,
  112, \dodoi{10.3847/1538-4357/ab0e0f}

\bibitem[{{Stanley} {et~al.}(2019){Stanley}, {Jolly}, {K{\"o}nig}, \&
  {Knudsen}}]{stanley19}
{Stanley}, F., {Jolly}, J.~B., {K{\"o}nig}, S., \& {Knudsen}, K.~K. 2019, \aap,
  631, A78, \dodoi{10.1051/0004-6361/201834530}

\bibitem[{{Taylor}(2005)}]{topcat}
{Taylor}, M.~B. 2005, Astronomical Society of the Pacific Conference Series,
  Vol. 347, {TOPCAT \& STIL: Starlink Table/VOTable Processing Software}, ed.
  P.~{Shopbell}, M.~{Britton}, \& R.~{Ebert}, 29

\bibitem[{{Vallini} {et~al.}(2017){Vallini}, {Ferrara}, {Pallottini}, \&
  {Gallerani}}]{vallini17}
{Vallini}, L., {Ferrara}, A., {Pallottini}, A., \& {Gallerani}, S. 2017,
  \mnras, 467, 1300, \dodoi{10.1093/mnras/stx180}

\bibitem[{{Venemans} {et~al.}(2020, accepted){Venemans}, {Walter}, {Neeleman},
  {Novak}, {Otter}, \& {Decarli}}]{venemans20}
{Venemans}, B., {Walter}, F., {Neeleman}, M., {et~al.} 2020, accepted, \apj

\bibitem[{{Venemans} {et~al.}(2007){Venemans}, {McMahon}, {Warren},
  {Gonzalez-Solares}, {Hewett}, {Mortlock}, {Dye}, \& {Sharp}}]{venemans07}
{Venemans}, B.~P., {McMahon}, R.~G., {Warren}, S.~J., {et~al.} 2007, \mnras,
  376, L76, \dodoi{10.1111/j.1745-3933.2007.00290.x}

\bibitem[{{Venemans} {et~al.}(2019){Venemans}, {Neeleman}, {Walter}, {Novak},
  {Decarli}, {Hennawi}, \& {Rix}}]{venemans19}
{Venemans}, B.~P., {Neeleman}, M., {Walter}, F., {et~al.} 2019, \apjl, 874,
  L30, \dodoi{10.3847/2041-8213/ab11cc}

\bibitem[{{Venemans} {et~al.}(2017{\natexlab{a}}){Venemans}, {Walter},
  {Decarli}, {Ferkinhoff}, {Wei{\ss}}, {Findlay}, {McMahon}, {Sutherland}, \&
  {Meijerink}}]{venemans17b}
{Venemans}, B.~P., {Walter}, F., {Decarli}, R., {et~al.} 2017{\natexlab{a}},
  \apj, 845, 154, \dodoi{10.3847/1538-4357/aa81cb}

\bibitem[{{Venemans} {et~al.}(2017{\natexlab{b}}){Venemans}, {Walter},
  {Decarli}, {Ba{\~n}ados}, {Carilli}, {Winters}, {Schuster}, {da Cunha},
  {Fan}, {Farina}, {Mazzucchelli}, {Rix}, \& {Weiss}}]{venemans17c}
---. 2017{\natexlab{b}}, \apj, 851, L8, \dodoi{10.3847/2041-8213/aa943a}

\bibitem[{{Walter} {et~al.}(2008){Walter}, {Brinks}, {de Blok}, {Bigiel},
  {Kennicutt}, {Thornley}, \& {Leroy}}]{walter08}
{Walter}, F., {Brinks}, E., {de Blok}, W.~J.~G., {et~al.} 2008, \aj, 136, 2563,
  \dodoi{10.1088/0004-6256/136/6/2563}

\bibitem[{{Walter} {et~al.}(2009){Walter}, {Riechers}, {Cox}, {Neri},
  {Carilli}, {Bertoldi}, {Weiss}, \& {Maiolino}}]{walter09}
{Walter}, F., {Riechers}, D., {Cox}, P., {et~al.} 2009, \nat, 457, 699,
  \dodoi{10.1038/nature07681}

\bibitem[{{Walter} {et~al.}(2003){Walter}, {Bertoldi}, {Carilli}, {Cox}, {Lo},
  {Neri}, {Fan}, {Omont}, {Strauss}, \& {Menten}}]{walter03}
{Walter}, F., {Bertoldi}, F., {Carilli}, C., {et~al.} 2003, \nat, 424, 406,
  \dodoi{10.1038/nature01821}

\bibitem[{{Wang} {et~al.}(2019{\natexlab{a}}){Wang}, {Wang}, {Fan}, {Wu},
  {Yang}, {Neri}, \& {Yue}}]{wangfeige19}
{Wang}, F., {Wang}, R., {Fan}, X., {et~al.} 2019{\natexlab{a}}, \apj, 880, 2,
  \dodoi{10.3847/1538-4357/ab2717}

\bibitem[{{Wang} {et~al.}(2018){Wang}, {Yang}, {Fan}, {Yue}, {Wu}, {Schindler},
  {Bian}, {Li}, {Farina}, {Ba{\~n}ados}, {Davies}, {Decarli}, {Green}, {Jiang},
  {Hennawi}, {Huang}, {Mazzucchelli}, {McGreer}, {Venemans}, {Walter}, \&
  {Beletsky}}]{wangfeige18}
{Wang}, F., {Yang}, J., {Fan}, X., {et~al.} 2018, \apjl, 869, L9,
  \dodoi{10.3847/2041-8213/aaf1d2}

\bibitem[{{Wang} {et~al.}(2013){Wang}, {Wagg}, {Carilli}, {Walter}, {Lentati},
  {Fan}, {Riechers}, {Bertoldi}, {Narayanan}, {Strauss}, {Cox}, {Omont},
  {Menten}, {Knudsen}, {Neri}, \& {Jiang}}]{wang13}
{Wang}, R., {Wagg}, J., {Carilli}, C.~L., {et~al.} 2013, \apj, 773, 44,
  \dodoi{10.1088/0004-637X/773/1/44}

\bibitem[{{Wang} {et~al.}(2019{\natexlab{b}}){Wang}, {Shao}, {Carilli},
  {Jones}, {Walter}, {Fan}, {Riechers}, {Decarli}, {Bertoldi}, {Wagg},
  {Strauss}, {Omont}, {Cox}, {Jiang}, {Narayanan}, {Menten}, \&
  {Venemans}}]{wang19}
{Wang}, R., {Shao}, Y., {Carilli}, C.~L., {et~al.} 2019{\natexlab{b}}, \apj,
  887, 40, \dodoi{10.3847/1538-4357/ab4d4b}

\bibitem[{{Willott} {et~al.}(2015){Willott}, {Bergeron}, \&
  {Omont}}]{willott15}
{Willott}, C.~J., {Bergeron}, J., \& {Omont}, A. 2015, \apj, 801, 123,
  \dodoi{10.1088/0004-637X/801/2/123}

\bibitem[{{Willott} {et~al.}(2010){Willott}, {Delorme}, {Reyl{\'e}}, {Albert},
  {Bergeron}, {Crampton}, {Delfosse}, {Forveille}, {Hutchings}, {McLure},
  {Omont}, \& {Schade}}]{willott10a}
{Willott}, C.~J., {Delorme}, P., {Reyl{\'e}}, C., {et~al.} 2010, \aj, 139, 906,
  \dodoi{10.1088/0004-6256/139/3/906}

\bibitem[{{Yang} {et~al.}(2020){Yang}, {Wang}, {Fan}, {Hennawi}, {Davies},
  {Yue}, {Banados}, {Wu}, {Venemans}, {Barth}, {Bian}, {Boutsia}, {Decarli},
  {Farina}, {Green}, {Jiang}, {Li}, {Mazzucchelli}, \& {Walter}}]{yang20}
{Yang}, J., {Wang}, F., {Fan}, X., {et~al.} 2020, \apjl, 897, L14,
  \dodoi{10.3847/2041-8213/ab9c26}

\bibitem[{{York} {et~al.}(2000){York}, {Adelman}, {Anderson}, {Anderson},
  {Annis}, {Bahcall}, {Bakken}, {Barkhouser}, {Bastian}, {Berman}, {Boroski},
  {Bracker}, {Briegel}, {Briggs}, {Brinkmann}, {Brunner}, {Burles}, {Carey},
  {Carr}, {Castander}, {Chen}, {Colestock}, {Connolly}, {Crocker}, {Csabai},
  {Czarapata}, {Davis}, {Doi}, {Dombeck}, {Eisenstein}, {Ellman}, {Elms},
  {Evans}, {Fan}, {Federwitz}, {Fiscelli}, {Friedman}, {Frieman}, {Fukugita},
  {Gillespie}, {Gunn}, {Gurbani}, {de Haas}, {Haldeman}, {Harris}, {Hayes},
  {Heckman}, {Hennessy}, {Hindsley}, {Holm}, {Holmgren}, {Huang}, {Hull},
  {Husby}, {Ichikawa}, {Ichikawa}, {Ivezi{\'c}}, {Kent}, {Kim}, {Kinney},
  {Klaene}, {Kleinman}, {Kleinman}, {Knapp}, {Korienek}, {Kron}, {Kunszt},
  {Lamb}, {Lee}, {Leger}, {Limmongkol}, {Lindenmeyer}, {Long}, {Loomis},
  {Loveday}, {Lucinio}, {Lupton}, {MacKinnon}, {Mannery}, {Mantsch}, {Margon},
  {McGehee}, {McKay}, {Meiksin}, {Merelli}, {Monet}, {Munn}, {Narayanan},
  {Nash}, {Neilsen}, {Neswold}, {Newberg}, {Nichol}, {Nicinski}, {Nonino},
  {Okada}, {Okamura}, {Ostriker}, {Owen}, {Pauls}, {Peoples}, {Peterson},
  {Petravick}, {Pier}, {Pope}, {Pordes}, {Prosapio}, {Rechenmacher}, {Quinn},
  {Richards}, {Richmond}, {Rivetta}, {Rockosi}, {Ruthmansdorfer}, {Sand ford},
  {Schlegel}, {Schneider}, {Sekiguchi}, {Sergey}, {Shimasaku}, {Siegmund},
  {Smee}, {Smith}, {Snedden}, {Stone}, {Stoughton}, {Strauss}, {Stubbs},
  {SubbaRao}, {Szalay}, {Szapudi}, {Szokoly}, {Thakar}, {Tremonti}, {Tucker},
  {Uomoto}, {Vanden Berk}, {Vogeley}, {Waddell}, {Wang}, {Watanabe},
  {Weinberg}, {Yanny}, {Yasuda}, \& {SDSS Collaboration}}]{york00}
{York}, D.~G., {Adelman}, J., {Anderson}, John~E., J., {et~al.} 2000, \aj, 120,
  1579, \dodoi{10.1086/301513}

\end{thebibliography}

\appendix

\section{Optimizing Gaussian line detection}
\label{sec:app_optimalsn}

\begin{figure}
	\centering
	\includegraphics[height=6cm]{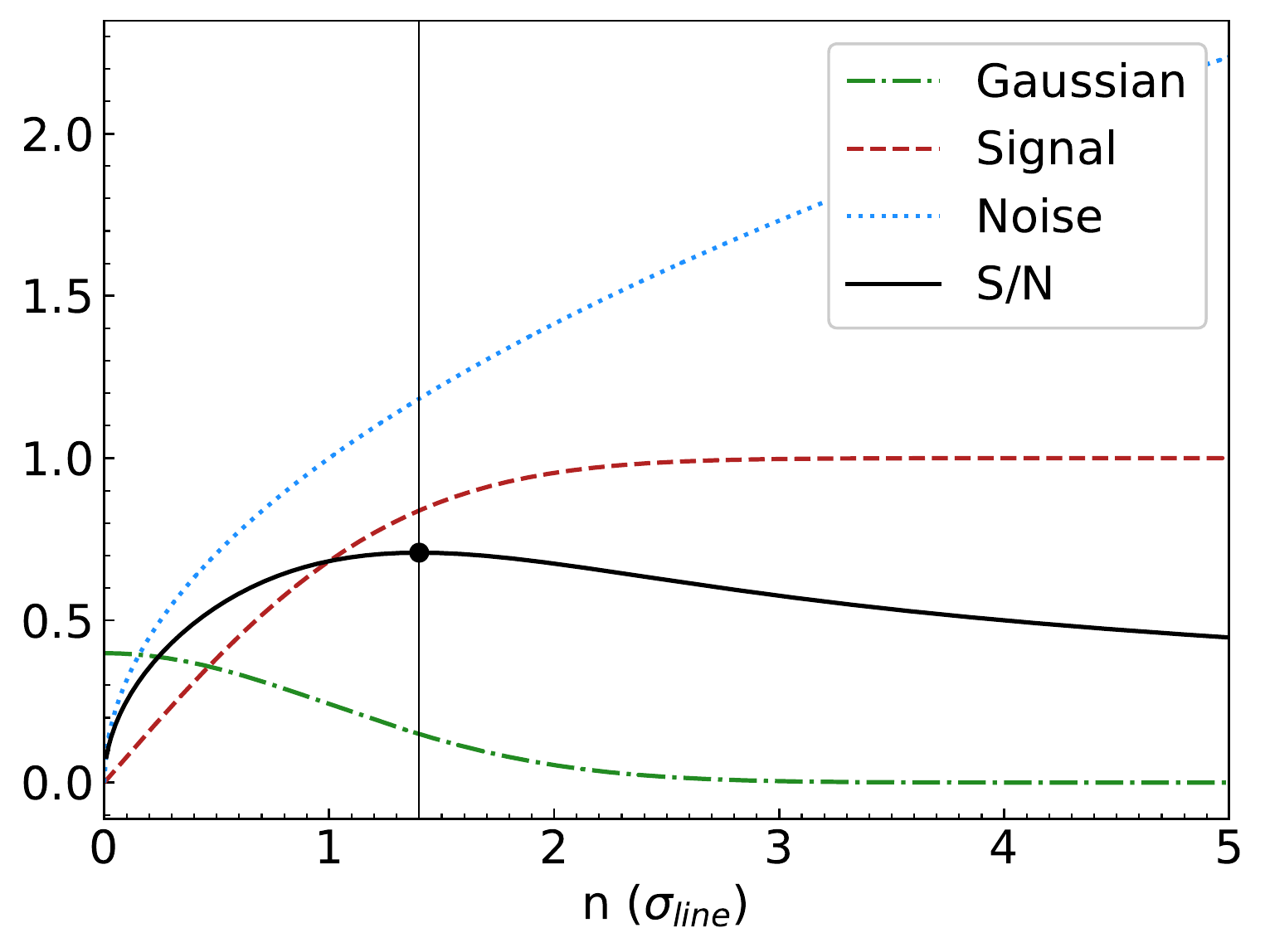}
	\caption{The cumulative signal and noise obtained by integrating a Gaussian  profile over an interval of $[-n\sigma_{\mathrm{line}}, +n\sigma_{\mathrm{line}}]$. The $y$-axis scaling for both the signal and the noise curve is arbitrary, and the value of $n$ corresponding to the S/N peak does not depend on it.}
	\label{fig:optimalsn}
\end{figure}

In order to calculate the optimal bandwidth across which to integrate a line, we first
assume that the emission line has a single Gaussian profile with a line width $\sigma_{\mathrm{line}}$, and that the noise in individual velocity bins is Gaussian and not correlated between neighboring channels.  Integrating from the line peak outwards, the collected flux grows as $\mathrm{Signal}= F_{\mathrm{line}} \times \mathrm{erf}(n/\sqrt{2})$, where $F_{\mathrm{line}}$ is the total line flux and $n$ represents the number of  $\sigma_{\mathrm{line}}$  to integrate over in both positive and negative velocity directions. The uncertainties from  individual velocity channels are added in quadrature, therefore the noise increases with the number of channels in the sum as $\mathrm{Noise}=\sqrt{(2n\sigma_{\mathrm{line}})/ \Delta v} \times \sigma_{\mathbf{rms}}$, where $2n\sigma_{\mathrm{line}}$ is the total integration width and  $ \sigma_{\mathbf{rms}}$ is the noise measured in a velocity bin $\Delta v$\,(\kms) wide. From this, it follows that  $S/N \propto \mathrm{erf}(n/\sqrt{2})/\sqrt{n}$. Numerical evaluation of this function yields the maximum value at $n=1.4$, see Figure~\ref{fig:optimalsn}. Therefore, the total integration width that optimizes the S/N of a Gaussian line is equal to $2.8 \sigma_{\mathrm{line}}$. For a Gaussian, the  $\mathrm{FWHM} = 2 \sqrt{2 \ln 2} \sigma_{\mathrm{line}}$, and the total width translates to $\approx1.2\times\mathrm{FWHM}$, which accounts for 84\% of the line flux.

\section{Residual scaling}
\label{sec:app_residualscale}

The derivation of the residual scaling method is rooted in the fact that we should be able to measure the true emission independent of the cleaning threshold. We denote this true flux as $F$. If we clean the map down to two different thresholds, the following must hold $F=C+\epsilon R=C_2+\epsilon R_2$, where $C$ and $R$ refer to aperture fluxes measured in the clean components and the residual maps, respectively (the subscript refers to a different cleaning threshold). The factor $\epsilon$ is the clean beam to dirty beam volume ratio which ensures that the summands have equal units (hence the name residual scaling). A special case is obtained if no cleaning is applied, then $C_2=0$ and $R_2=D$, where $D$ is the aperture flux measured inside the dirty map (i.e.\ the residual after zero iterations of clean). Solving the two equations yields $\epsilon=C/(D-R)$ and $F=\epsilon D$. The essence of the method lies in measuring the proper dirty beam volume, valid only inside a specific aperture. Because the  integral of the dirty beam approaches zero (sum of finite number of cosine waves over all area), its volume is only meaningful inside a finite spatial region. This is not the case with the clean beam, which is a 2D Gaussian whose integral always converges to a finite non-zero value.

\section{Cleaning without multi-scales} \label{sec:app_ms}

\begin{figure*}
	\centering
	\includegraphics[width=\linewidth]{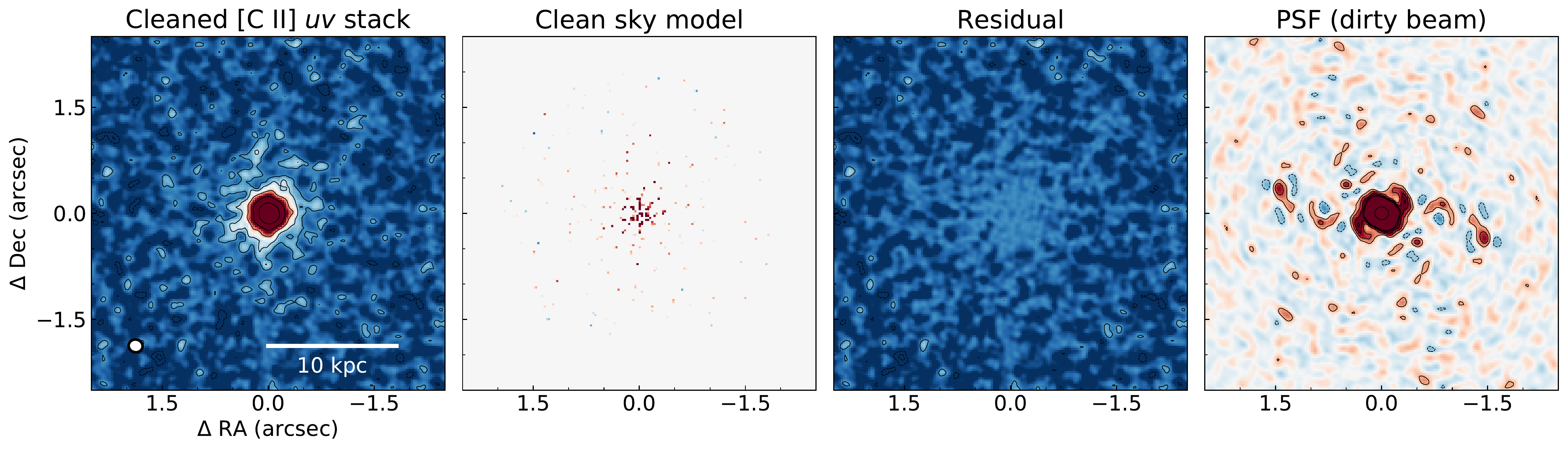}\\
	\includegraphics[height=6.8cm]{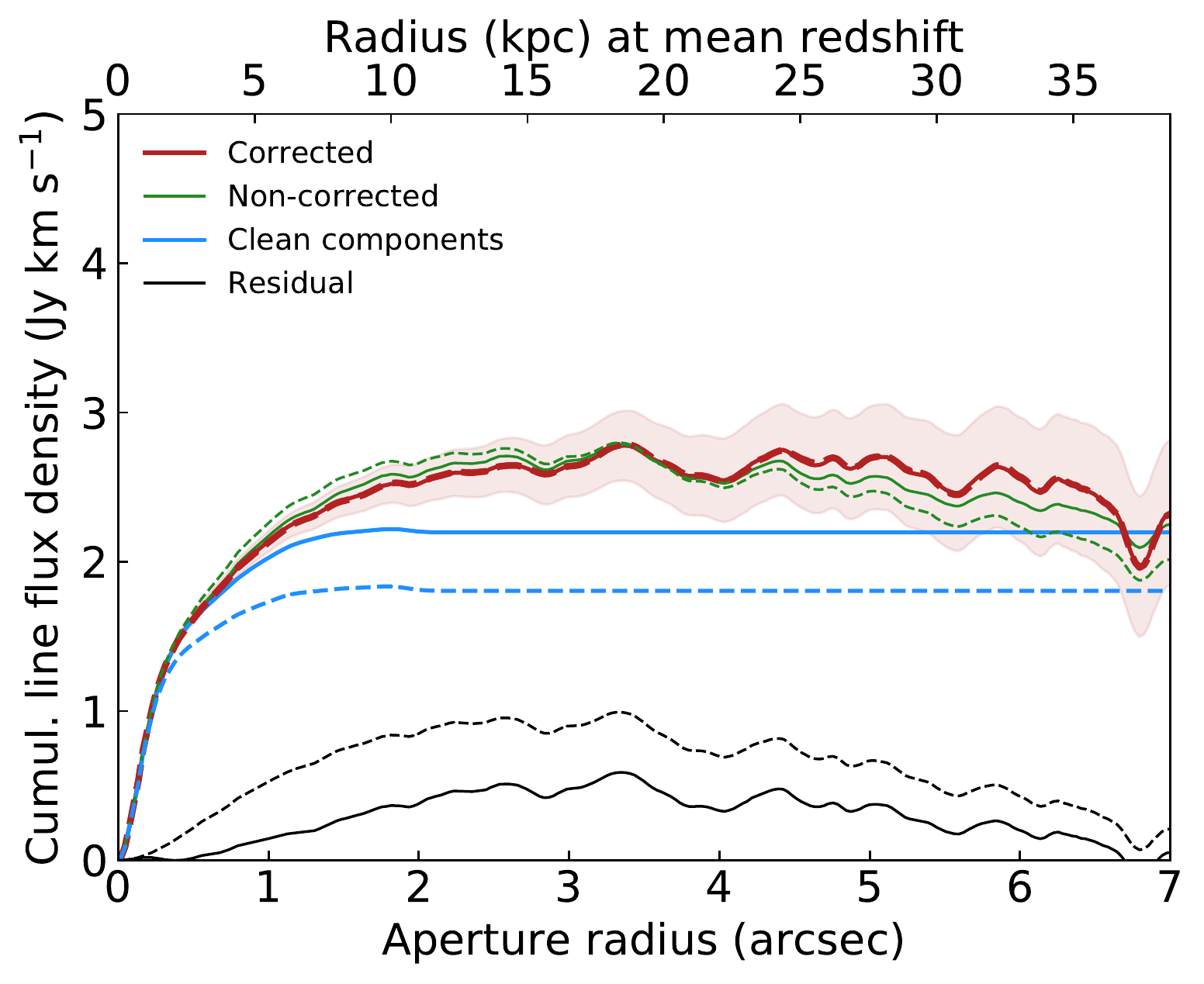}
	\includegraphics[height=6.8cm]{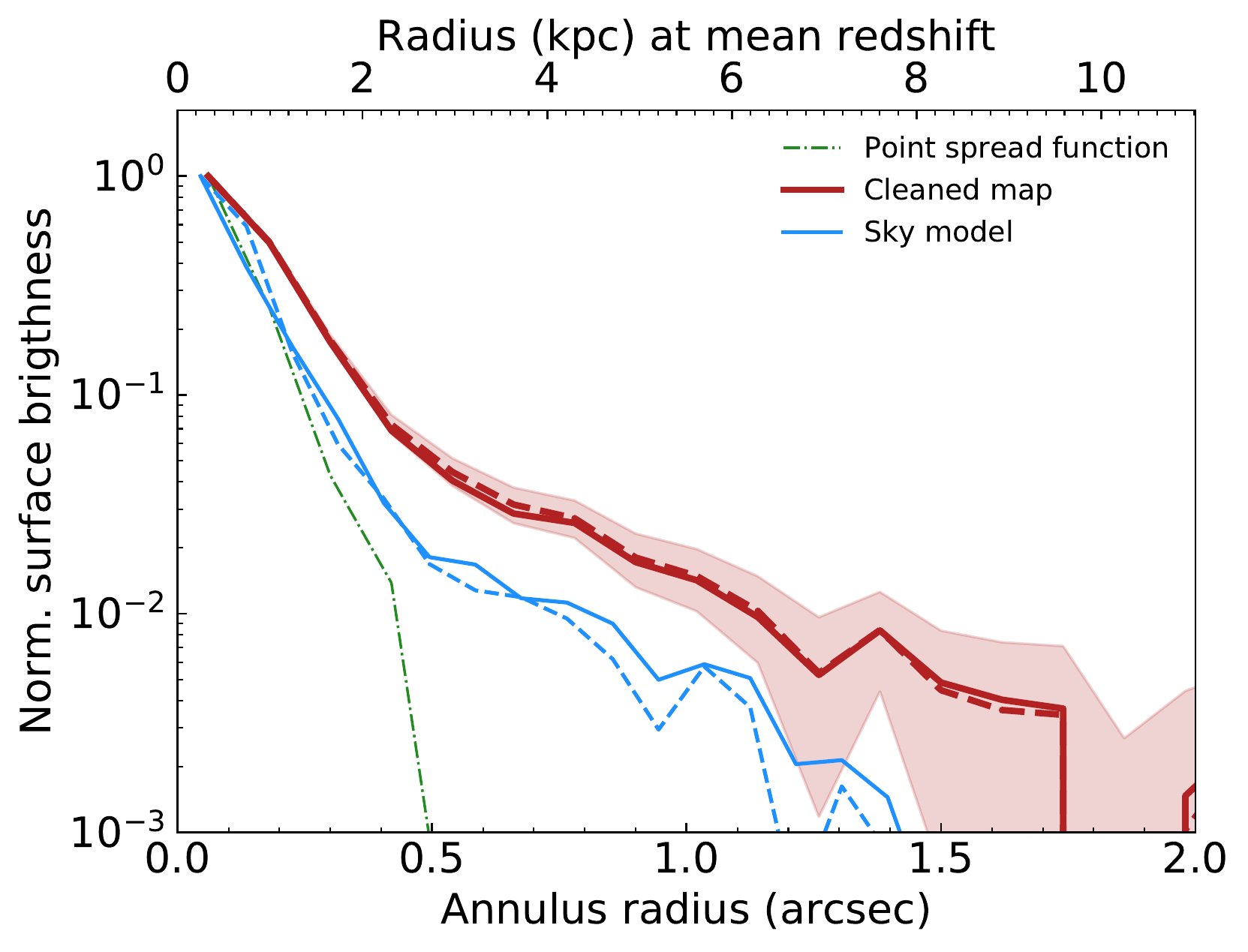}
	\caption{Emission of the \cii\ line measured from the imaged $uv$-stack, similar to Figure~\ref{fig:uvstackcii}, but with a different cleaning algorithm utilized. 
		{\it Upper panels} show the results of deconvolution without multiple scales. 
		{\it Lower panels} show the comparison between multi-scale cleaned maps (solid lines) and single-scale cleaned maps (dashed lines). Despite obviously different ratios of cleaned to non-cleaned flux, and different sky models, the final corrected flux measurements do not depend on the cleaning strategy.
	}
	\label{fig:cleancompare}
\end{figure*}

The process of deconvolution, i.e.\ cleaning, relies on extrapolating the sky model from available visibilities, where the missing Fourier components must be somehow filled in by the algorithm. Therefore, the final sky model is only one of many possible representations that are consistent with the observed data.
Throughout this work we utilize the multi-scale clean algorithm as described in Section~\ref{sec:uvstackmethod}.
If we are using extended source sky model components, the question arises whether we are actually forcing the algorithm to indeed create extended structures. To quantify this effect we re-imaged the $uv$-stack from Section~\ref{sec:results_ciiextent} without multi-scales, using only delta functions to populate the sky model. The results are shown in Figure~\ref{fig:cleancompare}.

The final cleaned map shows no significant differences compared to the multi-scale version, and the dirty beam does not change as imaging weights were kept the same. However, the clean sky model is now comprised of individual pixel sources (delta functions), and there is considerably more flux in the residual on large scales (all below $2\sigma$). Nevertheless, both the curves of growth (after applying residual scaling correction) and surface brightness profiles are consistent between the two cleaning approaches.

\section{Fitting the stacked data in the uv-plane} \label{sec:app_uvfit}

\begin{figure*}
	\begin{flushleft}
		\includegraphics[height=6cm]{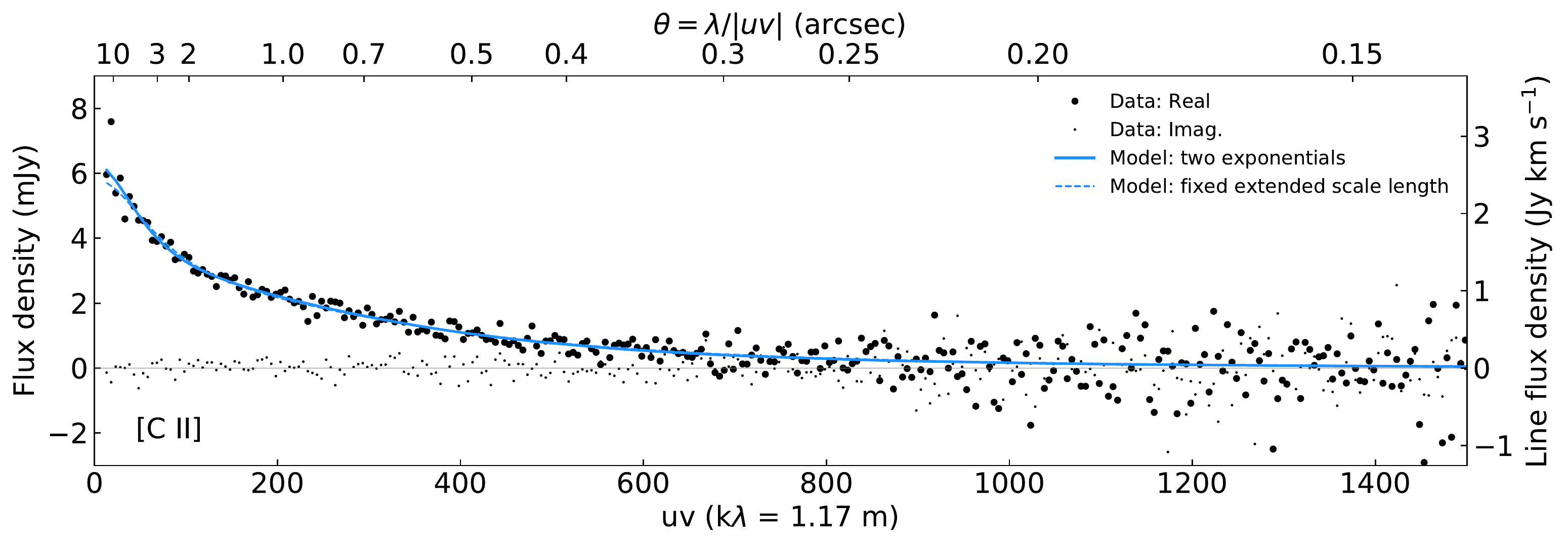}\\
		\includegraphics[height=6cm]{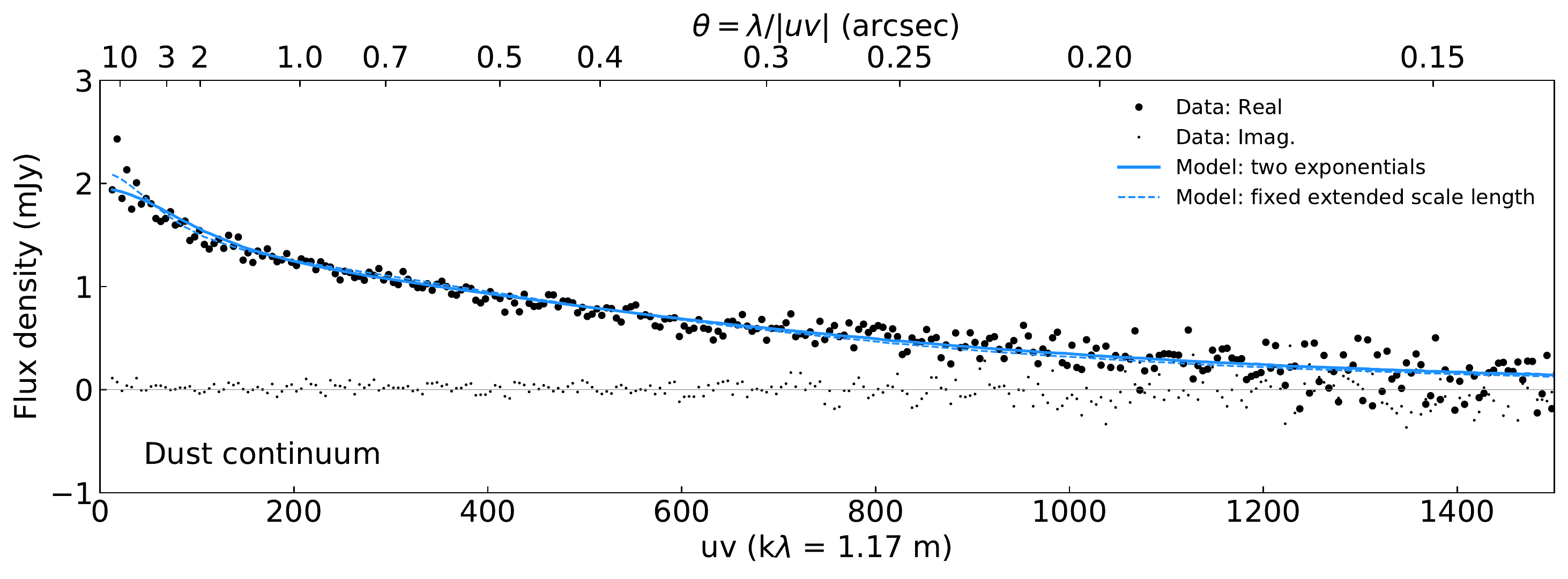}
	\end{flushleft}
	\caption{Weighted average flux densities in annuli of $uv$-distances, drawn from the stacked clean sample of $z\gtrsim6$ suasar host galaxies (sample 4 from the Section~\ref{sec:samples}). 
	The average velocity bandwidth of 420\,\kms\ is used to scale the right $y$-axis of the \cii\ measurements. The real part of the visibilities is used as a proxy for the amplitudes, while the imaginary part is centered around zero, indicating that the emission is in the phase center. The solid blue line indicates the best fit two-component model obtained with the {\sc uvmultifit} package. The dashed blue line shows the second model, where an additional constraint was imposed, by fixing one exponential scale length to 2\,kpc. Both models are consistent with the data within scatter.
	}
	\label{fig:uvfit}
\end{figure*}

We fit the $uv$-data of our clean $uv$-stacking sample using the software package {\sc uvmultifit} \citep{uvmultifit}. The fitting model was a linear combination of two radially symmetric exponential brightness profiles (Lorentzians in the $uv$-plane), each with four free parameters: two positional offsets, the total flux, and the major axis FWHM 
(related to the effective or half-light radius as $R_{\mathrm{eff}}=1.678\times l$, where $l$ is the scale length of the exponential profile, $l=\mathrm{FWHM}/(2\ln2)$).
We compare the visibilities binned in $uv$-distances of 5\,k$\lambda$ and the fitted model in Figure~\ref{fig:uvfit}. There is no spatial offset between the two fitted components, but they are both shifted by $\sim17$\,mas with respect to the phase center, which was chosen to be the centroid position of the \cii\ emission for each individual source during the $uv$-stacking process.

The fit yielded the following flux densities, $F$, and exponential scale lengths, $l$, for the compact and the extended components:
$F_{\cii}^{\mathrm{comp.}} = 2.8 \pm 0.1$\,mJy with $l=0.43 \pm 0.02$\,kpc,
$F_{\cii}^{\mathrm{ext.}} = 3.4 \pm 0.2$\,mJy with $l=2.6 \pm 0.2$\,kpc,
$F_{\mathrm{dust}}^{\mathrm{comp.}} = 1.19 \pm 0.03$\,mJy with $l=0.20 \pm 0.01$\,kpc,
and $F_{\mathrm{dust}}^{\mathrm{ext.}} = 0.76 \pm 0.03$\,mJy with $l=1.3 \pm 0.1$\,kpc.
We fitted a second model with one less free parameter, by choosing to fix the scale length of one exponential function to 2\,kpc, the value derived from image plane analysis (see Sections~\ref{sec:results_ciiextent} and \ref{sec:results_dustextent}). This fit is shown in Figure~\ref{fig:uvfit} with the dashed line, and is still consistent with the observed data points.
Fitting in the $uv$-plane circumvents issues with the clean/dirty beam, and yields the conclusions consistent with those presented in the main text.

\section{Subsamples and  biases}
\label{sec:app_subsamples}

\begin{figure*}
	\centering
	\includegraphics[height=6.8cm]{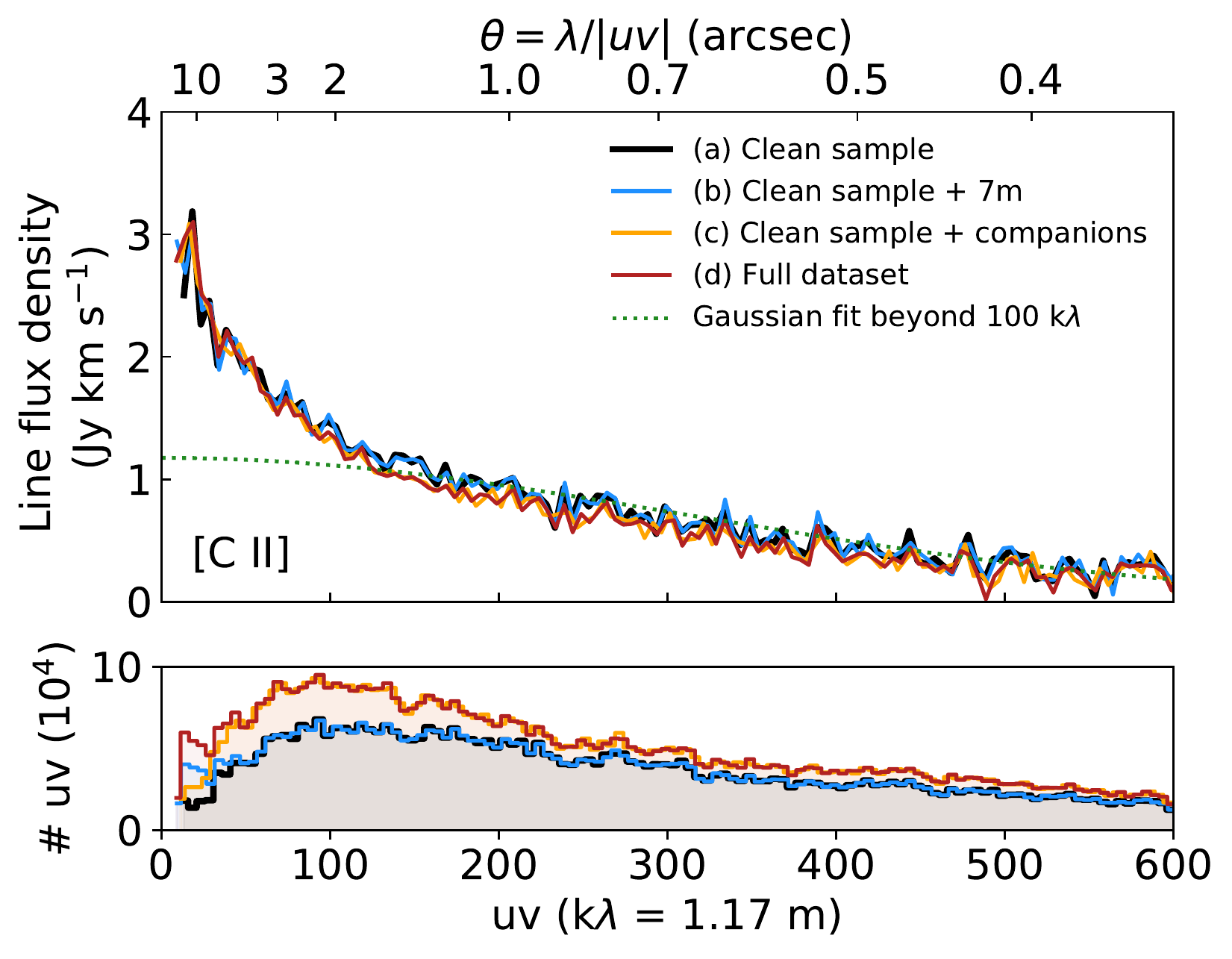}
	\includegraphics[height=6.8cm]{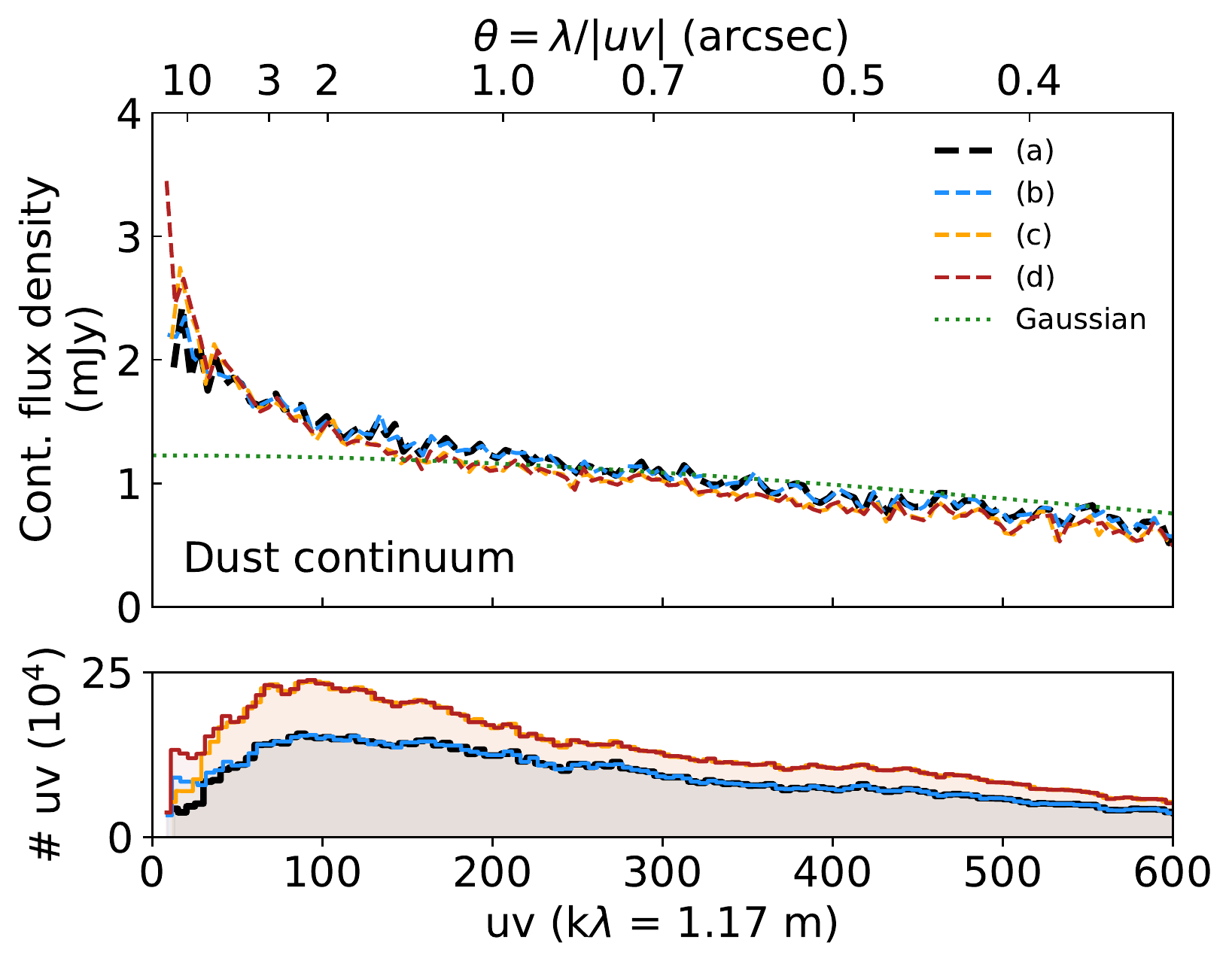}\\
	\includegraphics[height=6.8cm]{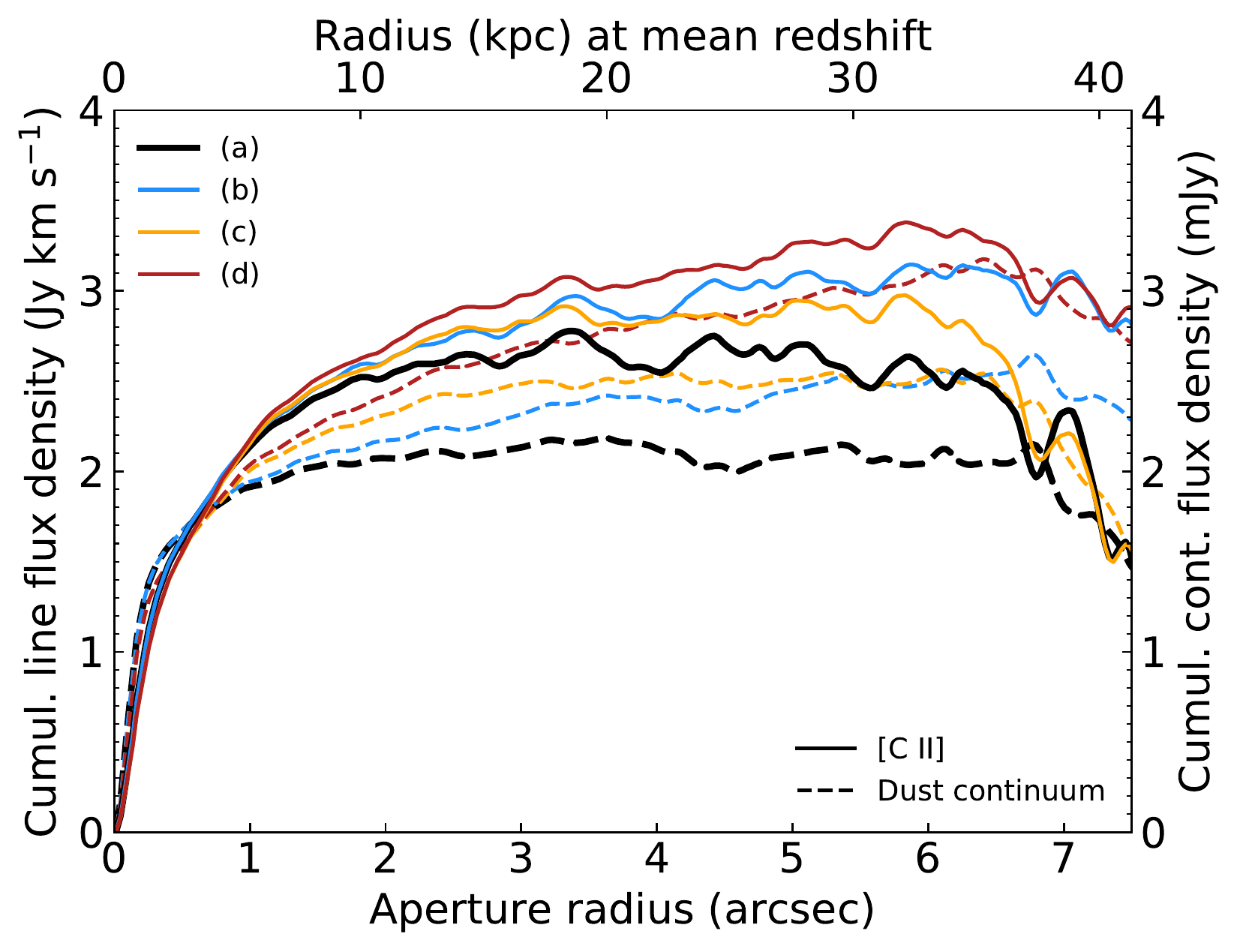}
	\includegraphics[height=6.8cm]{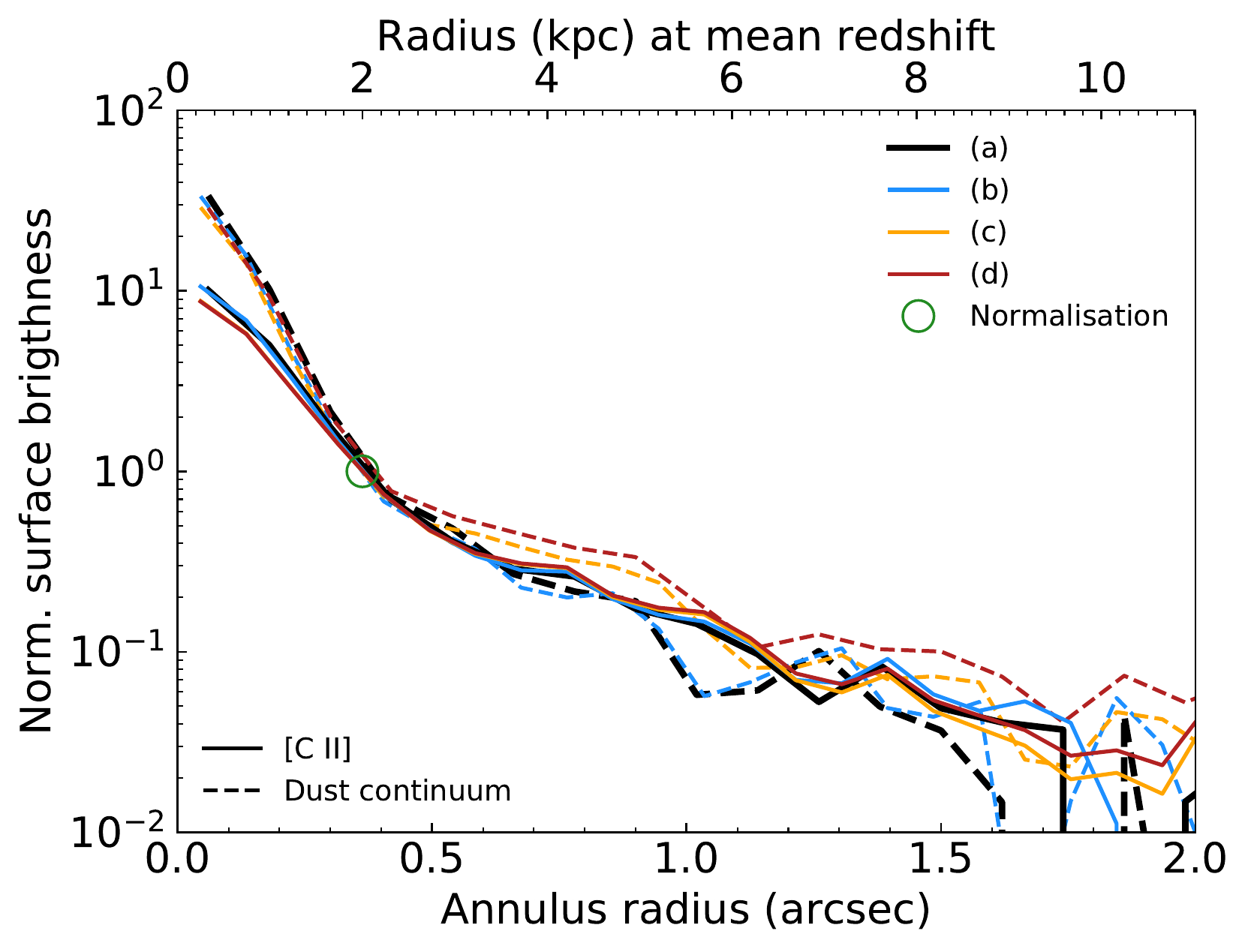}
	
	\caption{Comparison between the  \cii\  (solid lines) and the dust continuum $\nu_{\mathrm{obs}}=256$\,GHz emission (dashed lines) for different data  samples. {\it The upper left panel}  shows the mean \cii\ flux in annuli of $uv$-distances, as well as the histogram of available number of visibilities in each bin. Excess amplitudes at short $uv$-distances indicate possible extended emission.
	{\it The upper right panel} is same as the left one, but measured on the dust continuum. {\it The lower left panel}  shows the curves of growth. Both the continuum and \cii\ saturate at similar radii (subsamples a and c). Including the 7\,m data requires integrating over larger radii (subsamples b and d).
	{\it The lower right panel}  shows normalized surface brightness profiles. All subsamples demonstrate a more compact dust continuum component, compared to \cii,  and an extended tail in both cases. As expected, including companions and merger systems increases the flux measured in the 3 -- 5\,kpc range (more pronounced in the case of the dust continuum, see dashed lines of subsamples c and d).
	}
	\label{fig:uvsamples}
\end{figure*}

\begin{figure*}
	\centering
	\includegraphics[height=6cm]{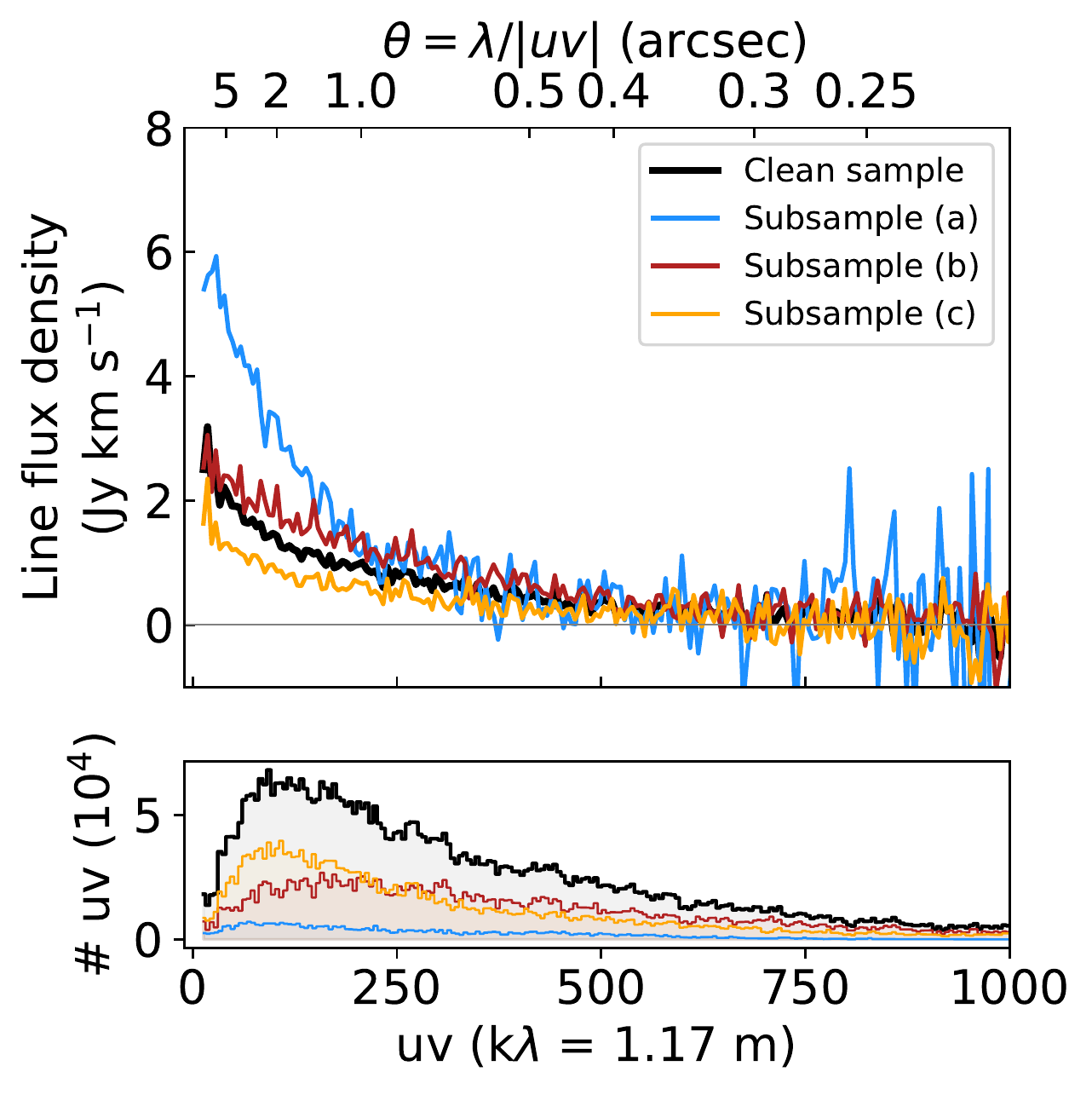}
	\includegraphics[height=6cm]{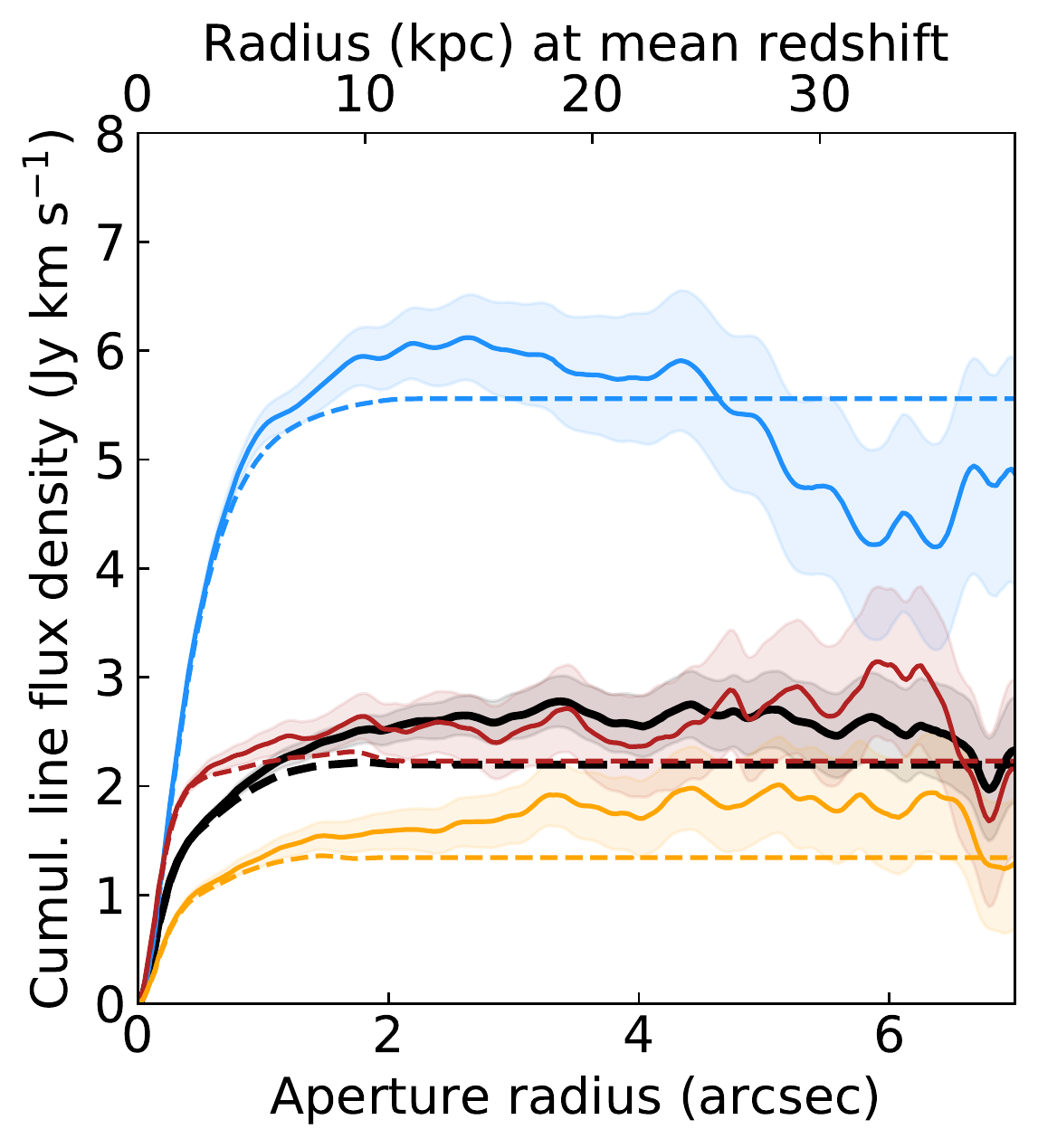}
	\includegraphics[height=6cm]{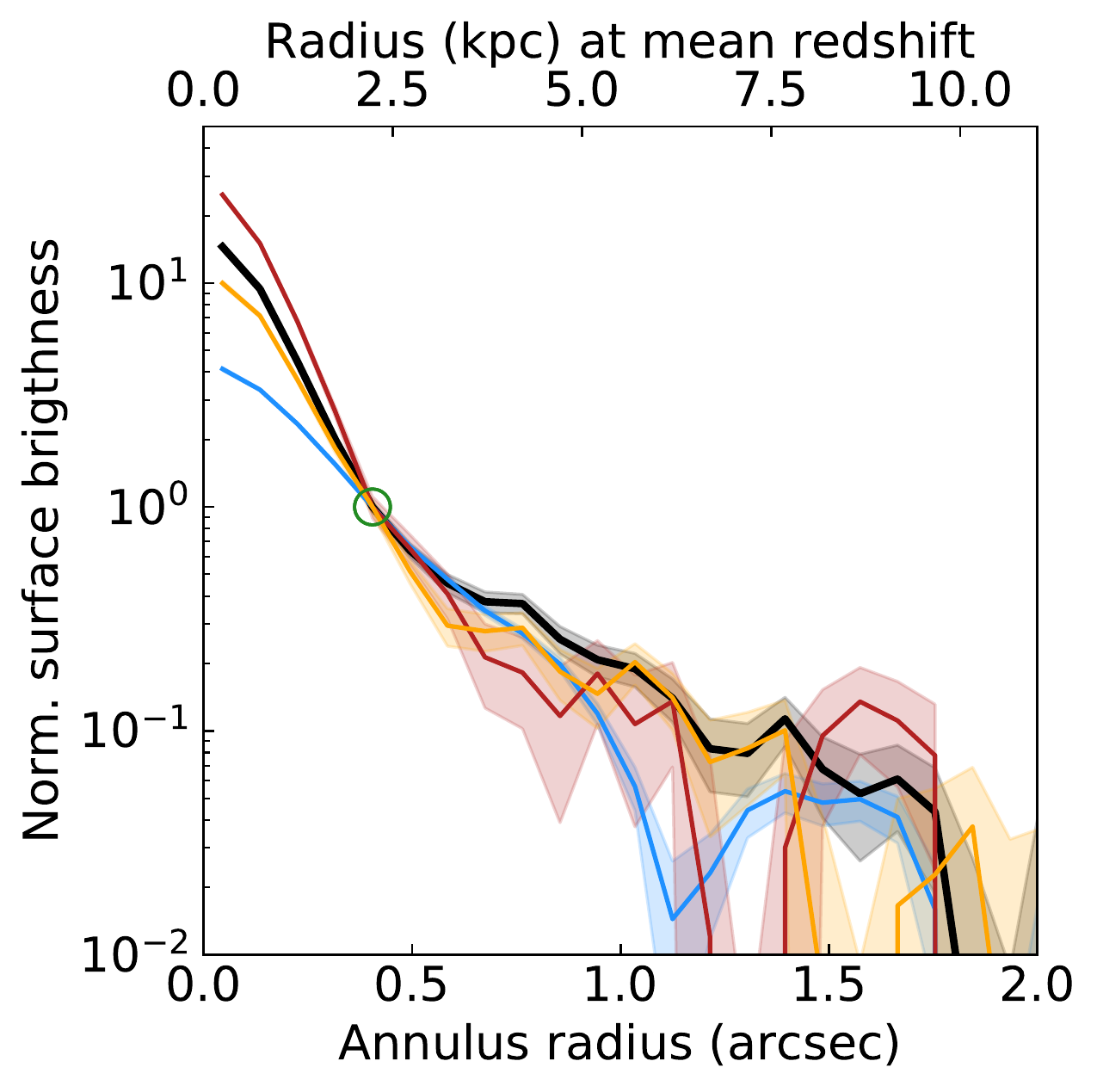}
	
	\caption{Analysis of \cii\ emission for smaller subsamples (three colored lines as listed in the legend), compared to the the clean sample (black line). {\it The left panel} shows line flux as a function of $uv$-distance, and the number of visibility data points for each bin in the histogram below.  {\it The middle panel} shows the curves of growth measured in the maps (solid lines), and the values from the clean components only (dashed lines). {\it The right panel} shows the surface brightness profile normalized at the  radius indicated with the green circle. In all of the cases, the \cii\ curves of growth saturate around $\sim$10\,kpc, and a two component surface brightness profile is recovered.}
	\label{fig:uvsubsamples}
\end{figure*}

In the main text, for $uv$-stacking, we have been using the clean sample, in which all systems with close companions and obvious mergers were removed, in addition to any 7\,m array data. Here, we evaluate the robustness of our results by considering smaller and larger data samples.

We first enlarge our clean sample by considering systems with companion galaxies and available 7\,m array observations.
The 7\,m array data are provided by the ALMA compact array (ACA), where smaller dishes allow for denser spacing of antennas, thus allowing measurements of visibilities on shorter baselines compared to the 12\,m array. 
There are only three quasar host galaxies with such observations available in our sample, and one of them is designated as a companion system as well. Hence, the largest spatial scales data in the $uv$-stack will be provided by only two to three objects. Including them in the stack could potentially bias the mean emission measurement. To investigate this effect we define four different samples, which we analyze and present separately in Figure~\ref{fig:uvsamples}. The four tested samples are: 

\begin{enumerate}[label=(\alph*),itemsep=-3pt]
\item Clean $uv$-stacking sample (19 objects, 32 datasets), as in the main text. 
\item Clean $uv$-stacking sample with  7\,m data added (19 objects, 34 datasets).
\item Clean $uv$-stacking sample with companion and merger systems added (26 objects, 45 datasets).
\item Full dataset\footnote{Six datasets are always excluded from $uv$-stacking (see Table~\ref{tab:data}).} containing companion and merger systems, and all the 7\,m data  (26 objects, 48 datasets).
\end{enumerate}

The two top panels of Figure~\ref{fig:uvsamples} demonstrate that amplitudes cannot be described by a single Gaussian in any of the samples, for both \cii\ and dust continuum emission.
Between different samples, the dust continuum measurements show somewhat larger dispersion at short baselines than the \cii\ values. The 7\,m array data double the number of available visibilities below 30\,k$\lambda$ (blue vs. black and red vs. yellow lines).
The curves of growth shown in the lower left panel of Figure~\ref{fig:uvsamples}  differ for the four sets of data considered. 
The inclusion of companion systems increases the mean flux, as one would expect, but the value still saturates at $\sim$10\,kpc. 
The inclusion of the 7\,m data requires integration over larger radii to encompass the data from the shortest baselines and results in $\sim$20\% larger flux densities of both the \cii\ and the dust continuum. Because this contribution arises from three or fewer sources, it likely does not represent the mean behavior of full sample of observed $z\sim6$ quasar host galaxies.
At small radii the residual scaling is numerically stable, but because of division with a small number (aperture flux densities measured in the residual and the dirty map approach each other), it becomes numerically unstable beyond 7$\arcsec$.
The last panel of Figure~\ref{fig:uvsamples} shows the surface brightness profiles, normalized at 2\,kpc. All four samples exhibit similar functional forms.

For the remainder of this section we analyze smaller subsamples.
Our full sample contains sources that span approximately one order of magnitude in \cii\ line flux densities ($\sim$1 -- 10\,\jykms) and galaxies on the fainter end of this range could have intrinsically different properties from those on the brighter end. 
Also, the addition of objects at various flux levels in the $uv$-plane can translate into erroneous spatial structure. To quantify possible biases we split our clean sample into three smaller subsamples, based on the flux values measured at the lowest baselines (first available bin of 15\,k$\lambda$):

\begin{enumerate}[label=(\alph*),itemsep=-3pt]
\item Bright subsample, \cii\ flux density above 3\,\jykms\ (3 objects,  5 datasets).
\item Medium bright subsample, between 2 -- 3\,\jykms\   (8 objects, 14 datasets).
\item Faint subsample, below 2\,\jykms\ (8 objects, 13 datasets).
\end{enumerate}

These are presented in Figure~\ref{fig:uvsubsamples}.
Because we are significantly reducing our sample size this way, the S/N becomes poorer. Nevertheless, all our main conclusions hold within the errors. The spatial structure is non-Gaussian, the total \cii\ flux saturates at $\sim$10\,kpc, and surface brightness profiles exhibit  extended exponential tails.
The bright subsample is composed of only three sources, namely J025--33, P009--10, and P183+05, which all show extended morphology and significant emission within 5\,kpc (for details on individual objects see Appendix~\ref{sec:app_sources}), thus the surface brightness profile at smaller radii is less steep then the remaining two subsamples.

As a final consistency check we ensured a consistent overlap in $uv$-coverage  between all objects by selecting only the  high-resolution cycles (sample 1 from Section~\ref{sec:samples}).  The analysis performed on such a selection yielded no new systematics.
In summary, our main results regarding the spatial extent of \cii\ and dust continuum emission remain unchanged across all considered samples.

\section{Velocity stacks}
\label{sec:app_velstack}

\begin{figure}
	\centering
	\includegraphics[width=0.48\linewidth]{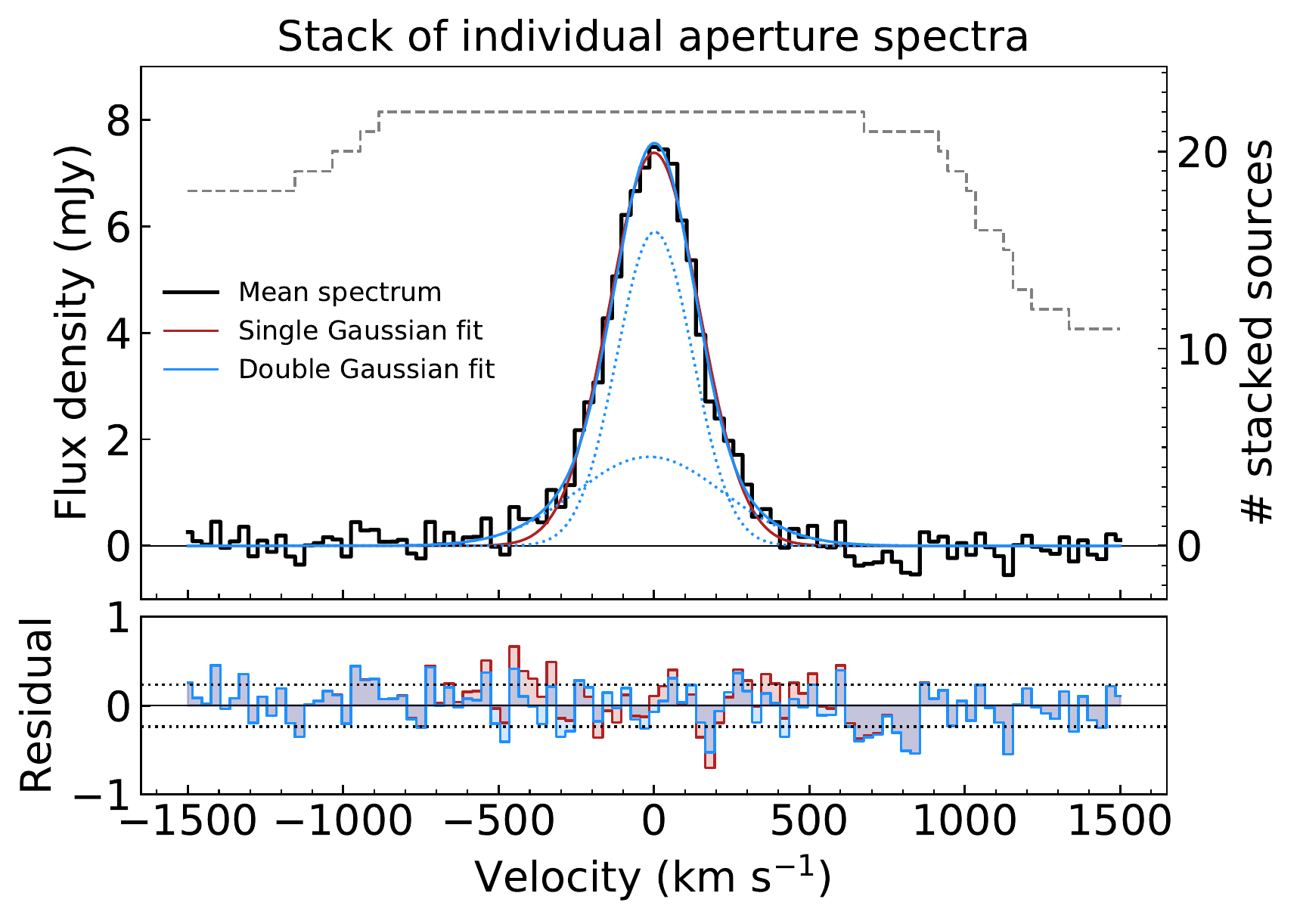}\\
	\includegraphics[width=0.48\linewidth]{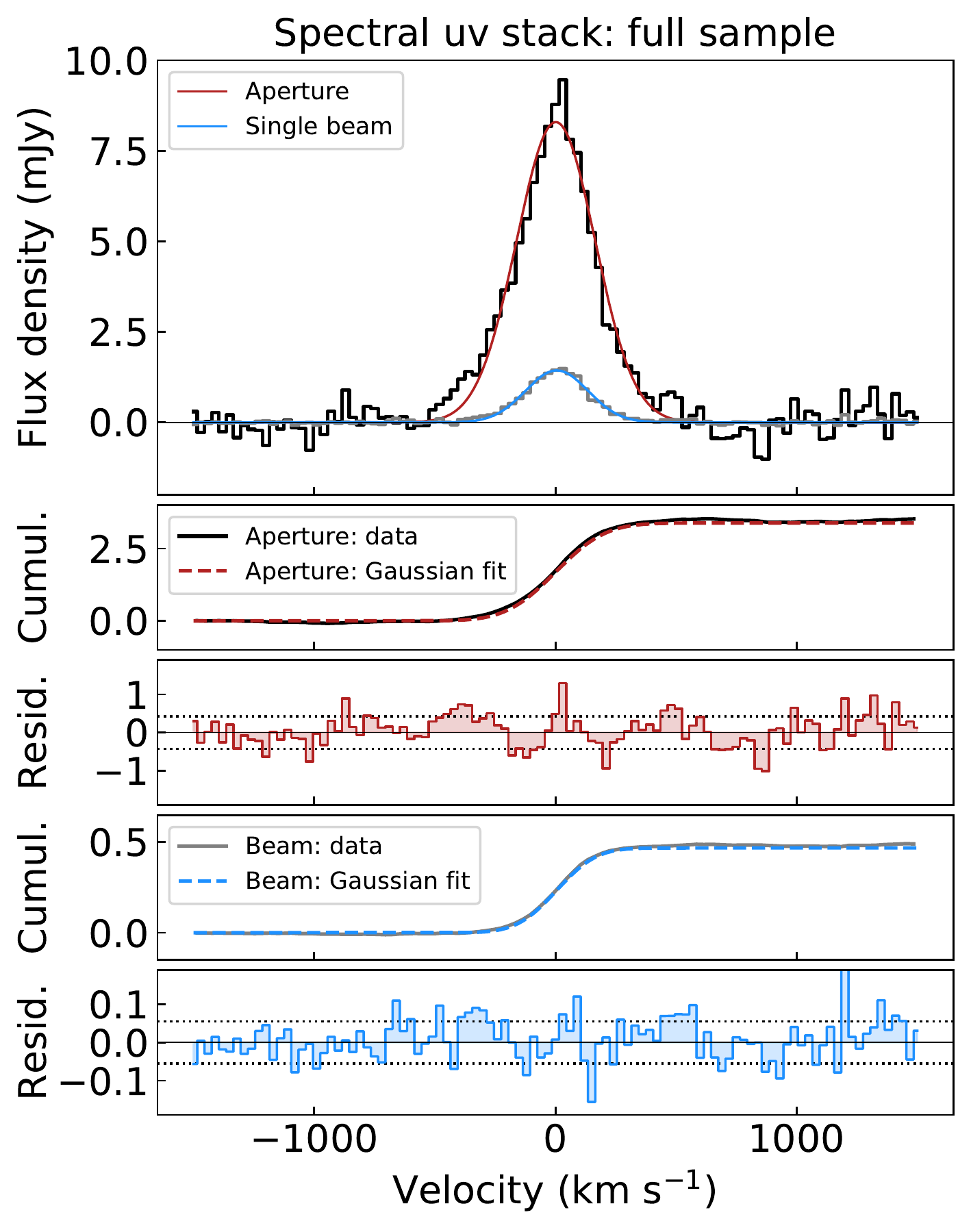}

	\caption{Stacked \cii\ spectrum without linewidth normalization.
		{\it Top: } Aperture spectra stacking, same as Figure~\ref{fig:aperspecstack}. {\it Bottom: } $uv$-plane spectral stacking, same as right panels in Figure~\ref{fig:uvspecstack}, with the cumulative flux distribution units changed to \jykms, whereas the residuals remain in the units of mJy.
 }
	\label{fig:aperspecstackvel}
\end{figure}

Stacking spectra that are individually well described by single Gaussians of different linewidths will result in a spectral shape with at least one narrow and one broad component. To demonstrate this effect, we repeated the \cii\ stacking from Section~\ref{sec:results_outflow} without line-width renormalization, in velocity space, and present the results in Figure~\ref{fig:aperspecstackvel}.
After subtracting a single fitted Gaussian from the stacked spectrum, positive residuals can be seen around velocities $\sim400$\,\kms\ in both the central beam and the larger aperture measurement. The total observed flux density excess compared to a single Gaussian fit remains low, less than 4\%, as evident from the cumulative distribution in Figure~\ref{fig:aperspecstackvel}. 
The specific shape of the residuals (a peak at zero velocity, followed by one negative and one broader positive feature toward larger velocities) can also be reproduced by averaging pure Gaussian functions, one per object, that correspond to measured \cii\ lines (in terms of their total fluxes and FWHMs), and then subtracting a single Gaussian fit from the average. The similar shape of the residual is also visible in Figure~A.2 from \cite{bischetti19}, indicating that at least some excess emission is potentially due to averaging spectra of various line widths.

\section{Individual source analysis}
\label{sec:app_sources}

In this section we provide more details on individual objects. We presented multiple \cii\ flux diagnostics throughout Section~\ref{sec:methods}. We show similar analysis for every object of our sample in Figure~\ref{fig:individual}.
We also highlight several particular galaxies as potential interesting follow-up candidates.
Three objects, listed in Section~\ref{sec:comp} as having complex morphology, deserve a mention here.
System J0100+2802 exhibits a factor of two increase in \cii\ flux between 3 and 10\,kpc. This source was excluded from our $uv$-stacking analysis and its interpretation is unclear. 
It was studied in detail by \cite{wangfeige19}, and a possible lensing scenario was proposed by \cite{fujimoto19b}. 
Systems J025--33 and P009--10\footnote{We note that the large discrepancy between fluxes measured in  two cycles is due to partial line coverage in the earlier cycle, where only half of the \cii\ line is observed.} are both among the brightest in our sample and show a non-Gaussian amplitude distribution in the $uv$-plane. However, it is unclear whether a  companion is present, or whether we observe more of the underlying extended emission due to higher surface brightness.
Another object where we observe a factor of $\sim2$ increase in flux densities between the central 3\,kpc and extended 10\,kpc is J0842+1218, as evident in two independent cycles. 
This system is already known for having a companion galaxy at a separation of 50\,kpc \citep[see][]{decarli17} and a second, fainter companion at 31\,kpc \citep{neeleman19}.
Similarly, J1044--0125 shows evidence for enhanced emission on large scales, which is particularly interesting as it is hinted at in the 7\,m data.
Moreover, P359--06 shows some evidence of extended emission in two separate cycles, although the effect is not as  pronounced as in the previous two sources.
Finally, we note that P183+05 and P007+04 are identified as proximate damped Ly$\alpha$ absorption systems \citep[see][respectively]{banados19b, farina19}, which may have some effect on our measurements if the foreground galaxy is aligned with the quasar host.

\begin{figure*}
	\includegraphics[width=\linewidth]{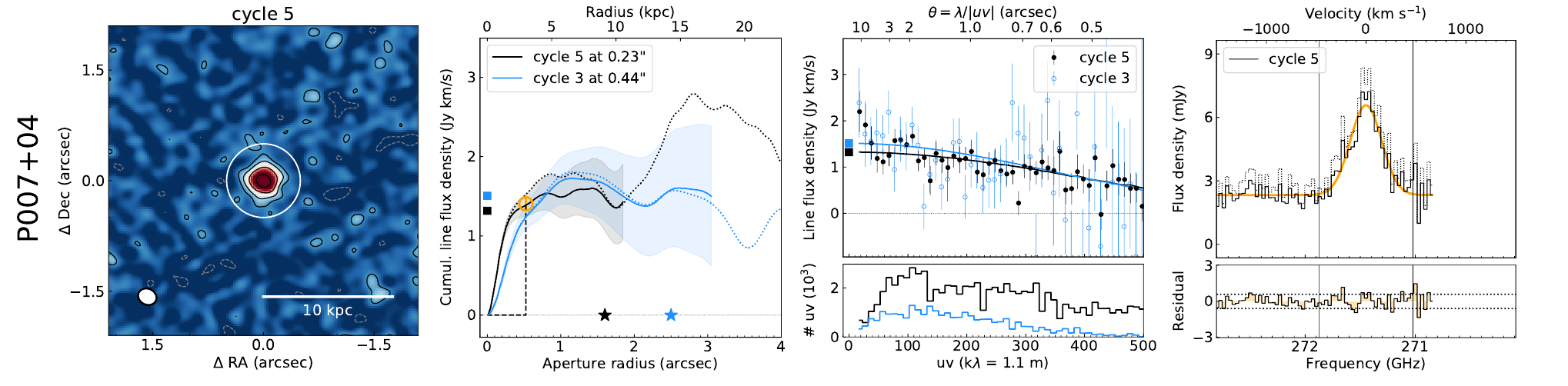}     \\
	\includegraphics[width=\linewidth]{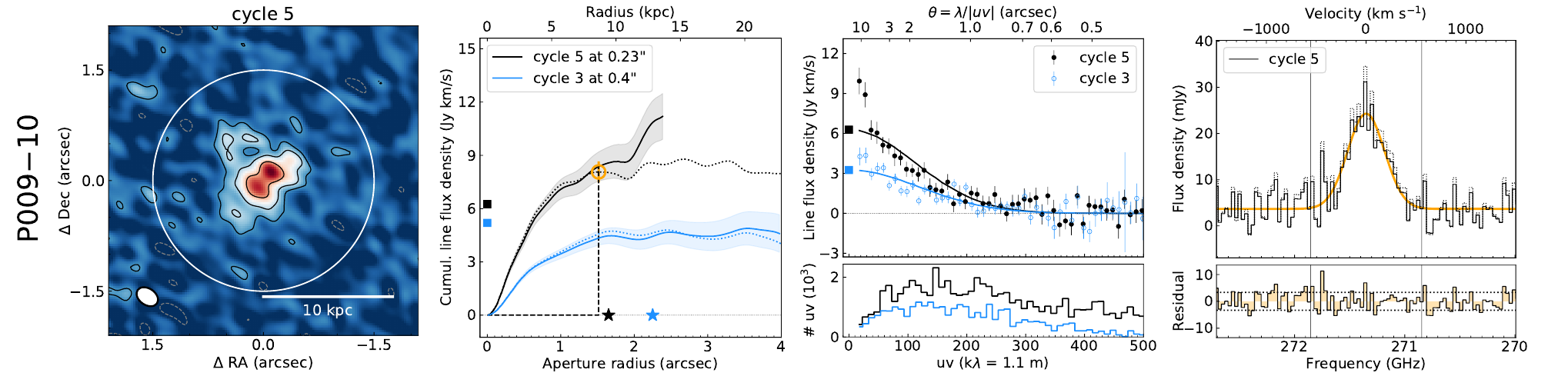}     
	\caption{ Emission line flux diagnostics of individual sources drawn from the full sample of 27 quasar host galaxies.
		{\it The first panel:} \cii\ intensity map imaged from the nominal high-resolution dataset using $1.2\times\mathrm{FWHM}$ of the line width. Contours are logarithmic starting from $2\sigma$ in powers of 2. Synthesized beam FWHM is shown in the corner. Manually chosen aperture used to extract the spectrum is shown with the white circle. 
		{\it The second panel:}  Curves of growth measured from the intensity maps of various cycles as indicated in the legend. Full lines show residual corrected fluxes, dotted lines show uncorrected values. Dashed line corresponds to the chosen aperture size. Stars on the horizontal zero line indicate the MRS of the dataset (color coded by cycle). Squares on the zero radius line are the same as in the third panel (color coded by cycle). The orange circle is the flux measured from the 1D Gaussian fit of the spectrum from the fourth panel multiplied by 0.84 to offset the missing tails from the intensity map collapse.
		{\it The third panel:} Flux measured in annuli of $uv$-distances of various cycles is shown with points. Solid line represents the model corresponding to a single 2D Gaussian fit obtained with the {\sc uvmodelfit} task. Squares at $uv=0$ show the total flux estimated from the fit. Histograms in the lower panel show the number of visibilities available in a given annulus after data averaging. The upper $x$-axis shows the resolution of a given baseline length.
		{\it The fourth panel:} Spectrum extracted from the manually chosen aperture (same as in the first panel). Full lines show residual corrected fluxes, dotted lines show uncorrected values. Orange line is a single Gaussian fit to the corrected spectrum, with the residual shown in the lower panel (also in units of mJy). Dotted horizontal lines outline $\pm1\sigma$ variations in the residual. Vertical lines outline $\pm3\sigma$ of the line width to assist in the search of possible broad components.
}
	\label{fig:individual}
\end{figure*} 

\begin{figure*}\ContinuedFloat
\includegraphics[width=\linewidth]{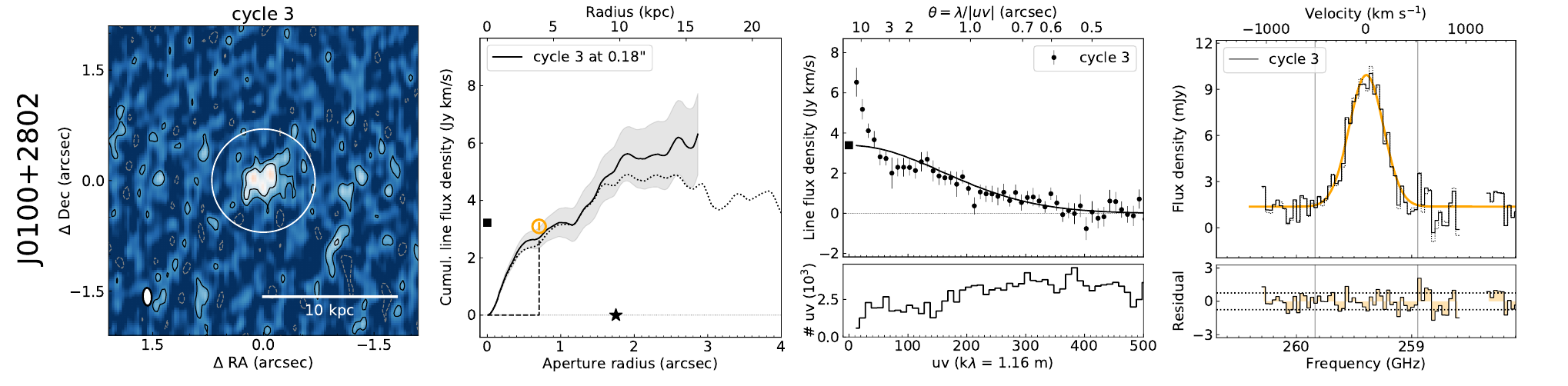}  \\
\includegraphics[width=\linewidth]{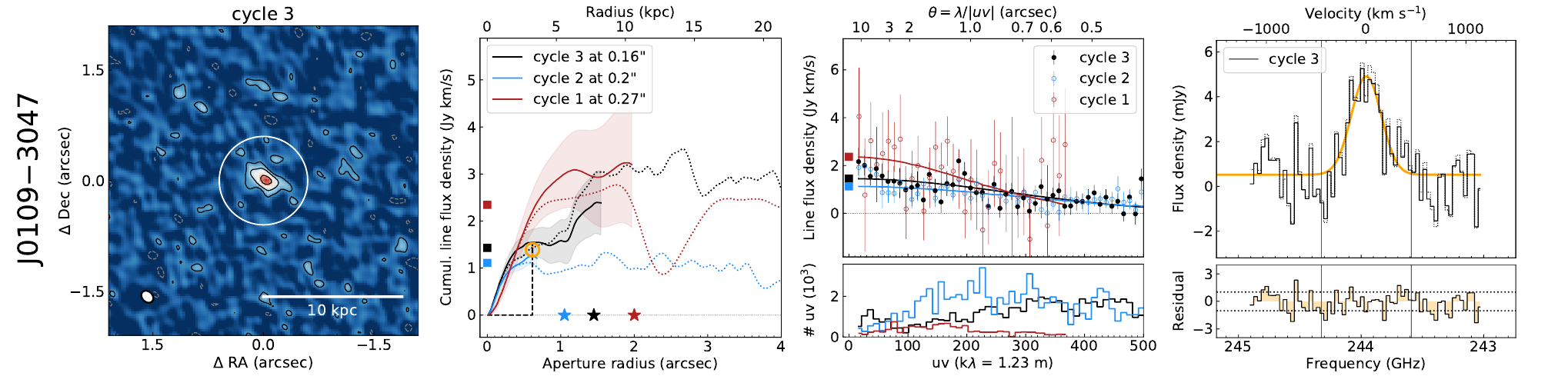} \\
\includegraphics[width=\linewidth]{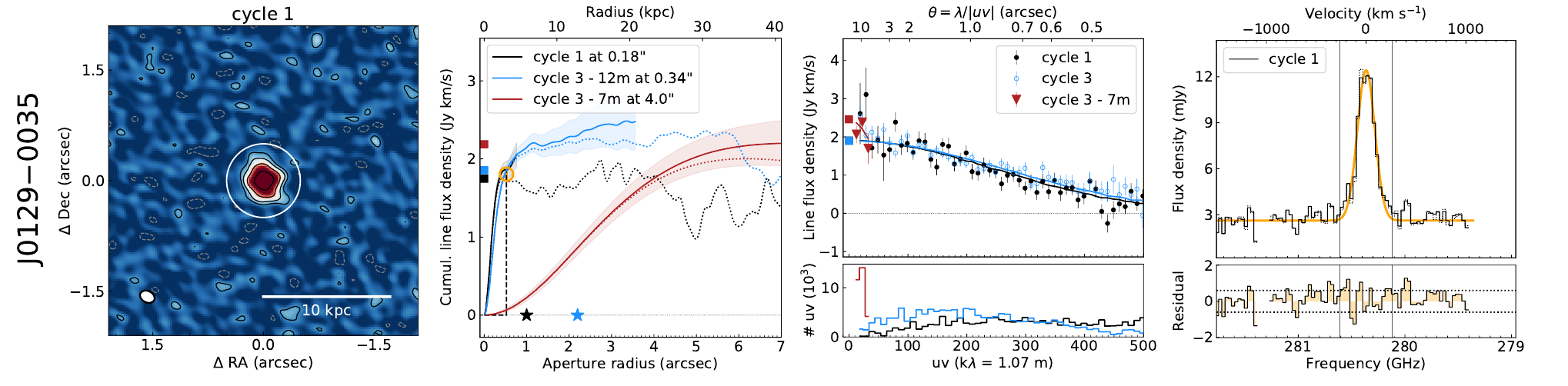}  \\
\includegraphics[width=\linewidth]{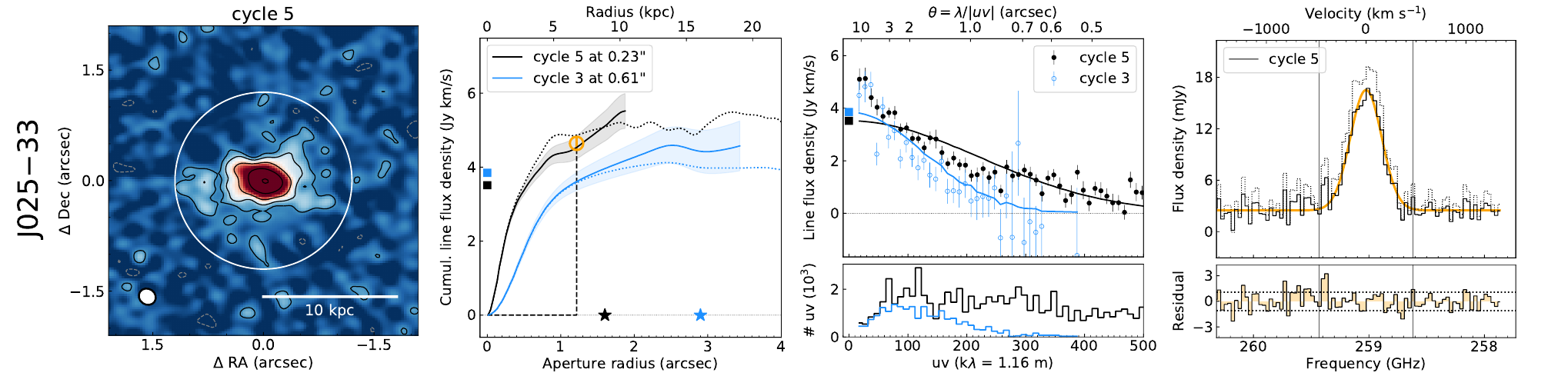}     \\
\includegraphics[width=\linewidth]{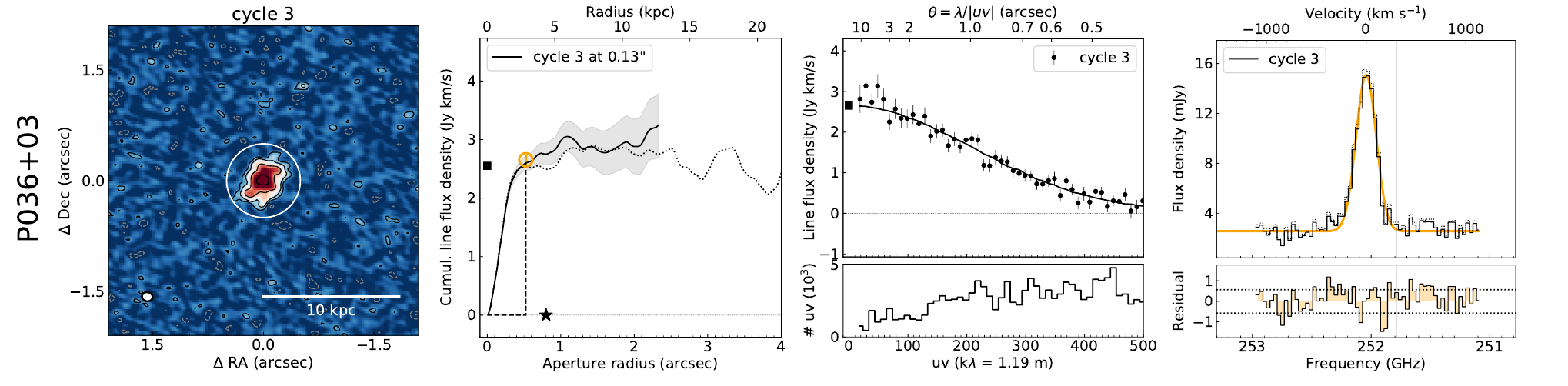}     
	\caption{Continued.}
\end{figure*} 

\begin{figure*}\ContinuedFloat
\includegraphics[width=\linewidth]{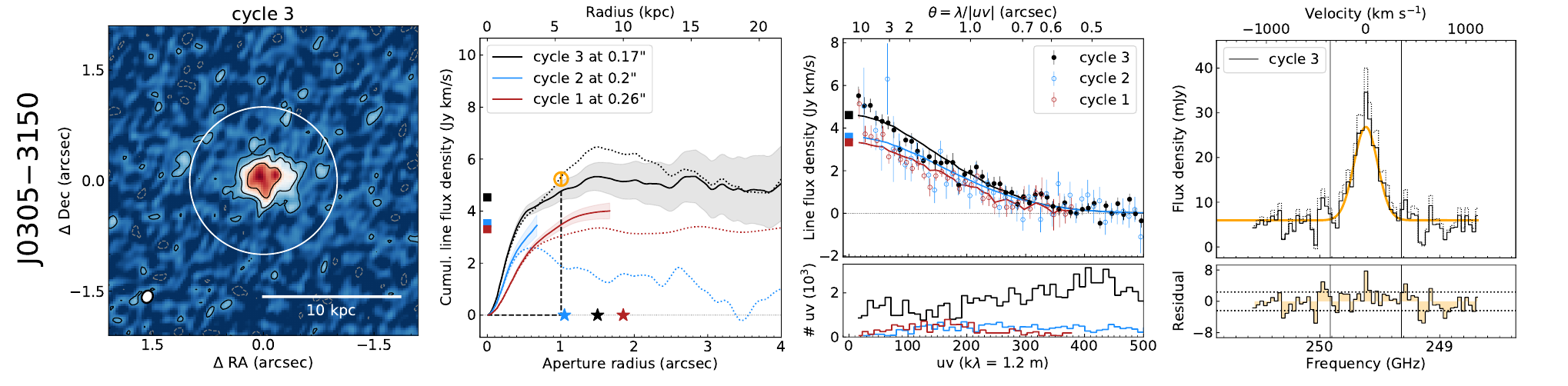}  \\
\includegraphics[width=\linewidth]{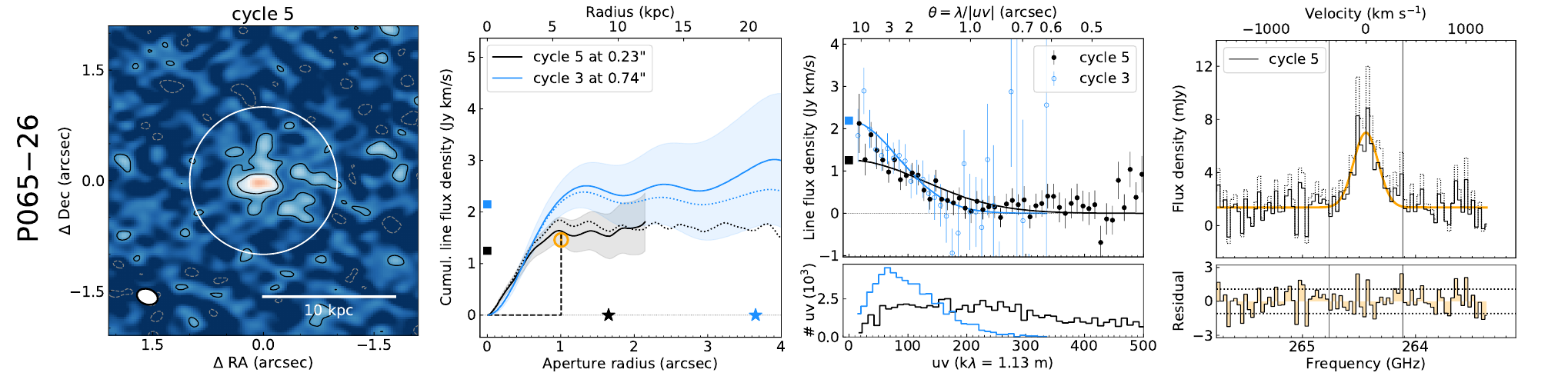}     \\
\includegraphics[width=\linewidth]{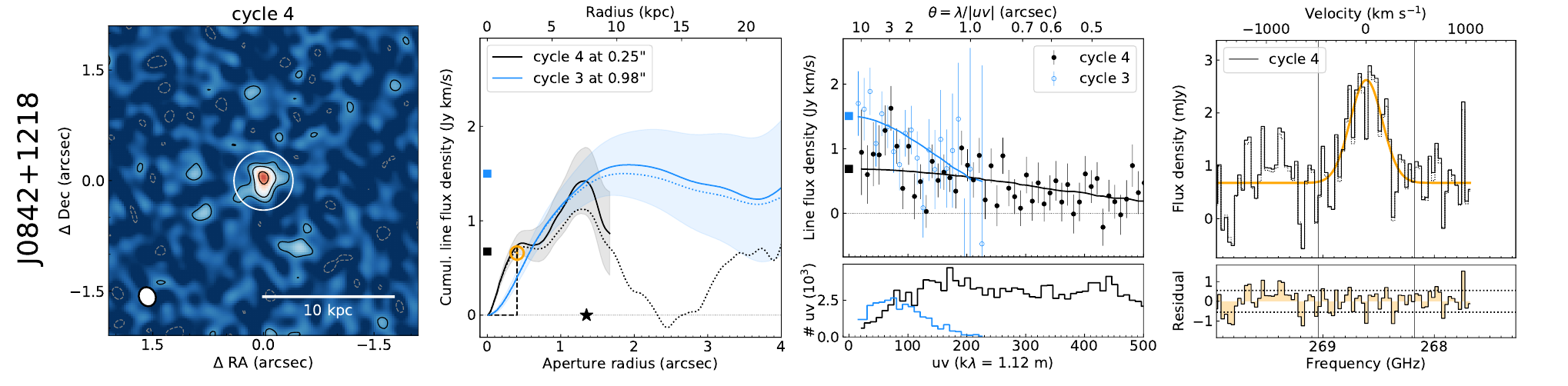}  \\
\includegraphics[width=\linewidth]{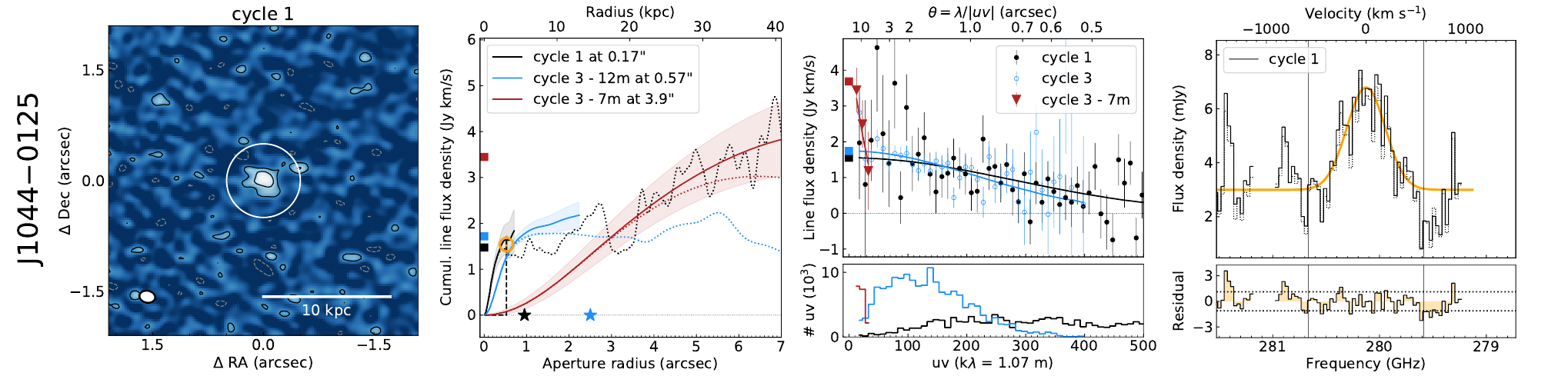} \\
\includegraphics[width=\linewidth]{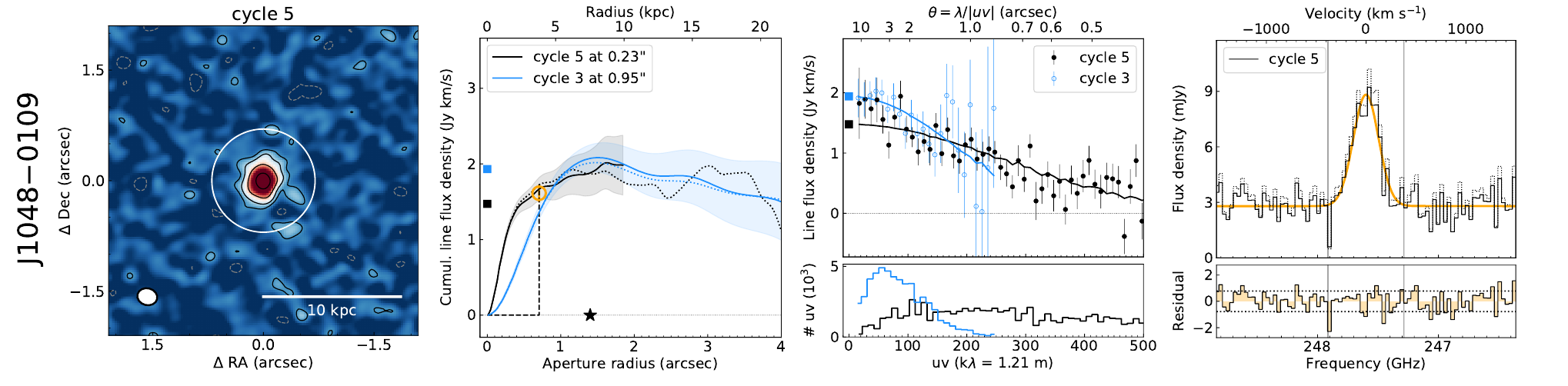}  
	\caption{Continued.}
\end{figure*} 

\begin{figure*}\ContinuedFloat
\includegraphics[width=\linewidth]{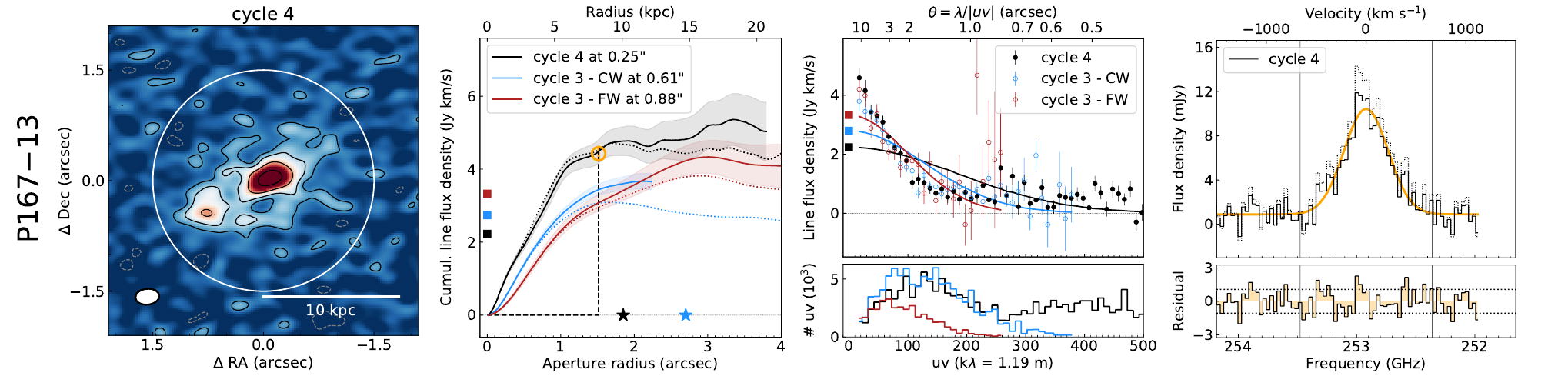}   \\
\includegraphics[width=\linewidth]{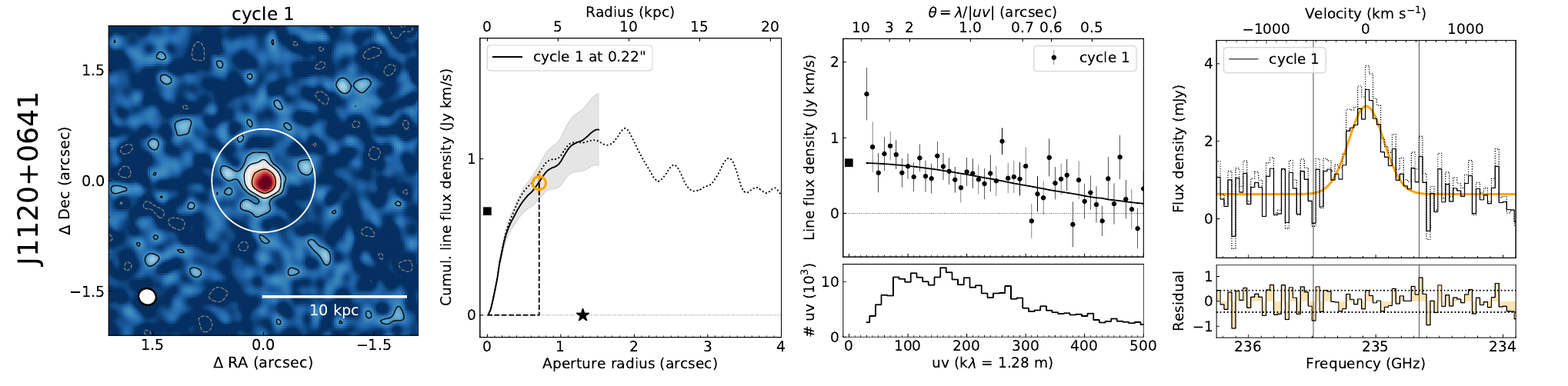}  \\
\includegraphics[width=\linewidth]{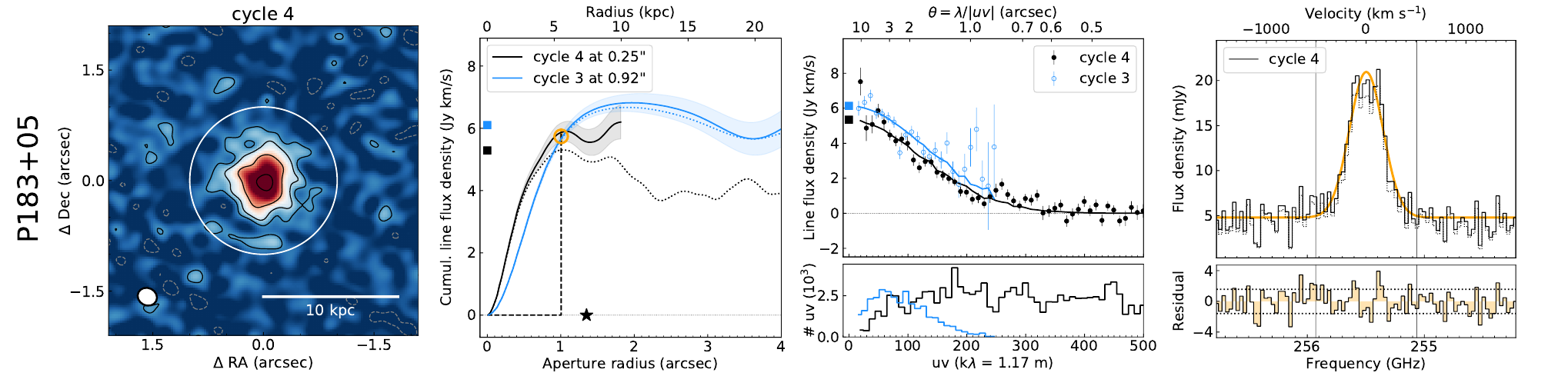}     \\
\includegraphics[width=\linewidth]{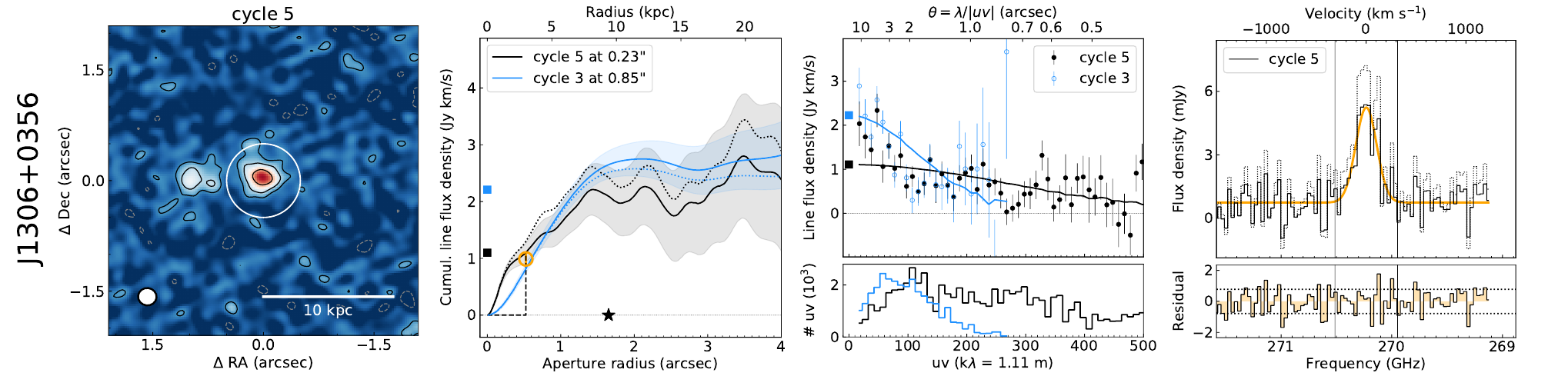}  \\
\includegraphics[width=\linewidth]{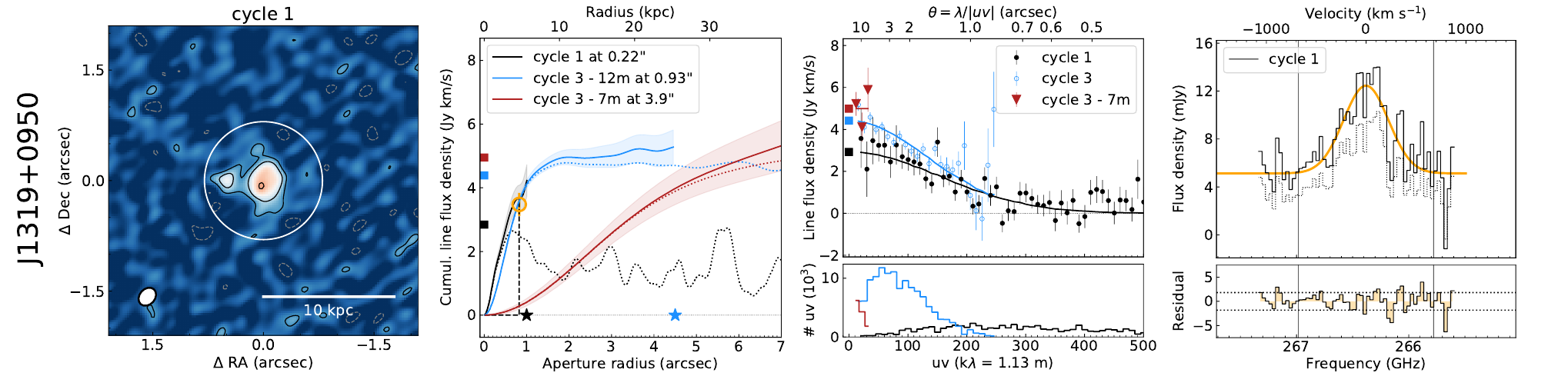}  
	\caption{continued.}
\end{figure*} 

\begin{figure*}\ContinuedFloat
\includegraphics[width=\linewidth]{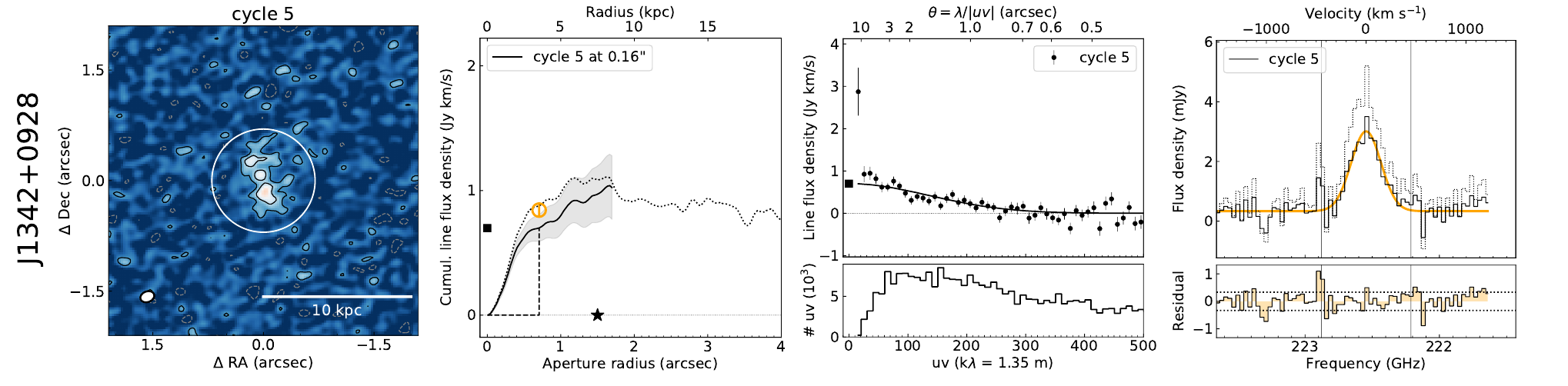} \\
\includegraphics[width=\linewidth]{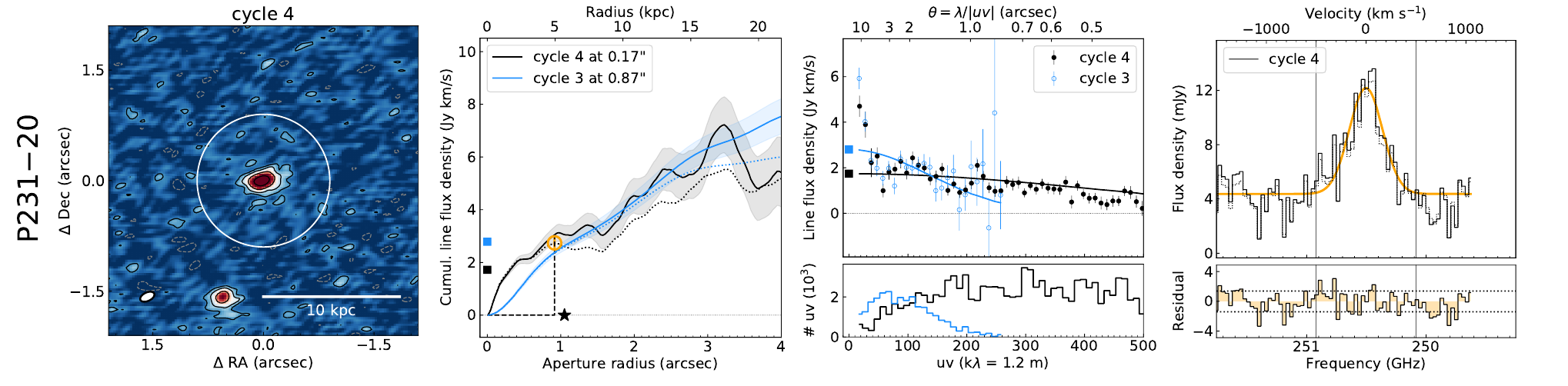}     \\
\includegraphics[width=\linewidth]{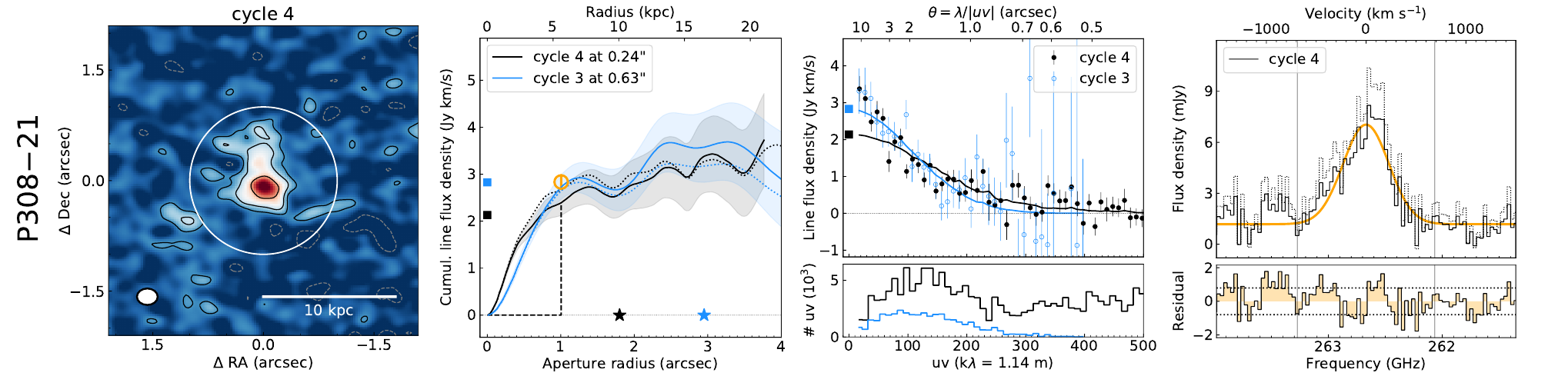}     \\
\includegraphics[width=\linewidth]{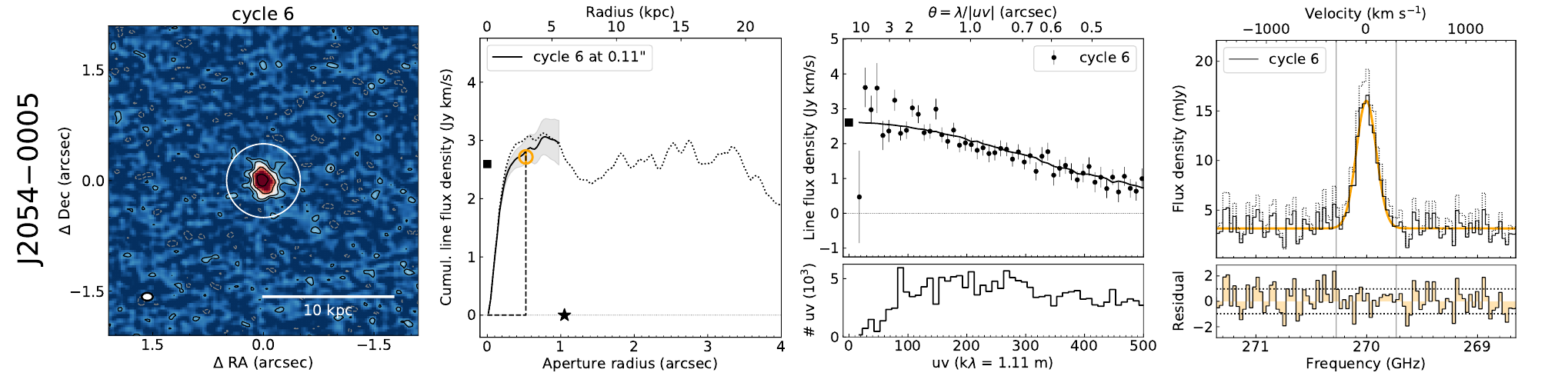}  \\
\includegraphics[width=\linewidth]{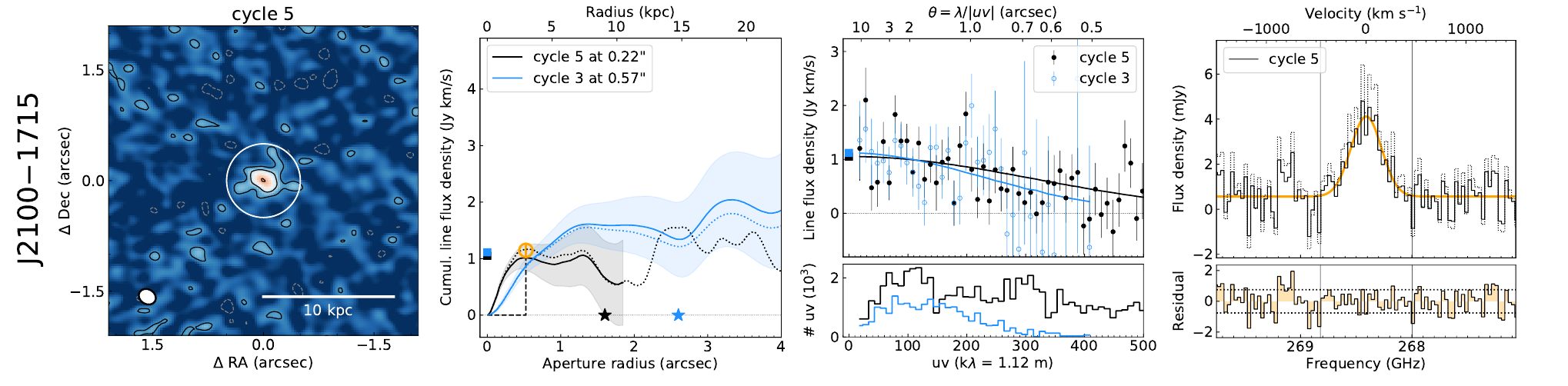}  
	\caption{Continued.}
\end{figure*} 

\begin{figure*}\ContinuedFloat
\includegraphics[width=\linewidth]{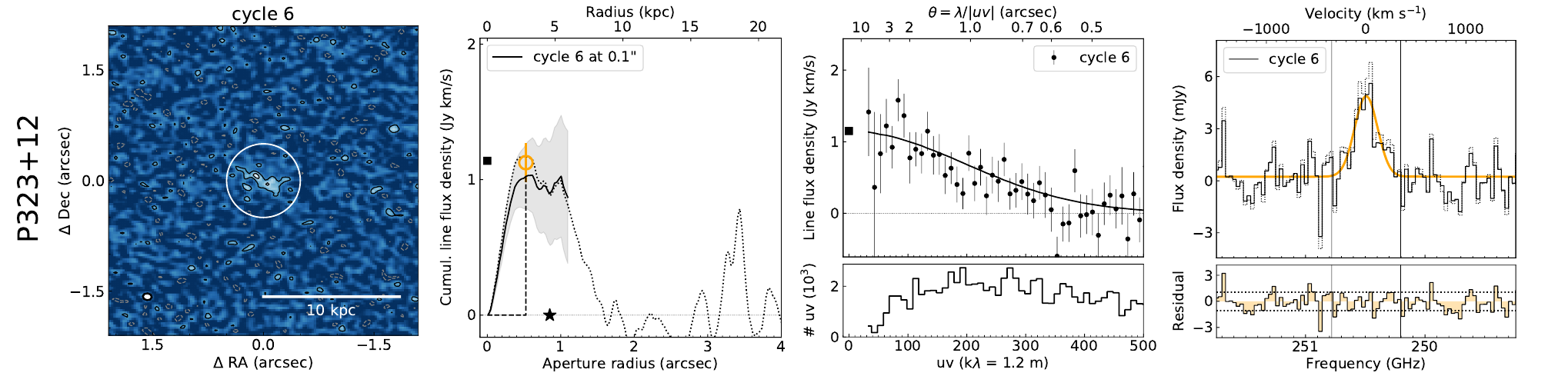}     \\
\includegraphics[width=\linewidth]{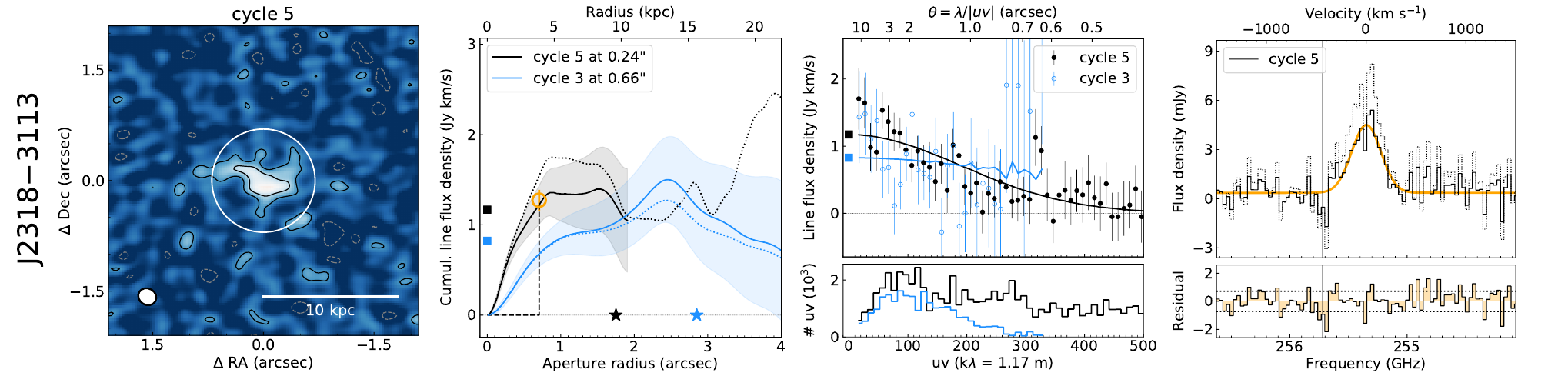}  \\
\includegraphics[width=\linewidth]{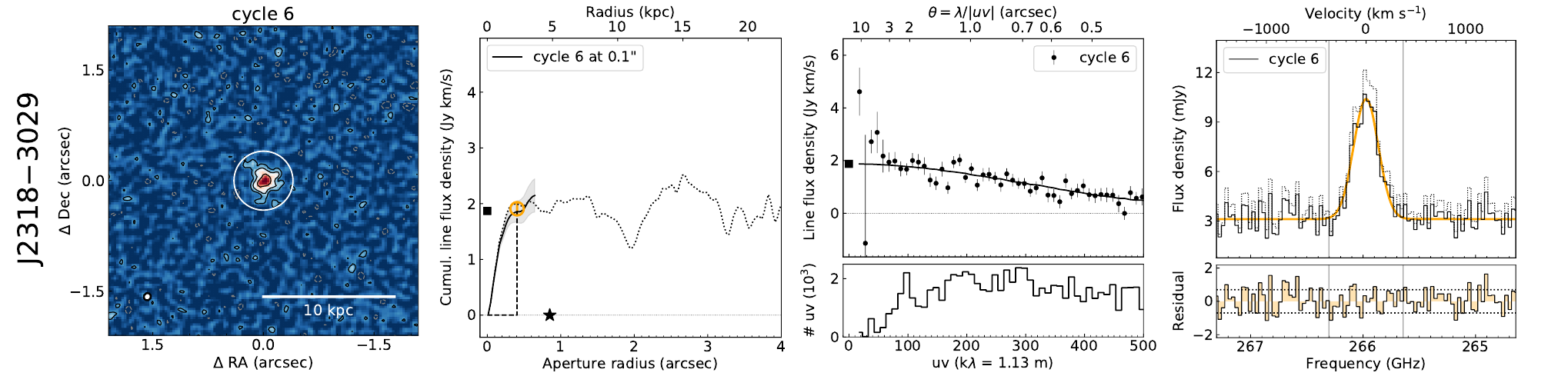}  \\
\includegraphics[width=\linewidth]{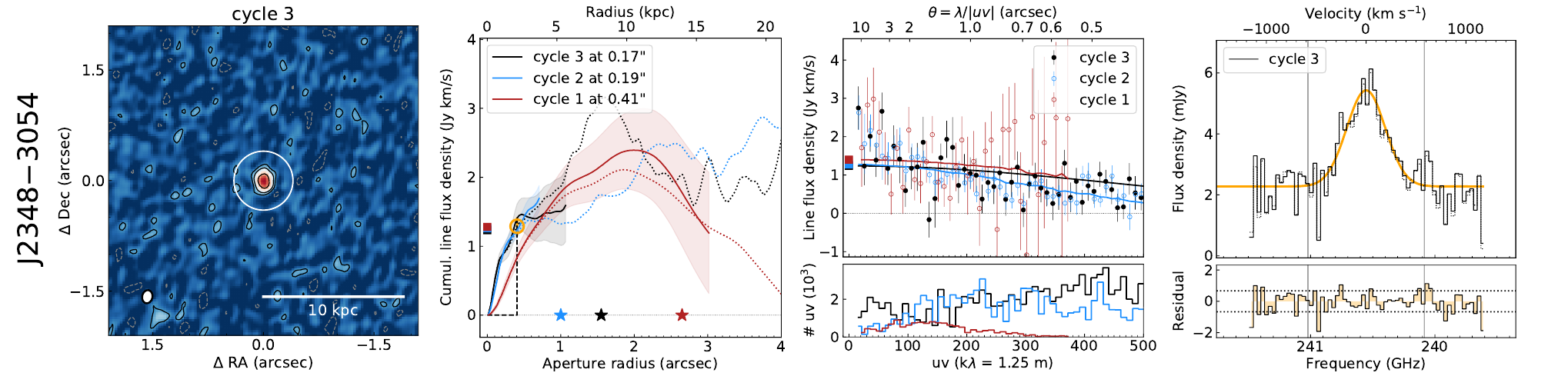} \\
\includegraphics[width=\linewidth]{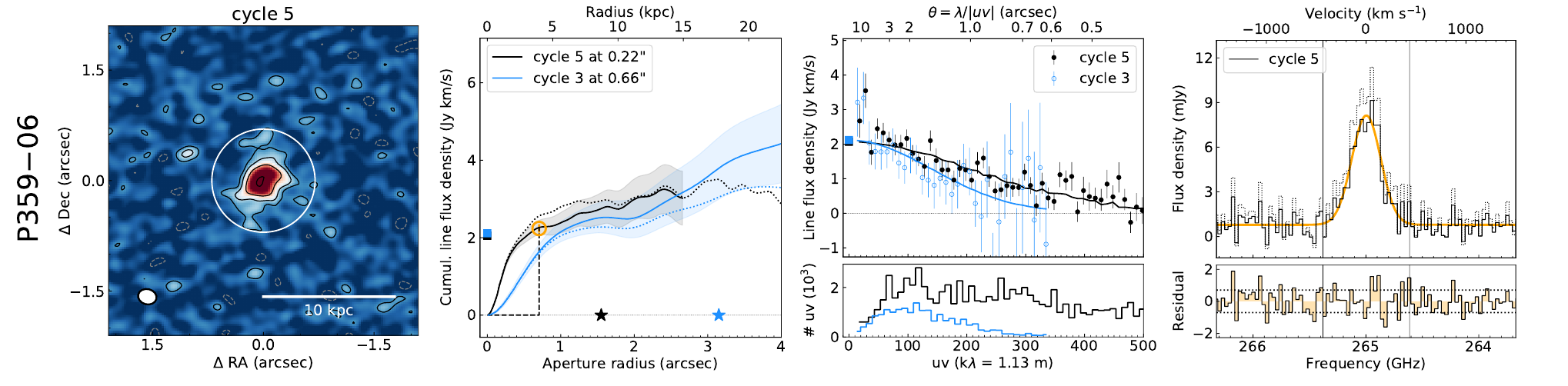}  

	\caption{Continued.}
\end{figure*} 

\end{document}